\documentclass[12pt]{article}
\usepackage[pdftex]{graphicx}
\usepackage{epstopdf}
\usepackage{amsmath}
\usepackage{amssymb}
\usepackage{authblk}
\usepackage{mathrsfs}
\usepackage{slashed}
\usepackage{subcaption}
\usepackage{tikz-cd}
\usepackage{blkarray}
\usepackage{enumerate}
\usepackage{cite}

\makeatletter
 
  \@addtoreset{equation}{section}
 \makeatother

\newcommand{\be}{\begin{equation}}
\newcommand{\ee}{\end{equation}}
\newcommand{\bea}{\begin{eqnarray}}
\newcommand{\eea}{\end{eqnarray}}
\newcommand{\beann}{\begin{eqnarray*}}
\newcommand{\eeann}{\end{eqnarray*}}

\newcommand{\ba}{\begin{array}}
\newcommand{\ea}{\end{array}}

\DeclareMathOperator{\Tr}{Tr}

\DeclareMathOperator{\diag}{diag}
\DeclareMathOperator{\Vol}{Vol}
\DeclareMathOperator{\Sdet}{Sdet}
\DeclareMathOperator{\Res}{Res}

\newcommand{\e}{\epsilon}

\newcommand{\M}{{\cal M}}
\newcommand{\zb}{{\bar{z}}}

\newcommand{\A}{{\cal A}}

\newcommand{\del}{\partial}

\newcommand{\C}{\mathbb{C}}
\newcommand{\Z}{\mathbb{Z}}
\newcommand{\R}{\mathbb{R}}

\newcommand{\Phib}{\bar{\Phi}}
\newcommand{\phib}{\bar{\phi}}
\newcommand{\Hb}{\bar{H}}

\newcommand{\psib}{\bar{\psi}}
\newcommand{\Tb}{\bar{T}}
\newcommand{\rhob}{\bar{\rho}}
\newcommand{\delb}{\bar{\del}}

\newcommand{\nub}{\bar{\nu}}
\newcommand{\Ab}{\bar{A}}
\newcommand{\ab}{\bar{a}}
\newcommand{\lambdab}{\bar{\lambda}}
\newcommand{\cb}{\bar{c}}
\newcommand{\ws}{\wedge{*}}

\makeatletter
\newcommand\mathcircled[1]{%
  \mathpalette\@mathcircled{#1}%
}
\newcommand\@mathcircled[2]{%
  \tikz[baseline=(math.base)] \node[draw,circle,inner sep=1pt] (math) {$\m@th#1#2$};%
}
\makeatother

\title{The Volume of the Quiver Vortex Moduli Space}
\author[1]{Kazutoshi Ohta\thanks{kohta@law.meijigakuin.ac.jp}}
\author[2]{Norisuke Sakai\thanks{norisuke.sakai@gmail.com}}
\affil[1]{\small\it Institute of Physics, Meiji Gakuin University, Yokohama, Kanagawa 244-8539, Japan}
\affil[2]{\small\it Department of Physics, and Research and
Education Center for Natural Sciences,
Keio University, 4-1-1 Hiyoshi, Yokohama, Kanagawa 223-8521, Japan,
and
iTHEMS, RIKEN,
2-1 Hirasawa, Wako, Saitama 351-0198, Japan}
\date{}							

\begin{document}
\maketitle


\begin{center}
{\bf Abstract}
\end{center}

We study the moduli space volume of BPS vortices in 
quiver gauge theories on compact Riemann surfaces.
The existence of BPS vortices imposes constraints on the 
quiver gauge theories.
We show that the moduli space volume is given by a vev 
of a suitable cohomological operator (volume operator) in a 
supersymmetric quiver gauge theory, where BPS equations of the 
vortices are embedded.
In the supersymmetric gauge theory, the moduli space volume 
is exactly evaluated as a contour integral by using the 
localization.
Graph theory is useful to construct the supersymmetric quiver 
gauge theory and to derive the volume formula.
The contour integral formula of the volume (generalization of 
the Jeffrey-Kirwan residue formula) leads to the Bradlow bounds 
(
upper bounds on the vorticity 
by the area of the Riemann surface divided by the 
intrinsic size of the vortex). 
We give some examples of various quiver gauge theories and discuss 
properties of the moduli space volume in these theories.
Our formula are applied to the volume of the vortex moduli space 
in the gauged non-linear sigma model with 
$\C P^N$ target space, which is obtained by a strong coupling limit 
of a parent quiver gauge theory.
We also discuss a non-Abelian generalization of the quiver gauge 
theory and ``Abelianization'' of the volume formula.

\newpage

\section{Introduction}

Vortices are co-dimension two solitons and play an important 
role for non-perturbative effects in gauge theories. 
In particular, the Bogomol'nyi-Prasad-Sommerfield (BPS) vortices 
appear as solutions to the BPS differential equations 
\cite{Bogomolny:1975de,Prasad:1975kr} which minimize the energy 
of the Yang-Mills-Higgs system in three spacetime dimensions.

If the vortex equations are considered on compact 
Riemann surfaces $\Sigma_h$ with the genus $h$, the number of 
the vortices (vorticity) is restricted by an upper bound 
which is given by the finite area of $\Sigma_h$ divided by 
the intrinsic size of the vortex. 
This bound is called the Bradlow bound \cite{Bradlow:1990ir,Manton:2010sa}.

Parameters of the vortex solutions are called moduli, 
and thier space is called the moduli space.
(See for review \cite{MantonSutcliffe,Eto:2006pg}.)
The structure of the moduli space is important to understand 
properties of the vortices themselves.
The volume of the moduli space appears in the thermodynamics of 
the vortices \cite{MantonSutcliffe,Manton:1993tt,Manton:1998kq,Eto:2007aw}.
Since the thermodynamical partition function is proportional to 
the volume of the vortex moduli space, 
we can derive the free energy or equation of state from the volume. 
Although an integration of the volume form on the moduli space 
should give the volume of the moduli space, it is generally difficult 
to know the geometry of the moduli space, including the K\"ahler 
metric, except for some special cases \cite{Fujimori:2010fk}.
(See also \cite{MantonSutcliffe} for details.)
On the other hand, the volume of the moduli space can be evaluated 
exactly without a detailed knowledge of the metric.

There are various way to obtain the volume of the moduli space.
One way is to take advantage of the property of the moduli space 
as a K\"ahler manifold \cite{Manton:1998kq,MantonSutcliffe}.
To evaluate the volume by using the properties of the K\"ahler 
manifold, we need to know a topological structure of the moduli 
space like cohomologies or boundary divisors.
The other way is to embed the BPS equation into supersymmetric 
gauge theory and to utilize the ``localization'' 
\cite{Moore:1997dj,Gerasimov:2006zt,Miyake:2011yr,Miyake:2011fq,Ohta:2018leq}.
The localization method gives the volume as simple contour integrals 
even without knowing the geometry 
of the moduli space.
In this sense, the localization method is universal and can be 
applied to any kind of the BPS equations in principle.
In previous works \cite{Miyake:2011yr,Miyake:2011fq,Ohta:2018leq},
the volume of the moduli space of the vortex with a single 
$U(N_c)$ gauge group and $N_f$ matters in the fundamental 
representation has been evaluated. 

There have been a number of studies of vortices in gauge theories 
on curved manifolds, namely gauged nonlinear sigma models (GNLSM) 
\cite{Yang:1998qca,Baptista:2004rk,Baptista:2007ap,Baptista:2008ex,
Baptista:2010rv,Romao:2018egg
}. 
The GNLSM can be obtained if one considers a product of two 
gauge groups and matters charged under both of these gauge groups 
and takes a strong coupling limit of one of the gauge groups. 
Before taking the limit, we have linear gauged sigma model with a 
product of gauge groups and can be considered as a parent theory 
of GNLSM. 
Vortices in such theories with product gauge groups have also 
been studied before \cite{Schroers:1996zy}. 
Gauge theories with a product of gauge groups and matters in 
bi-fundamental representations between two gauge groups are 
called quiver gauge theories.
If we take a decoupling limit of a gauge group in the quiver 
gauge theory, where a gauge coupling constant goes to zero, 
the decoupled gauge group behaves as a global symmetry for the 
matters.
So we can obtain the matters in the fundamental representation 
from the quiver gauge theory. 
Thus, the quiver gauge theory includes various types of the gauge 
theory in a very general form.
Once a general formula for the volume of the vortex moduli space 
for the quiver gauge theory is derived, the BPS vortex equations 
with various kinds of matters or target space can be obtained. 
This is a strong motivation to consider the quiver gauge theory.

The quiver gauge theory can be realized by using the graph theory.
The BPS vortices in the supersymmetric gauge theory on the graph
has been studied in \cite{Kan:2009tu}
inspired by the ``deconstruction''.
In addition, the quiver gauge theory naturally appears in the 
D-brane system of superstring theory.
Open strings between D-branes give the gauge fields and bi-fundamental 
matters in the quiver gauge theory.
Some of the quiver gauge theories can be realized as an effective 
theory on the D-branes at a tip of an orbifold.
The volume of the vortex moduli space of quiver gauge theory can 
play an important role for non-perturbative effects in superstring theory.

The purpose of our paper is to obtain a formula for the 
volume of the moduli space of BPS vortices in quiver gauge theories 
on compact Riemann surfaces.
We find that the graph theory in mathematical literature is 
useful to describe the quiver gauge theory.
The gauge groups and bi-fundamental matters are expressed 
in terms of a directed graph (quiver diagram), which consists 
of the vertices and arrows (edges) connecting between the vertices.
Each vertex represents a factor of the product gauge group, 
and each arrow gives the bi-fundamental matter, which transforms as
fundamental and anti-fundamental representation for the gauge group
at the source and target vertex of the arrow, respectively. 
Connection of vertices by edges in the graph is represented 
by a matrix called incidence matrix, which appears frequently 
in our construction of volume formulas for BPS vortices. 
Because of a zero left eigenvector of incidence matrix in a generic 
quiver gauge theories, the existence of BPS vortices imposes a 
stringent constraint on possible quiver gauge theories. 
We find two alternative solutions to the constraint. 
(i) All gauge groups have a common gauge coupling (universal coupling 
case). 
(ii) There is a gauge group whose gauge coupling vanishes (decoupled 
vertex case).

Embedding the vortex system into the supersymmetric quiver 
gauge theory, we define the supersymmetric transformation for the 
fields by a supercharge $Q$. 
Vacuum expectation values (vevs) of the cohomological operators, 
which is $Q$-closed but not $Q$-exact, is independent of the gauge 
coupling constants of the supersymmetric quiver gauge theory, 
since the action is $Q$-exact.
So we can control the gauge coupling constants of the supersymmetric 
gauge theory without changing the vevs of the
cohomological  operators. 
If we take the controllable gauge coupling constants to the same 
value as the physical coupling constants in the BPS equations to 
evaluate the volume of the moduli space,
the path integral is localized at the solution to the BPS equations.
At the fixed points of the BPS solution, the matter (Higgs) fields 
take non-trivial value and the supersymmetric quiver gauge theory 
is in the Higgs branch.
In the Higgs branch, we can show that the volume of the vortex 
moduli space is given by the vev of a suitable cohomological 
operator called the volume operator.

On the other hand, if we tune the controllable coupling constants 
to special values, the vevs of the Higgs fields vanish at the fixed 
points and the supersymmetric quiver gauge theory is in the Coulomb branch.
In the Coulomb branch, the evaluation of the vev of the volume 
operator reduces to simple contour integrals.
Using the coupling independence of the vev, we expect that the 
contour integrals also give the volume of the vortex moduli space 
as well as in the Higgs branch.
Thus we obtain the contour integral formula of the volume of 
the vortex moduli space in the quiver gauge theory.
As concrete examples to apply the contour integral formula 
for the volume of the vortex moduli space, we consider various 
quiver gauge theory with Abelian vertices.
We discuss the Abelian quiver gauge theory with two or three vertices. 
For the integral to converge, we need a suitable choice 
of contours, which reproduces exactly the Bradlow bounds.
The derivation of the Bradlow bounds from the contour integral 
can be regarded as a generalization of the Jeffrey-Kirwan (JK)
residue formula \cite{Jeffrey-Kirwan}.
A similar connection between the Bradlow bounds and the JK residue formula
is also considered and utilized in the calculation of the index on $S^1\times \Sigma_h$ \cite{Bullimore:2019qnt,Bullimore:2020nhv}.
In some examples of the quiver gauge theory with multiple Abelian vertices,
the moduli space becomes non-compact.
So we need to introduce regularization parameters, which can be regarded as the twisted mass of the matters.
After taking zero limits of the regularization parameters,
we can see the divergences of the volume of the moduli space
corresponding to the non-compactness of the moduli space.

We also apply the contour integral formula of the volume to 
a quiver gauge theory corresponding to the parent 
gauged {\it linear} sigma model of Abelian GNLSM with $\C P^N$ 
target space with $n$ flavors of charge scalar fields. 
When restricted to $N=n=1$, our result agrees with the previous 
result in \cite{Romao:2018egg}, which uses an entirely different 
method. 
We can also take a strong coupling limit of one of the gauge 
couplings, which gives the volume of the vortex moduli space 
of the GNLSM. 
Moreover, our contour integral formula provides a new results 
for the moduli space volume of the BPS vortex in the GNLSM with 
the target space $\C P^N$ and its parent GLSM with an arbitrary 
number $n$ of charged scalar fields.

The localization method can be extended to the case of 
non-Abelian quiver gauge theories. 
Since the non-Abelian gauge groups reduce to a product of $U(1)$'s 
in the Coulomb branch, the contour integral is expressed in 
terms of the Cartan part of the non-Abelian gauge groups, and 
the non-Abelian vertices in the quiver graph decompose into 
Abelian vertices. 
This ``Abelianization'' \cite{Blau:1994et,Blau:1995rs}
occurs in the localization formula and the 
quiver graph, because of the decomposition of the non-Abelian 
vertices into Abelian vertices in the quiver graph. 
Even with the Abelianization in the localization formula, 
the explicit evaluation of the contour integral becomes complicated 
due to the Vandermonde determinant which characterizes the 
non-Abelian case.
However, our formula gives in principle the volume of the vortex 
moduli space in any non-Abelian quiver gauge theories.
The non-Abelian generalization of the volume of the vortex moduli 
space in the GNLSM is also discussed.

The organization of this paper is as follows. 
In Sect.~\ref{BPS Vortex in Quiver Gauge Theory}, basics of the 
quiver gauge theory and graph theory are explained, and the BPS 
vortex equations are derived.
In Sect.~\ref{Embedding into Supersymmetric Quiver Gauge Theory}, 
the BPS vortex system is embedded into a supersymmetric quiver 
gauge theory, and the volume operator (a cohomological operator) 
is introduced to obtain the volume of the vortex moduli space 
in the Higgs branch.
The contour integral formula for the moduli space volume 
is obtained from localization in the Coulomb branch.
In Sect.~\ref{Volume of the Quiver Vortex Moduli Space},
we give various examples of the quiver gauge theory up to three 
Abelian vertices. Moduli space volumes in Abelian quiver gauge 
theories are evaluated explicitly by performing the contour integral.
In Sect.~\ref{GNLSM},
the vortex moduli space of a gauged linear sigma model 
(the parent theory of GNLSM) is obtained. 
That of the GNLSM is also obtained by taking the strong coupling limit.
In Sect.~\ref{Non-Abelian Generalization},
the contour integral formula is generalized to the non-Abelian 
cases, and 
the ``Abelianization'' of the non-Abelian quiver diagram is found. 
The Sect.~\ref{Conclusion and Discussions} is devoted to 
conclusion and discussions. 
In Appendix~\ref{sc:moduli_spce_metric}, path-integral 
measure for moduli is discussed.

\section{BPS Vortex in Quiver Gauge Theory}
\label{BPS Vortex in Quiver Gauge Theory}

\subsection{Quiver diagram and graph theory}

\begin{figure}
\begin{center}
\includegraphics{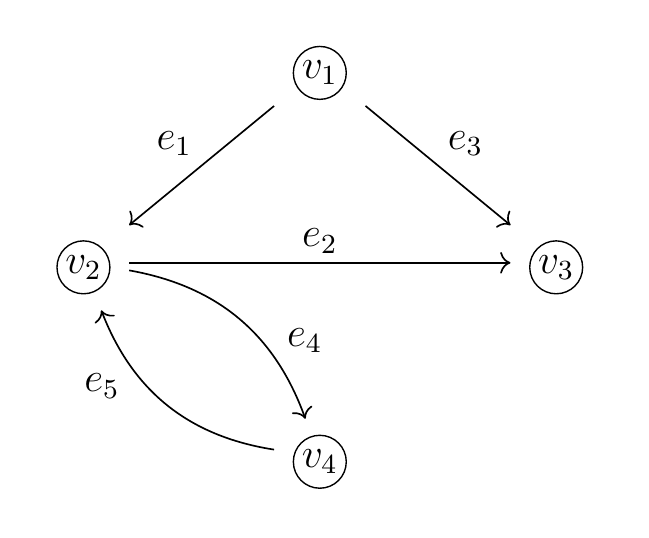}
\end{center}
\caption{An example of the quiver diagram.}
\label{quiver diagram}
\end{figure}

A quiver diagram is expressed by a directed graph $\Gamma=(V,E)$,
which consists of a set of vertices $V$ and a set of directed 
edges (arrows) $E$.
We denote elements of $V$ and $E$ by $v$ and $e$, respectively.
We depict an example of the quiver diagram in Fig.~\ref{quiver diagram}.

\begin{figure}
\begin{center}
\includegraphics{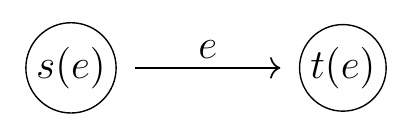}
\end{center}
\caption{A part of the quiver diagram of two vertices connected with an edge.
The source and target of the arrow (edge) $e$ are denoted by $s(e)$ and $t(e)$, respectively.}
\label{vertices and edge}
\end{figure}

We denote the total number of the vertices and edges by $n_E$ 
and $n_V$, respectively.
Each directed edge $e\in E$ connects from a source vertex $s(e)\in V$ 
to a target vertex $t(e)\in V$ (see Fig.~\ref{vertices and edge}), 
i.e.~each directed edge is specified by an ordered pair of two 
vertices $e=(s(e),t(e))$.

To describe the quiver diagram, it is useful to introduce language of graph theory.
The graph theory describes a structure of the (quiver) graph in terms of elements of matrices.
In graph theory, there are various kinds of the matrices or objects which describe and manipulate the graph
structure, but we here introduce some of them only, which will be used to construct the quiver gauge theory.

First of all, we introduce the incidence matrix. The incidence matrix $L$ maps from $V$ to $E$,
i.e.~$n_V \times n_E$ matrix, whose elements are defined by
\be
{L_v}^e = \begin{cases}
+1 & \text{if } s(e) = v \\
-1 & \text{if } t(e) = v\\
0 & \text{otherwise}
\end{cases}.
\ee
For example, if we make the incidence matrix for the 
graph depicted in Fig.~\ref{quiver diagram},
we obtain
\be
L=\left(
\begin{array}{rrrrr}
1 & 0 & 1 & 0 & 0\\
{-}1 & 1 & 0 & 1 & {-}1\\
0 & {-}1 & {-}1 & 0 & 0 \\
0 & 0 & 0 & {-}1 & 1
\end{array}
\right)
\ee
For generic quiver gauge theory, a sum of the elements in each 
column vanishes 
\be
\sum_v {L_v}^e = 0,
\label{eq:zero_vector}
\ee
namely the multiplication of a vector $(1, \cdots, 1)$ from 
the left annihilates the incidence matrix $L$. 
We will show that this will give a stringent constraint on quiver 
gauge theories admitting BPS vortices. 
Once the incidence matrix is given, 
we can reproduce the directed graph (quiver diagram).

If we assign variables $\vec{x}=(x_1,x_2,\ldots,x_{n_V})$ on 
each vertex, the incidence matrix multipied to the vector becomes 
a difference operator
\be
x^v {L_v}^e = x_{s(e)}-x_{t(e)},
\ee
where the repeated upper and lower indices are summed implicitly.
This property will be important to our formulation in the following.

Secondly, let us consider the Laplacian matrix $\Delta$ defined by
\be
\Delta_{vv'} = \begin{cases}
{\rm deg}(v) & \text{if } v = v' \\
-A_{vv'} & \text{if } v\neq v'
\end{cases},
\ee
where ${\rm deg}(v)$ represents the number of the edges which connect to
the vertex $v$ and $A_{vv'}$ is the number of edges from $v$ to $v'$.
($A_{vv'}$ is also called the adjacency matrix.)
The Laplacian matrix is also constructed from 
a square of the incidence matrix, i.e.~$\Delta \equiv LL^T$. 
Hence the Laplacian always has at least one zero eigenvalue 
with the eigenvector proportional to $(1, \cdots,1)$. 

Using the example in Fig.~\ref{quiver diagram}, the Laplacian matrix is given by
\be
\Delta=\left(
\begin{array}{rrrr}
2 & -1 & -1 & 0\\
-1 & 4 & -1 & -2\\
-1 & -1 & 2 & 0\\
0 & -2 & 0 & 2
\end{array}
\right).
\ee
We can notice that the Laplacian matrix is a generalization
of the Cartan matrix in the Lie algebra.

If we assign variables $\vec{x}=(x_1,x_2,\ldots,x_{n_V})$ on each vertex,
we can see that an inner product with the Laplacian matrix reduces
\be
\vec{x} \Delta \vec{x}^T  = \sum_{e\in E} (x_{s(e)}-x_{t(e)})^2,
\ee
which is a second order difference operator between vertices.
This is a reason why $\Delta$ is called the Laplacian on the graph.
The Laplacian matrix does not preserve the orientation of the edges.
So the Laplacian matrix cannot reproduce the whole structure of the quiver 
diagram including the orientation of the edges.

\subsection{Quiver gauge theory and vortices}

The quiver gauge theory is defined via a quiver diagram.
Unitary groups $U(N_v)$ are assigned to each vertex $v$, where $N_v$ is a rank of the unitary group.
The quiver gauge theory has a gauge symmetry of a product group $\prod_{v\in V}U(N_v)$ with the gauge couplings $g_v$.
The bi-fundamental matters (scalar fields) $H^e$ are associated with each edges $e$ and represented
by $({\bf N}_{s(e)},\bar{\bf N}_{t(e)})$, namely $N_{s(e)}\times N_{t(e)}$ complex matrices.

The vortex is a codimension two solitonic object, which appears as a classical static solution in $2+1$-dimensions.
The vortex solution minimizes the ``energy'' of the $2+1$-dimensional system.
To determine the energy,
we first consider $2+1$-dimensional quiver Yang-Mills-Higgs theory on $M_3=\R_t\times \Sigma_h$.
The metric on $M_3$ is given by
\be
ds_{2+1}^2 
=-dt^2 +
2g_{z\zb}dz \otimes d\zb.
\ee
On the Riemann surface $\Sigma_h$, 
there exists a volume form $\omega=\sqrt{g}dz\wedge d\zb$. 
An area of the Riemann surface $\A$ is given by an integral 
of the volume form 
\be
\A = \int_{\Sigma_h} \omega.
\ee

For each gauge vertex $v$,
there is gauge vector 1-form field
$A_{(3)}^v = A^v_0 dt + A^v_z dz + A^v_\zb d\zb$
on $M_3$.
The field strength on $M_3$ is given by
\be
F_{(3)}^v = d_{(3)} A_{(3)}^v + i A_{(3)}^v \wedge A_{(3)}^v,
\ee
where $d_{(3)}$ is the exterior derivative on three-dimensional manifold $M_3$.
On the other hand, on each edge $e$, we can assign a covariant derivative
of the scalar field $H^e$
\be
d^{(3)}_A H^e \equiv d_{(3)} H^e
+iA^{s(e)}H^e - i H^e A^{t(e)}.
\ee

The action is written in terms of the quiver diagram by
\be
\begin{split}
S_{(3)} &= -\int_{\R_t \times \Sigma_h}
\Bigg[
\sum_{v\in V}\Tr_{v}\bigg\{
\frac{1}{g_v^2}F^v_{(3)}\wedge *F^v_{(3)}\\
&\qquad\qquad\qquad\qquad
+\frac{g_v^2}{4}\Big(\zeta^v {\bf 1}_{N_v}-\sum_{e:\,s(e)=v}H^e\bar{H}^e+\sum_{e:\, t(e)=v}\bar{H}^e H^e\Big)^2
dt\wedge \omega
\bigg\}\\
&\qquad\qquad\qquad\qquad\qquad+\sum_{e\in E}
\Tr_{s(e)}d_A^{(3)} H^e \ws d_A^{(3)} \Hb^e
\Bigg],
\end{split}
\ee
where $\Tr_{v}$ stands for a trace over the rank $N_v$ gauge group at the vertex $v$, and the sum $\sum_{e:\,s(e)=v}$ ($\sum_{e:\,t(e)=v}$) is taken
over edges whose sources (targets) are given by $v$.

Taking a static configuration and  $A_0=0$ gauge,
the gauge vector field  reduces to (1,0)-form $A^v= A^v_zdz$
and (0,1)-form $\Ab^v=A_\zb d\zb$ on $\Sigma_h$,
where the field strength (magnetic field) is given by
\be
F^v = \del \Ab^v +\delb A^v+i (A^v\wedge \Ab^v+\Ab^v\wedge A^v),
\ee
where $\del$ and $\delb$ are the Dolbeault operators on $\Sigma_h$.
Introducing a ``metric''
\be
G_{vv'} =\frac{1}{g_v^2}\delta_{vv'},
\label{metric in original}
\quad
G_{ee'} =\delta_{ee'},
\ee
on $v$ and $e$, respectively, to raise and lower the indices,
the energy is given by
\be
\begin{split}
E&=\int_{\Sigma_h}
\Tr \Bigg[
\mu_v \wedge{*} \mu^v
+\frac{1}{2} \nu_e \ws \nub^e
 + g_v^2\zeta_v F^v \Bigg]\\
 &\geq 2\pi g_v^2\zeta_v  k^v ,
\end{split}
\label{Energy}
\ee
discarding the total divergence, where $\Tr$ is taken over suitable 
size of each term (gauge groups), and magnetic flux (first Chern class) 
for each gauge vertices is defined as
\be
\frac{1}{2\pi}\int_{\Sigma_h}\Tr F^v = k^v \in \Z.
\label{k flux}
\ee
We have introduced here
\bea
\mu^v &=& F^v -\frac{g_v^2}{2}\Big(\zeta^v {\bf 1}_{N_v}
- \sum_{e:\,s(e)=v} H^e\Hb^e + \sum_{e:\,t(e)=v} \Hb^e H^e\Big)
\omega,\label{original moment map 1}\\
\nu^e & =& 2\del_A \Hb^e,\label{original moment map 2}\\
\nub^{e} &=& 2\delb_A H^e,\label{original moment map 3}
\eea
where ${\bf 1}_{N_v}$ stands for a $N_v\times N_v$ unit matrix, and
\be
\begin{split}
\del_A \Hb^e & \equiv \del \Hb^e - i \Hb^e A^{s(e)} + i A^{t(e)} \Hb^e,\\
\delb_A H^e & \equiv \delb H^e + i \Ab^{s(e)} H^e - i H^e \Ab^{t(e)}.
\end{split}
\ee
The energy is saturated at a solution to the so-called BPS equations
\be
\mu^v = \nu^e = \nub^e =0.
\label{BPS equations}
\ee
We call the solution of the above differential equations on 
$\Sigma_h$ as the BPS vortex in the quiver gauge theory.
Once we get the BPS equations, which minimize the energy (\ref{Energy}),
the equations (\ref{BPS equations}) can be regarded just as
the differential equations on two-dimensional $\Sigma_h$ and disregard 
the time direction. 
If we are interested in the solution to eqs.~(\ref{BPS equations}) 
and in the parameters of these solutions (moduli space),
it is sufficient to consider the two-dimensional field theory on $\Sigma_h$ 
to embed the BPS equations.
From this point of view, we will embed the BPS equations into 
the two-dimensional supersymmetric gauge theory.

Provided $g_v^2\not=0$, we can take a linear combination of 
$\mu^v$ weighted by $1/g_v^2$ to obtain 
\be
0=\sum_{v \in V}\frac{\mu^v}{g_v^2} 
=\sum_{v \in V}\left( \frac{F^v}{g_v^2} -\frac{\zeta^v}{2}
 {\bf 1}_{N_v}
\omega\right),
\label{eq:constraint_quiver_moment-map}
\ee
because of the zero vector for incidence matrix in generic quiver 
gauge theory in Eq.~\eqref{eq:zero_vector}. 
If we take the trace and integral over $\Sigma_h$, 
we find
\be
\sum_{v\in V}\left( \frac{2\pi k^v}{g_v^2}
-N_v\frac{\zeta^v \A}{2}\right)=0. 
\label{constraint}
\ee
We will see that the integral formula for the volume of the 
moduli space gives a constraint identical to \eqref{constraint}. 
Since $k^v$ are integer valued, this condition 
cannot be satisfied for the generic $g_v$ and $\zeta^v$.
This means that the BPS vortices (solution) with $k^v\neq 0$ 
cannot exist on $\Sigma_h$ for the generic $g_v$ and $\zeta^v$.
In fact, Eq.~(\ref{constraint}) gives a stringent restriction 
nolt only on parameters such as $g_v^2$ and $\zeta_v$ of the 
theory and also on allowed vorticity $k^v$ of BPS vortices. 

First, the FI parameters $\zeta^v$ of the theory need to satisfy 
\be
\sum_{v\in V}N_v\zeta^v =0,
\label{constraint_zeta}
\ee
in order for vacuum ($k^v=0$ for all $v$) to exists. 

In order to allow BPS states with nonzero vorticity, 
two types of solutions are available 
\begin{enumerate}[(i)]
\item
Universal coupling: 

\be 
g_1=g_2=\cdots=g. 
\label{eq:common_coupling}
\ee
The local constraint \eqref{eq:constraint_quiver_moment-map} 
at each point in space reduces in this case to 
\be
\sum_{v \in V}F^v = 0. 
\label{constraint2}
\ee
Therefore the vorticity of $N_v-1$ gauge groups are no longer 
constrained. 
However, the gauge field of one of the gauge groups is completely 
determined (up to vacuum gauge field) by those of other gauge groups. 

As a more general solution with the universal coupling, we can 
consider the case $g_v^2=n_v g^2$ with $n_v \in {\mathbb Z}_+$. 
This solution allows not all but multiple of $n_v$ vorticity 
for each gauge group $v$. 

In sect.\ref{Volume of the Quiver Vortex Moduli Space} 
we consider the case of universal coupling. 

\item
Decoupled vertex:

Another solution for the quiver gauge theories admitting BPS vortices 
is the case when there is at least one decoupled vertex $g_{v'}=0$. 
The decoupled vertex gives only a global 
symmetry and no BPS condition arises for the $v'$ vertex. 
The incidence matrix no longer posseses a zero vector, 
and the constraint \eqref{eq:constraint_quiver_moment-map} 
is absent. 
Hence we can have arbitrary coupling and FI parameters for 
other gauge groups, provided there is a decoupled vertex in 
the quiver diagram. 
We will consider such a case in sect.\ref{GNLSM}. 
\end{enumerate}


\section{Embedding into Supersymmetric Quiver Gauge Theory}
\label{Embedding into Supersymmetric Quiver Gauge Theory}

We would like to consider the volume of the moduli space of the 
quiver BPS vortices, which are the solutions to the equations 
(\ref{BPS equations}).
It is useful to embed the system of the quiver BPS vortices 
into a supersymmetric quiver gauge theory, whose partition function 
is localized at the BPS solution.

The BPS vortex solution to the quiver BPS equations 
(\ref{BPS equations}) involves the given gauge coupling constants 
$g_v$ as parameters. 
On the other hand, the embeded supersymmetric gauge theory has 
a gauge coupling constants $g_{0,v}$.
The coupling constants $g_{0,v}$ appear as overall constants 
of the action of the supersymmetric quiver gauge theory and 
we will see that the partition function and vevs are independent 
of them, thanks to the localization theorem. 
Therefore we can choose $g_{0,v}$ differently from the 
``physical'' gauge coupling $g_{v}$ in the BPS vortex. 
We will find that the coupling constants $g_{0,v}$ in the 
supersymmetric quiver gauge theory are controllable parameters 
which interpolate between the Higgs and Coulomb branch picture. 

In the following subsections, we concentrate on the quiver gauge theory having only the Abelian vertices for a while,
since it is sufficient to see the localization theorem and derivation of the volume of the moduli space.
We will consider non-Abelian quiver gauge theories later. 
We will find that they can be treated by means of a decomposition 
of the non-Abelian vertices into the Abelian vertices.

\subsection{Abelian vertices}

Let us consider the supersymmetric quiver gauge theory
which contains the Abelian vertices only, i.e.~the total gauge group is $G=U(1)^{n_V}$.


On each vertex, there exist bosonic scalar fields $\phi^v$,
gauge vector fields $A^v$, $\Ab^v$ and auxiliary fields $Y^v$,
which are 0-forms, (1,0)-forms, (0,1)-forms
and 2-forms on $\Sigma_h$, respectively.
There also exist their superpartner fermions $\eta^v$,
$\lambda^v$, $\lambdab^v$ and $\chi^v$, which are
Grassmann-valued 0-forms, (1,0)-forms, (0,1)-forms and 2-forms on $\Sigma_h$,
respectively.
These bosons and fermions form vector multiplets of the Abelian
gauge theory on each vertex.

The supersymmetric transformations between the vector multiplets
are given by
\be
\begin{array}{lcl}
Q\phi^v = 0, &&\\
Q\phib^v = 2\eta^v, && Q\eta^v = 0,\\
QA^v = \lambda^v, && Q\lambda^v = -\del\phi^v,\\
Q\Ab^v = \lambdab^v, && Q\lambdab^v = -\delb\phi^v,\\
QY^v = 0, && Q\chi^v = Y^v,
\end{array}
\ee
where $\phib^v$ is a complex conjugate of $\phi^v$.
We note here that if we apply the $Q$ transformations twice 
on the fields, it generates a gauge transformation with a
gauge parameter $\phi^v$, i.e.~$Q^2=\delta_{\phi^v}$.

We also have chiral superfields on each edge. The chiral superfield consists
of a complex scalar field $H^e$ and its fermionic partner  $\psi^e$ of 0-form,
and an auxiliary field $\Tb^e$ and its fermionic partner $\rhob^e$ of (0,1)-form, on the edge $e$.
The chiral fields generally transform in the bi-fundamental representation, which
means that they possess positive charges of a gauge group at the source of the edge $s(e)$
and negative charges at the target $t(e)$, for the Abelian gauge theory.

For those chiral superfields,
the supersymmetric transformations are given by
\be
\begin{array}{lcl}
QH^e = \psi^e, && Q\psi^e = i\phi^{v}{L(H)_v}^e,\\
Q\Tb^e = i{\phi^{v}L(\rhob)_v}^e, && Q \rhob^e = \Tb^e,
\end{array}
\ee
which also satisfy $Q^2=\delta_{\phi^v}$.
Here ${L(H)_v}^e$ is defined from the incidence matrix ${L_v}^e$ as
\be
\begin{split}
{L(H)_v}^e&\equiv {L_v}^e H^e\\
&=
\begin{cases}
+H^e & \text{if } v=s(e)\\
-H^e & \text{if } v=t(e)\\
0 & \text{others}
\end{cases},
\end{split}
\ee
and similarly ${L(\rhob)_v}^e\equiv {L_v}^e \rhob^e$,
where we do not take the sum for the repeated index $e$.

For their complex conjugate fields,
which contain 0-form $(\Hb^e,\psib^e)$, and (1,0)-form $(T^e,\rho^e)$, we have
\be
\begin{array}{lcl}
Q\Hb^e = \psib^e, && Q\psib^e = -i{L^T(\Hb)^e}_v \phi^{v}  ,\\
QT^e = - i{L^T(\rho)^e}_v \phi^{v} , && Q \rho^e = T^e,
\end{array}
\ee
where we defined ${L^T(\Hb)^e}_v\equiv \Hb^e{{L^T}^e}_v$ and ${L^T(\rho)^e}_v\equiv \rho^e{{L^T}^e}_v$,
without summing over the repeated edge index $e$,
by using the transpose of the incidence matrix.

For later convenience, we introduce a norm between
forms $\alpha$ and $\beta$ on $\Sigma_h$
\be
\left\langle
\alpha,\beta
\right\rangle
\equiv
\int_{\Sigma_h}
\alpha \ws \bar{\beta}.
\ee
Using this norm, the action for the vector multiplets on $v\in V$ is written as a $Q$-exact form
\be
S_V = Q\Xi_V,
\label{action for vector}
\ee
where
\be
\Xi_V = - \left[
\langle
\lambda_v,
\del \phi^v
\rangle
+
\langle
\bar{\lambda}_v,
\bar{\del} \phi^v
\rangle
+
\langle
\chi_v ,Y^v-2\mu_0^v
\rangle
\right].
\ee
A part of the BPS equations appears in the action (\ref{action for vector});
\be
\mu_0^v = F^v -\frac{g_{0,v}^2}{2}\Big(\zeta^v
- \sum_{e:\,s(e)=v} H^e\Hb^e + \sum_{e:\,t(e)=v} \Hb^e H^e\Big)
\omega,\label{moment map 1-1}
\ee
which comes from the D-term constraint and is the same as $\mu^v$ in (\ref{original moment map 1})
for the original vortex system if we replace the coupling constants $g_{0,v}$ with $g_v$.
We however need to distinguish
between $\mu_0^v$
in the supersymmetric action
and $\mu^v$ in the original quiver vortex BPS equations (\ref{original moment map 1}),
since the solution includes the different coupling constants.

In the $Q$-exact action (\ref{action for vector}),
the repeated lower-upper indices are summed implicitly 
and the raising or lowering of the indices,
such as $\phi_v=G_{0,vv'}\phi^{v'}$,
is given by a ``metric'' on the
vertices
\be
G_{0,vv'} = \frac{1}{g_{0,v}^2}\delta_{vv'},
\ee
which contains the gauge couplings $g_{0,v}$
in contrast to the metric (\ref{metric in original}).
In this sense, the action (\ref{action for vector}) has the gauge couplings $g_{0,v}$
as an overall factor $1/g_{0,v}^2$.

Using the metric $G_{0,vv'}$, we can rewrite eq.~(\ref{moment map 1-1}) as
\be
\mu_0^v = F^v -\frac{1}{2}\Big(g_{0,v}^2\zeta^v
- {L(H)^v}_e \Hb^e
\Big)\omega,
\label{moment map 1-2}
\ee
where
${L(H)^v}_e
=G_0^{vv'}\delta_{ee'}{L(H)_{v'}}^{e'}
=g_{0,v}^2\delta^{vv'}\delta_{ee'}{L(H)_{v'}}^{e'}$,
so ${L(H)^v}_e$ contains the coupling constants $g_{0,v}$
unlike ${L(H)_v}^e$.

For the chiral superfields, we can construct a $Q$-exact action given by
\be
S_C =  Q\Xi_C,
\label{Q-exact chiral action}
\ee
where
\be
\Xi_C = \frac{1}{2}
\bigg[
\langle
\psi_e,i\phi^v{L(H)_v}^e
\rangle
-
\langle
\psib_e,
i{L^T(\Hb)^e}_v \phi^{v}
\rangle
-\frac{1}{2}\langle
\rho_e, T^e-2\nu^e
\rangle
-\frac{1}{2}\langle
\rhob_e,
\Tb^e-2\nub^e
\rangle
\bigg].
\ee
The constraints for the F-term conditions
appear in the chiral superfields action (\ref{Q-exact chiral action}) as;
\bea
\nu^e & =& 2\del_A \Hb^e,\label{moment map 2}\\
\nub^e &=& 2\delb_A H^e,\label{moment map 3}
\eea
where
\be
\begin{split}
\del_A \Hb^e &= \del \Hb^e-i{L^T(\Hb)^e}_v A^v,\\
\delb_A H^e &= \delb H^e+i\Ab^v{L(H)_v}^e,
\end{split}
\ee
for the Abelian theory.
Eqs.~(\ref{moment map 2}) and (\ref{moment map 3})
are the same as the original ones (\ref{original moment map 2}) and (\ref{original moment map 3})
since they do not depend on the gauge couplings.

The raising or lowering of the indices of the edge $e$ is just given by $\delta_{ee'}$, thus 
we can see the $Q$-exact action (\ref{Q-exact chiral action}) does not contain any coupling constant $g_{0,v}$.

The total supersymmetric action is given by the sum of the vector and chiral multiplet parts
\be
S =  S_V + S_C.
\ee
By definition, the total action is also written in a 
$Q$-exact form
\be
S=Q(\Xi_V+\Xi_C).
\ee
If we rescale the total action like
\be
S \to t S,
\ee
the partition function or the vev of the supersymmetric operator ${\cal O}$, which satisfies $Q{\cal O}=0$,
is independent of $t$, since the derivative with respect to $t$ reduces to the vev of the $Q$-exact operator
and vanishes, i.e.~
\be
\frac{\del }{\del t}\left\langle{\cal O}\right\rangle_t
=-\left\langle{\cal O}S\right\rangle_t=-\left\langle Q({\cal O}\Xi)\right\rangle_t=0,
\ee
where $\langle \cdots \rangle_t$ stands for the vev with the rescaled action $tS$.
Note that we also find by a similar argument
that the partition function or the vev of the supersymmetric
operator is independent of the gauge coupling constants $g^2_{0,v}$ in $S_V$.

If we extract the bosonic part of the action from $S_V$ and $S_C$, we obtain
\be
\begin{split}
\left. S_V\right|_B
&=
\langle
\del\phi_v , \del\phi^v
\rangle
+
\langle
\delb\phi_v , \delb\phi^v
\rangle
-\langle
Y_v ,Y^v
\rangle
+ 2\langle
Y_v, \mu_0^v
\rangle,\\
\left. S_C\right|_B
&=
\frac{1}{2}\bigg[
\langle
\phi^v L(H)_{ve},\phi^v{L(H)_v}^e
\rangle
+
\langle
L^T(\Hb)_{ev}\phi^v,{L^T(\Hb)^e}_v\phi^v
\rangle
-\langle T_e, T^e\rangle
+\langle T_e ,\nu^e\rangle
+\langle \nu_e , T^e \rangle
\bigg].
\end{split}
\ee
After integrating out the auxiliary fields $Y^v$ and $T^e$,
we find
\be
\begin{split}
\left. S_V\right|_B
&=
\langle
\del\phi_v , \del\phi^v
\rangle
+
\langle
\delb\phi_v , \delb\phi^v
\rangle
+\langle
\mu_{0,v} ,\mu_0^v
\rangle,\\
\left. S_C\right|_B
&=
\frac{1}{2}\bigg[
\langle
\phi^v L(H)_{ve},\phi^v{L(H)_v}^e
\rangle
+
\langle
L^T(\Hb)_{ev}\phi^v,{L^T(\Hb)^e}_v\phi^v
\rangle
+\langle \nu_e, \nu^e\rangle
\bigg].
\end{split}
\ee
From the coupling independence, the path integral
is localized at the fixed points which are determined by
the equations
\bea
&&\mu_0^v = \nu^e=\nub^e=0,\label{fixed point eq 1}\\
&&\del \phi^v = \delb\phi^v = 0,\label{fixed point eq 2}\\
&& \phi^v{L(H)_v}^e = {L^T(\Hb)^e}_v\phi^v=0,\label{fixed point eq 3}
\eea
The equations in the first line (\ref{fixed point eq 1}) are 
the BPS equation for the quiver vortex at the gauge coupling $g_{0,v}$.
The second line (\ref{fixed point eq 2}) and third line 
(\ref{fixed point eq 3}) show that the scalar fields 
$\phi^v$ take constant values on $\Sigma_h$, and $\phi^v$ and 
${L(H)_v}^e$ are ``orthogonal'' with each other, respectively, 
at the fixed points.
The orthogonality conditions (\ref{fixed point eq 3}) 
are solved either by $\langle\phi^v\rangle=0$ and $\langle H^e\rangle\neq 0$ 
(the Higgs branch point) or by $\langle\phi^v\rangle \neq 0$ and 
$\langle H^e\rangle =0$ (Coulomb branch point), leading to two distinct 
branches.
Eqs.~(\ref{fixed point eq 3}) can also contain special solutions
in mixed branches ($\langle\phi^v\rangle\neq 0$ and $\langle H^e\rangle\neq0$),
where $\langle\phi^v\rangle$ is proportional to the vector annihilated by the incidence matrix.
This mixed branch will be closely related to the constraints (\ref{eq:constraint_quiver_moment-map})
in the derivation of the volume formula.


Finally, we here write down the fermionic part of the action
\be
\begin{split}
\left.S_V\right|_F
&=
2\langle\lambda_v  , \del\eta^v\rangle
+2\langle \lambdab_v , \delb\eta^v\rangle
-2\left\langle \chi_v
,\del \lambdab^v+\delb \lambda^v
 +\frac{1}{2}\left(\psi^e {L^T(\Hb)_e}^v +  {L(H)^v}_e \psib^e\right)
\right\rangle,\\
\left. S_C\right|_F
&=
i\langle
\psi_e,\eta^v{L(H)_v}^e
\rangle
-i
\langle
\psib_e,
{L^T(\Hb)^e}_v\eta^v
\rangle
\\
&\qquad\qquad
+\frac{i}{2}
\langle\psi_e,
\phi^v{L(\psi)_v}^e\rangle
-\frac{i}{2}
\langle\psib_e,
{L^T(\psib)^e}_v \phi^v
\rangle
\\
&\qquad\qquad\qquad
+\frac{i}{4}\langle \rho_e, {L^T(\rho)^e}_v\phib^v \rangle
-\langle \rho_e, \del_A \psib^e
-i{L^T(\Hb)^e}_v \lambda^v \rangle
 \\
&\qquad\qquad\qquad\qquad
-\frac{i}{4}\langle \rhob_e,   \phib^v{L(\rhob)_v}^e \rangle
-\langle  \rhob_e, \delb_A \psi^e +i\lambdab^v {L(H)_v}^e
 \rangle,
\end{split}
\ee
for later discussions.

\subsection{Higgs branch localization}

Firstly, we consider the localization of the supersymmetric gauge theory
in the Higgs branch,
where the scalar fields $\phi^v$ vanish and the Higgs scalar $H^e$ and gauge fields $A^v$
take non-vanishing values in general.

Using the coupling independence of the supersymmetric theory,
we can choose the controllable couplings to be
$g_{0,v}=g_v$,
where $g_v$ are the coupling constants appeared in the BPS quiver
vortex equation which we would like to consider.
After choosing the couplings in the Higgs branch,
the fixed point equations (\ref{fixed point eq 1}) reduce to the BPS equations for the quiver vortex.
Thus solutions to the localization fixed point equations are
given by configurations of the quiver vortex.

We denote one of the quiver vortex solutions by $\hat{A}^v$, 
$\hat{\Ab}^v$, $\hat{H}^e$ and $\hat{\Hb}^e$.
In the Higgs branch, the fields are expanded around this solution (fixed point) as
\be
\begin{array}{lcl}
A^v = \hat{A}^v + \frac{1}{\sqrt{t}} \tilde{A}^v,&&
\Ab^v = \hat{\Ab}^v + \frac{1}{\sqrt{t}} \tilde{\Ab}^v,\\
H^e = \hat{H}^e + \frac{1}{\sqrt{t}} \tilde{H}^e,&&
\Hb^e = \hat{\Hb}^e + \frac{1}{\sqrt{t}} \tilde{\Hb}^e.
\end{array}
\ee
Other bosonic and fermionic fields are expanded around vanishing backgrounds.
We just rescale these fields like $\phi^v \to \phi^v/\sqrt{t}$.

We now introduce Faddeev-Popov ghosts $c^v$ and $\cb^v$ and Nakanishi-Lautrup (NL)
field $B^v$ on the vertices to fix the gauge.
The BRST transformation, which is nilpotent $\delta_B^2=0$, is
given by
\be
\begin{split}
&\delta_B \cb^v = 2 B^v,\\
&\delta_B c^v=\delta_B B^v=0,
\end{split}
\ee
for the ghosts and NL fields,
\be
\begin{split}
\delta_B \tilde{A}^v &= -\del c^v,\\
\delta_B \tilde{\Ab}^v &= -\delb c^v,\\
\delta_B \tilde{H}^e &= ic^v {L(\hat{H})_v}^e,\\
\delta_B \tilde{\Hb}^e &= -i {{L^T(\hat{\Hb})}^e}_v c^v,
\end{split}
\ee
for the fluctuations of
the bosonic fields, and similarly for the fermions.

To be compatible with the supersymmetric transformation,
we need to choose a gauge fixing function by
\be
f^v = \del^\dag \tilde{A}^v + \delb^\dag \tilde{\Ab}^v
+\frac{i}{2}\left(
\tilde{H}^e {L^T(\hat{\Hb})_e}^v
-{L(\hat{H})^v}_e\tilde{\Hb}^e
\right)-\frac{1}{2}B^v,
\ee
where we have introduced co-differentials
\be
\del^\dag \equiv -{*}\delb{*},
\quad
\delb^\dag \equiv -{*}\del{*},
\ee
which give the divergences of the gauge field.


The gauge fixing action is given by a $\delta_B$-exact form
\be
\begin{split}
S_{\text{GF+FP}} &= \delta_B \left\langle
\cb_v ,f^v
\right\rangle\\
&=2\left\langle
B_v ,f^v
\right\rangle
+\left\langle
\del c_v ,
\del c^v
\right\rangle
+\left\langle
\delb c_v ,
\delb c^v
\right\rangle\\
&\qquad
+\frac{1}{2}\left\langle
c^v L(\hat{H})_{ve},c^v {L(\hat{H})_v}^e
\right\rangle
+\frac{1}{2}\left\langle
L^T(\hat{\Hb})_{ev} c^v ,
{L^T(\hat{\Hb})^e}_v c^v
\right\rangle.
\end{split}
\ee
So the gauge fixed total action is given by
\be
S' = S_V + S_C +S_{\text{GF+FP}}.
\ee
Precisely speaking, the supersymmetric gauge fixing term is written in terms of a linear combination
of $Q$ and $\delta_B$ ($Q_B\equiv Q+\delta_B$).
We can show that $Q_B$ is nilpotent and the total actions $S'$
including the gauge fixing term is written as a $Q_B$-exact form \cite{Ohta:2018leq}.
The localization works for the nilpotent operator $Q_B$.
However, this $\delta_B$-exact gauge fixing term is sufficient for our later discussions.

It is useful to intruduce combined vector notations 
\be
\vec{\cal B}\equiv (\tilde{\Hb}^e,\tilde{A}^v)^T,\quad
\vec{\cal Y}\equiv (B^v,T^e/\sqrt{2},Y^v)^T
\ee
for bosonic fields, and
\be
\vec{\cal F}\equiv (\psib^e,\lambda^v)^T,\quad
\vec{\cal X}\equiv (\eta^v,\rho^e/\sqrt{2},\chi^v)^T
\ee
for fermionic fields.
Thus we can regard $\eta^v$ as a superpartner of the Nakanishi-Lautrup field $B^v$,
and the degrees of the freedom between the bosons and fermions are balanced
with each other under
the $Q_B$-symmetry.

Let us now rescale the gauge fixed total action by an overall parameter $t$ like $S'\to tS'$.
Using the vector notation,  the rescaled action
reduces to
\be
t\left. S'\right|_B = -\left\langle
\vec{\cal Y}^T,\vec{\cal Y}
\right\rangle
+\left\langle
(\hat{D}_H \vec{\cal B})^T,\vec{\cal Y}
\right\rangle
+\left\langle
(\hat{D}^\dag_H \vec{\cal Y})^T,\vec{\cal B}
\right\rangle+{\cal O}(1/\sqrt{t}),
\label{quadratic boson}
\ee
for bosons, and
\be
t\left. S'\right|_F =
\left\langle
(\hat{D}_H \vec{\cal F})^T,\vec{\cal X}
\right\rangle
-\left\langle
(\hat{D}^\dag_H \vec{\cal X})^T,\vec{\cal F}
\right\rangle+{\cal O}(1/\sqrt{t}),
\label{quadratic fermion}
\ee
for fermions.
We here denote the quadratic terms explicitly and cubic or higher order terms are represented by
${\cal O}(1/\sqrt{t})$,
which vanish in the $t\to \infty$ limit.
We have also defined a first order differential operator by
\be
\hat{D}_H=\begin{pmatrix}
-i {L(\hat{H})^v}_e & 2\del^\dag \\
\sqrt{2}\del_{\hat{A}} & -i\sqrt{2}{L^T(\hat{\Hb})^e}_v \\
{L(\hat{H})^v}_e & 2\delb
\end{pmatrix}.
\ee

Since we can take the $t\to \infty$ limit (WKB or 1-loop approximation)
thanks to the coupling independence of the supersymmetric theory,
the above quadratic part of the action
is sufficient to perform the path integral and reproduce the exact results.

It is easy to integrate out all the fluctuations in the quadratic terms (\ref{quadratic boson}) and (\ref{quadratic fermion}),
except for zero modes (the kernel of the operator $\hat{D}_H$).
After integrating out all the non-zero modes of the fluctuations, we obtain a 1-loop determinant
\be
\text{(1-loop det)} = \frac{\det' \hat{D}_H^\dag \hat{D}_H}
{\det' \hat{D}_H^\dag\hat{D}_H}=1,
\label{1-loop det}
\ee
where $\det'$ stands for the determinants except for the zero modes.
The determinants of the denominator and numerator are canceled with each other
between contributions from the bosons and fermions, respectively.

After integrating out the non-zero modes, there exist the residual integrals over the zero modes.
The bosonic zero modes satisfy
\be
\hat{D}_H \vec{\cal B}_0=0,
\label{zero mode eq}
\ee
which is a linearized equation of the BPS quiver vortex. So the bosonic zero modes span the cotangent space
of the vortex moduli space and we find
\be
\dim \ker \hat{D}_H = \dim_\C \M_{k^v},
\ee
where $\M_{k^v}$ is the moduli space of the quiver vortex of $k^v$-flux sector given by (\ref{k flux}).
The residual integrals over the bosonic zero modes 
simply reduce to the integrals over the moduli space of 
the vortex and just give the volume of the moduli space, which 
is our purpose.

On the other hand, there also exists a residual integral over the fermionic zero modes, which satisfy
\be
\hat{D}_H \vec{\cal F}_0=0.
\ee
This means that the path integral should vanish due to Grassmann integrals of the zero modes.
Thus we need to insert an appropriate supersymmetric operator
in order to compensate the fermionic zero modes.
If we consider the vev of this supersymmetric operator,
we can obtain the volume of the vortex moduli space from the integral of the bosonic zero modes.


\subsection{$Q$-cohomological Volume Operator}

In order to compensate the fermionic zero modes, we now introduce an operator which contains
the fermion as bi-linear terms. It also must be $Q$-closed (but not $Q$-exact trivially) to preserve the localization arguments
(supersymmetry).

To construct the non-trivial $Q$-closed operator, we first define the following $n$-form operator ${\cal O}_n$ by
\be
{\cal O}_0 \equiv W(\phi),\quad
{\cal O}_1 \equiv \frac{\del W(\phi)}{\del \phi^v}(\lambda^v+\lambdab^v),\quad
{\cal O}_2 \equiv \frac{\del W(\phi)}{\del \phi^v} F^v
- \frac{\del^2 W(\phi)}{\del \phi^v\del\phi^{v'}}\lambda^v \wedge \lambdab^{v'},
\ee
through an arbitrary function $W(\phi)$ of $\phi^v$. These operators obey
the so-called descent equations;
\be
\begin{split}
&Q {\cal O}_0 = 0,\\
& Q {\cal O}_1 = -d {\cal O}_0,\\
& Q {\cal O}_2 =d{\cal O}_1.
\end{split}
\ee
Thus a possible non-trivial $Q$-closed operator can be constructed from an integral of the above 2-form
operator
\be
{\cal I} = \int_{\Sigma_h} {\cal O}_2,
\ee
since the Riemann surface does not have the boundary.

Choosing $W(\phi^v) = \frac{1}{2}(\phi^v)^2$ in particular and adding some $Q$-exact terms, we find an operator 
\be
{\cal I}_V(g_v) = \int_{\Sigma_h}
\bigg[
\phi_v\mu^v(g_v)
-\lambda_v\wedge \lambdab^v
+\frac{i}{2}\psi_e\psib^e \omega
\bigg],
\ee
is still $Q$-closed,
where $\mu^v(g_v)$ is 
given by\footnote{
The raising and lowering of the indices $v$ are also done by the metric $G_{vv'}=\frac{1}{g_v^2}\delta_{vv'}$.
}
\be
\mu^v(g_v) = F^v -\frac{1}{2}\left(g_{v}^2\zeta^v - {L(H)^v}_e\Hb^e\right)
\omega,
\ee
at the coupling $g_v$

In the Higgs branch, where the coupling constants are tuned to be $g_{0,v}=g_v$,
let us consider a vev of an exponential of the operator ${\cal I}_V(g_v)$
\be
\left\langle
e^{i\beta {\cal I}_V(g_v)}
\right\rangle^{g_{0,v}=g_v}_{k_v},
\label{Higgs branch vev}
\ee
where a parameter $\beta$ is introduced. 
The vev in the path integral around the Higgs branch background
is denoted as
$\langle \cdots \rangle^{g_{0,v}=g_v}_{k^v}$
with turning the coupling constants to be $g_{0,v}=g_v$
and fixing\footnote{
Since we would like to see the volume of the vortex moduli space
with the given magnetic flux $k^v$,  topological sectors of the magnetic
flux is not summed in our path integral.
} the magnetic flux (vorticity) as $k^v$.
The above vev can be evaluated at the fixed points 
because of the localization in the Higgs branch,
since the operator $e^{i\beta {\cal I}_V(g_v)}$ also belongs 
to the $Q$-cohomological operator,
and does not spoil the localization argument.

The localization fixed point in the Higgs branch is given by 
a solution to $\mu^v(g_v)=0$.
Thus the vev (\ref{Higgs branch vev}) reduces to
\be
\left\langle
e^{i\beta {\cal I}_V(g_v)}
\right\rangle^{g_{0,v}=g_v}_{k^v}
=\left\langle
e^{-i\beta \int_{\Sigma_h} \left(
\lambda_v\wedge \lambdab^v-\frac{i}{2}\psi_e\psib^e \omega
\right)}
\right\rangle^{g_{0,v}=g_v}_{k^v}.
\label{Higgs branch vev 2}
\ee
The fermion bi-linears just compensate the fermionic zero modes as expected.
Since the number of the fermionic zero modes is equal to the
complex dimension of the moduli space,
the vev of (\ref{Higgs branch vev 2}) is proportional to $\beta^{\dim_\C {\cal M}_{k^v}}$
after integrating overall fermionic zero modes.

After integrating the fermionic zero modes, the integral over the bosonic zero mode
$\vec{\cal B}_0$,
which is a solution to eq.~(\ref{zero mode eq}),
still remains.
As mentioned above, eq.~(\ref{zero mode eq}) has $\dim_\C \M_{k^v}$ linearly independent solutions.
If we denote the solution (kernel) as $\vec{\cal B}^{(i)}_0$ ($i=1,\ldots,\dim_\C \M_{k^v}$),
which are normalized by
\be
\left\langle \vec{\cal B}^{(i)T}_0, \vec{\cal B}^{(j)}_0\right\rangle = \delta^{ij},
\ee
then the bosonics zero mode is expanded as
\be
\vec{\cal B}_0 =\sum_{i=1}^{\dim_\C \M_{k^v}} b^i \vec{\cal B}^{(i)}_0,
\ee
where $b^i$'s are complex variables.

Noting that $b^i$'s are functions of the complex moduli parameters $(m^i,\bar{m}^{\bar{\imath}})$
($i,\bar{\imath}=1,\ldots,\dim_\C {\cal M}_{k^v}$),
the metric on the moduli space can be read from the adiabatic expansion of the norm
\be
\begin{split}
\left\langle \delta\vec{\cal B}^T_0,\delta\vec{\cal B}_0 \right\rangle
&=\delta_{i\bar{\jmath}}\delta b^i \delta\bar{b}^{\bar{\jmath}}\\
&={\cal G}_{i\bar{\jmath}}\delta m^i \delta \bar{m}^{\bar{\jmath}},
\end{split}
\ee
where ${\cal G}_{i\bar{\jmath}}$ is the metric on the moduli space and given by\footnote{
We have assumed the metric of the moduli space is Hermitian (K\"ahler).}
\be
{\cal G}_{i\bar{\jmath}} \equiv 
\delta_{k\bar{l}}\frac{\del b^k}{\del m^i}\frac{\del \bar{b}^{\bar{l}}}{\del \bar{m}^{\bar{\jmath}}}.
\ee
The integral over the bosonic zero modes is given by
\be
\label{eq:zero_mode_measure}
\begin{split}
\left\langle
e^{i\beta {\cal I}_V(g_v)}
\right\rangle^{g_{0,v}=g_v}_{k^v}
&={\cal N}_H \beta^{\dim_\C {\cal M}_{k^v}} \int \prod_{i=1}^{\dim_\C {\cal M}_{k^v}} d^2 b^i\\
&={\cal N}_H \beta^{\dim_\C {\cal M}_{k^v}} \int_{{\cal M}_{k^v}} \prod_{i=1}^{\dim_\C {\cal M}_{k^v}} d^2 m^i
\, \sqrt{{\cal G}},
\end{split}
\ee
where ${\cal N}_H$ is a numerical factor depending on a definition 
of the path integral measure in the Higgs branch,
and $\sqrt{{\cal G}}$ appears as the Jacobian in changing the 
integral variables to the moduli parameters. 
In Appendix~\ref{sc:moduli_spce_metric}, we derive path-integral 
measure for moduli from 
the viewpoint of effective Lagrangian in the case of the general real corrdinates. 

The integral measure over the moduli parameters
is nothing but the volume form on the moduli space of the vortex.
We finally find \cite{Ohta:2018leq}
\be
\left\langle
e^{i\beta {\cal I}_V(g_v)}
\right\rangle^{g_{0,v}=g_v}_{k^v}
={\cal N}_H \beta^{\dim_\C {\cal M}_{k^v}} \Vol({\cal M}_{k^v}).
\ee
Thus the $Q$-cohomological operator $e^{i\beta {\cal I}_V(g_v)}$ 
measures the volume of the moduli space
in the path integral.

Unfortunately the evaluation of the volume operator 
$e^{i\beta {\cal I}_V(g_v)}$ in the Higgs branch 
is difficult in general, since we do not have a precise knowledge 
on the metric of the moduli space.
If we however evaluate the same operator in the Coulomb branch, then
we will see the path integral reduces to a simple contour integral.
Using the coupling independence of the supersymmtric theory, we can evaluate the volume
of the moduli space in the Coulomb branch at the different coupling constants.

In the following, we will consider the Coulomb branch localization.

\subsection{Coulomb branch localization}

In the Coulomb branch, we tune the controllable coupling constants
into special values $g_{0,v} \to g_{c,v}$, which satisfy
\be
g_{c,v}^2 = \frac{4\pi k^v}{\zeta^v \A},
\ee
i.e.~the coupling constants are ajusted to be just at the Bradlow bound
for the given parameters $k^v$, $\zeta^v$ and $\A$.

Since the vevs (backgounds) of the Higgs fields should vanish in the Coulomb
branch, the solution
of the gauge fields $a^v$ and $\ab^v$ to 
$\mu_0^v=0$
at the critical couplings $g_{c,v}$ is given by
\be
F^v = \del \ab^v + \delb a^v
= \frac{g^2_{c,v} \zeta^v}{2} \omega.
\label{critical gauge field}
\ee
Using this solution, we can expand the gauge fields
around the backgrounds $a^v$ and $\ab^v$ as
\be
A^v = a^v  + \frac{1}{\sqrt{t}}\tilde{A}^v,
\quad
\Ab^v = \ab^v  + \frac{1}{\sqrt{t}}\tilde{\Ab}^v.
\ee

The scalar fields have vevs (backgrounds) in the Coulomb branch
and take constant values $\phi^v_0$ on $\Sigma_h$
as a consequence of the fixed point equation (\ref{fixed point eq 2}),
so we can expand the scalar fields as
\be
\phi^v = \phi_0^v + \frac{1}{\sqrt{t}}\tilde{\phi}^v,
\quad
\phib = \phib_0^v + \frac{1}{\sqrt{t}}\tilde{\phib}^v.
\ee

Using the index theorem in the Coulomb branch background,
we expect that there exist fermionic zero modes.
The number of the fermionic zero modes is determined by
the Betti numbers of the Riemann surface $\Sigma_h$.
First of all, there is one 0-form zero mode on each vertex
because of $\dim H^0=1$. These are the zero modes
of $\eta^v$, so we denote $\eta_0^v$.
Secondly, we have one 2-form zero mode $\chi_0^v$, related
to $\dim H^2 = 1$,
for each $\chi^v$.
We expand these fermionic fields as
\be
\eta^v = \eta_0^v + \frac{1}{\sqrt{t}}\tilde{\eta}^v,
\quad
\chi^v = \chi_0^v + \frac{1}{\sqrt{t}}\tilde{\chi}^v.
\ee

There are also 1-form zero modes $(\lambda_0^v,\lambdab_0^v)$ on $\Sigma_h$.
The (1,0)- and (0,1)-form can be expanded by
cohomology basis $\gamma^l \in H^{(1,0)}$ and $\bar{\gamma}^l \in H^{(0,1)}$ ($l=1,\cdots,h$), respectively, 
corresponding to each cycle of the Riemann surface $\Sigma_h$.
The cohomology bases are orthogonal with each other like
\be
\langle \gamma^l, \gamma^{l'}\rangle=\delta^{ll'}.
\ee
Thus $\lambda^v$ and $\lambdab^v$ are expanded as
\be
\lambda^v = \lambda^v_0  + \frac{1}{\sqrt{t}}\tilde{\lambda}^v,
\quad
\lambdab^v = \lambdab^v_0   + \frac{1}{\sqrt{t}}\tilde{\lambdab}^v,
\ee
where the zero modes are also expanded by the bases $\gamma^l$ and $\bar{\gamma}^l$ 
\be
\lambda^v_0 = \sum_{l=1}^h \lambda^v_{0,l}\gamma^l\ ,
\quad
\lambdab^v_0 = \sum_{l=1}^h \lambdab^v_{0,l}\bar{\gamma}^l \ ,
\ee
with Grassmann-valued coefficients $\lambda^v_{0,l}$ and $ \lambdab^v_{0,l}$.

Other fields are just rescaled by $1/\sqrt{t}$ as fluctuations, like $H^e \to H^e/\sqrt{t}$.
(We omit {\it tilde} on these fluctuations expanding around zero.)

We again rescale the whole action by $S\to tS$ and expand it around the background in the Coulomb branch.
The rescaled action becomes
\be
\begin{split}
tS
&=
\langle
\del\tilde{\phi}_v , \del\tilde{\phi}^v
\rangle
+
\langle
\delb\tilde{\phi}_v , \delb\tilde{\phi}^v
\rangle
-\langle
Y_v ,Y^v
\rangle
+ 2\langle
Y_v, \del\tilde{\Ab}^v+\delb\tilde{A}^v
\rangle\\
&\qquad
-2\left\langle \tilde{\chi}_v
,\del \tilde{\lambdab}^v+\delb \tilde{\lambda}^v\right\rangle
-2\left\langle\tilde{\eta}_v  , \del^\dag\tilde{\lambda}^v+\delb^\dag\tilde{\lambdab}^v\right\rangle
+\left\langle (\hat{D}_C\vec{\cal V})^T,\vec{\cal V}\right\rangle\\
&\qquad\qquad
+{\cal O}(1/\sqrt{t}),\\
\end{split}
\ee
up to the quadratic order of the fluctuations.
Here we have introduced a vector notation
\be
\vec{\cal V}\equiv (H^e,\psi^e,\rhob^e/\sqrt{2})^T
\ee
and a differential operator (supermatrix)
\be
\hat{D}_C\equiv
\begin{pmatrix}
2\delb_a^\dag\delb_a + |\phi_0^v{L_v}^e|^2
& -{{L^T}^e}_v(i\eta_0^v +{*}\chi_0^v)
&-i\sqrt{2}{{L^T}^e}_v\lambda_0^v\\
(i\eta_0^v-{*}\chi_0^v){L_v}^e
& i \phib_0^v{L_v}^e
& -\sqrt{2}\delb_a^\dag\\
 i\sqrt{2} \lambdab_0^v{L_v}^e
&\sqrt{2}\delb_a
&-i\phi_0^v{L_v}^e
\end{pmatrix},
\ee
which is given by the zero modes
and incidence matrix ${L_v}^e$ (charges of the bi-fundamental matters).
The first order differential operators $\delb_a$ and $\delb_a^\dag$ in $\hat{D}_C$
are covariant derivatives for the charged fields in the backgrounds of the gauge fields
$a^v$ and $\ab^v$ and acting on $\vec{\cal V}$; e.g.
\be
\delb_a H^e = \delb  H^e +i \ab^v {L(H)_v}^e.
\ee

In the Coulomb branch, we simply choose a Coulomb gauge by a gauge fixing function
\be
f^v = \del^\dag \tilde{A}^v + \delb^\dag \tilde{\Ab}^v-\frac{1}{2}B^v.
\ee
Then, the gauge fixing term and the action for the FP ghosts is given by
\be
\begin{split}
S_{\text{GF+FP}}&=\delta_B
\left\langle
\cb_v,f^v
\right\rangle\\
&=2\left\langle
B_v ,f^v
\right\rangle
+\left\langle
\del c_v ,
\del c^v
\right\rangle
+\left\langle
\delb c_v ,
\delb c^v
\right\rangle.
\end{split}
\ee
Using the rescaled action with gauge fixing 
\be
S'\to tS'=tS+tS_{\text{GF+FP}},
\ee
we can perform the path integral by the exact Gaussian 
integral (WKB approximation).
Then we obtain only a 1-loop determinant as an exact 
result of the residual zero mode integral
\be
\frac{1}{\Sdet \hat{D}_C},
\ee
where $\Sdet \hat{D}_C$ stands for a superdeterminant of $\hat{D}_C$,
since the Gaussian integrals are canceled with each other
between pairs; $(\phi^v,\phib^v)\leftrightarrow(c^v,\cb^v)$,
$(A^v,\Ab^v)\leftrightarrow(\tilde{\lambda}^v,\tilde{\lambdab}^v)$, and 
$(Y^v,B^v)\leftrightarrow(\tilde{\chi}^v,\tilde{\eta}^v)$.

Now if we introduce blocks of the supermatrix differential operator by
\be
\hat{D}_C = \begin{pmatrix}
A & B \\
C & D
\end{pmatrix},
\ee
where
\be
\begin{split}
A &\equiv 2\delb_a^\dag\delb_a + |\phi_0^v{L_v}^e|^2,\\
B &\equiv \begin{pmatrix}
-{{L^T}^e}_v(i\eta_0^v+{*}\chi_0^v)
& -i\sqrt{2}{{L^T}^e}_v\lambda_0^v
\end{pmatrix},\\
C &\equiv \begin{pmatrix}
(i\eta_0^v-{*}\chi_0^v){L_v}^e\\
i\sqrt{2}\lambdab_0^v{L_v}^e
\end{pmatrix},\\
D &\equiv \begin{pmatrix}
i \phib_0^v{L_v}^e
& -\sqrt{2}\delb_a^\dag\\
\sqrt{2}\delb_a
&-i\phi_0^v{L_v}^e
\end{pmatrix},
\end{split}
\ee
then the superdeterminant of $\hat{D}_C$ can be expressed by
\be
\frac{1}{\Sdet \hat{D}_C}=\frac{\det D}{\det A}e^{\Tr \log(1-X)},
\label{superdet}
\ee
where $X=D^{-1}CA^{-1}B$.

Firstly, the ratio of $\det A$ and $\det D$ are canceled with each other, except for the zero modes.
The number of the zero modes of $H^e$ and $\psi^e$ is the same, since both are the (0,0)-form fields.
On the other hand, the number of the zero modes of $H^e$ and $\rhob^e$ is different, since $\rhob^e$ is the (0,1)-form field,
whereas $H^e$ is the (0,0)-form field.
The difference of the number of zero modes is given by the Hirzebruch-Riemann-Roch theorem
\be
{\rm ind}\, \delb_a = \dim H^{(0,0)} - \dim H^{(0,1)} = k^v{L_v}^e+\frac{1}{2}\chi_h,
\ee
depending on the charges ${L_v}^e$ of the fields $H^e$ and $\rhob^e$,
background flux $k^v$ and Euler characteristic $\chi_h$ on $\Sigma_h$.
Thus we can evaluate explicitly the ratio of the determinant by
\be
\frac{\det D}{\det A}
=
\frac{1}{\prod_{e\in E}(-i\phi_0^v {L_v}^e)^{k^v{L_v}^e+\frac{1}{2}\chi_h}}.
\ee

Secondly, we can evaluate the exponent in (\ref{superdet}) at the 1-loop level, then we get
\be
\begin{split}
\Tr\log(1-X) &\simeq
-2i\sum_{e\in E}\Tr\frac{1}{(2\delb_a^\dag\delb_a + |\phi_0^v{L_v}^e|^2)^2}\\
&\qquad
\times\left\{(\eta_0^v{L_v}^e)(-i\phi_0^v{L_v}^e)({*}\chi_0^v{L_v}^e)
+
(\lambda_0^v{L_v}^e)(i\phib_0^v{L_v}^e)(\lambdab_0^v{L_v}^e)\right\}\\
&=-\frac{i}{2\pi}\sum_{e\in E}
\left\{
(\eta_0^v{L_v}^e)\frac{1}{i\phib_0^v{L_v}^e}({*}\chi_0^v{L_v}^e)
+(\lambda_0^v{L_v}^e)\frac{1}{-i\phi_0^v{L_v}^e}(\lambdab_0^v{L_v}^e)
\right\},
\end{split}
\ee
where we have used the heat kernel to evaluate the above infinite dimensional trace.

Let us now consider the vev of the volume operator $e^{i\beta {\cal I}_V(g_v)}$ in the Coulomb branch.
The controllable gauge coupling $g_{0,v}$ is now tuned to 
the critical value $g_{c,v}$,
which saturates the Bradlow bound, in the Coulomb branch.
The Coulomb branch solution satisfies $\mu^v_0(g_{c,v})=0$ but not $\mu^v(g_v)=0$.
Indeed, using the Coulomb branch solution (\ref{critical gauge field}) and $\langle H^e \rangle = \langle \Hb^e \rangle=0$,
we find
\be
\mu^v(g_v) = \left(\frac{2\pi k^v}{\A} -\frac{g_v^2 \zeta^v}{2}\right)\omega.
\ee
Thus we have
\be
{\cal I}_V(g_v) = -\sum_{v\in V}\left\{
2\pi \phi_{0}^v\left(\frac{\zeta^v \A}{4\pi} - \frac{k^v}{g_v^2}\right)
+\frac{1}{g_v^2}\sum_{l=1}^h \lambda_{0,l}^v  \lambdab_{0,l}^v
\right\}.
\ee

Now let us consider the vev of the volume operator
\be
\left\langle
e^{i\beta {\cal I}_V(g_v)}
\right\rangle^{g_{0,v}=g_{c,v}}_{k^v},
\ee
in the Coulomb branch by tuning the controllable parameter as $g_{0,v}=g_{c,v}$
and fixing the magnetic flux as $k^v$.
After integrating out all non-zero modes
and including all 1-loop corrections,
we obtain an integral over zero modes;
\be
\begin{split}
\left\langle
e^{i\beta {\cal I}_V(g_v)}
\right\rangle^{g_{0,v}=g_{c,v}}_{k^v}
&={\cal N}_C\int \prod_{v\in V}\left\{\frac{d\phi_0^v}{2\pi}\frac{d\phib_0^v}{2\pi}
d\eta_0^v d{*}\chi_0^v
\prod_{l=1}^h d\lambda_{0,l}^v d\lambdab_{0,l}^v\right\}
\frac{1}{\prod_{e\in E}(-i\phi_0^v {L_v}^e)^{k^v{L_v}^e+\frac{1}{2}\chi_h}}\\
&\qquad\qquad\qquad \times
\exp\left[-2\pi i \beta \sum_{v\in V}\phi_0^v
B^v
+\eta_0^v M_{vv'}{*} \chi_0^{v'}
-i\sum_{l=1}^h \lambda_{0,l}^v\Omega_{vv'} \lambdab_{0,l}^{v'}\right],
\end{split}
\label{zero mode integral}
\ee
where ${\cal N}_C$ is an irrelevant numerical constant depending on the path integral measure
of the non-zero modes, and we have defined
\bea
B^v &\equiv& \frac{\zeta^v \A}{4\pi} -  \frac{k^v}{g_v^2},\\
M_{vv'}&\equiv& \frac{1}{2\pi i}\sum_{e\in E} {L_v}^e\frac{1}{i\phib_0^{v''} {L_{v''}}^e}{{L^T}^e}_{v'},\\
\Omega_{vv'} &\equiv&\frac{\beta}{g_v^2}\delta_{vv'}
+\frac{1}{2\pi}\sum_{e\in E}{L_v}^e\frac{1}{-i\phi_0^{v''} {L_{v''}}^e}{{L^T}^e}_{v'}.
\eea

The integral over $\phib_0^v$, $\eta_0^v$ and ${*}\chi_0^v$ in (\ref{zero mode integral}) can be factorized
and irrelevant for the volume of the vortex moduli space,
since it does not contain any coupling or parameter like $g_v$, $k^v$, $\chi_h$ and $\beta$.
So we can renormalize the overall constant by
\be
\begin{split}
{\cal N}'_C
&\equiv
{\cal N}_C
\int \prod_{v\in V} \left\{\frac{d\phib_0^v}{2\pi}
d\eta_0^v d{*}\chi_0^v \right\}\,
e^{ \eta_0^v M_{vv'} {*}\chi_0^{v'}}\\
&={\cal N}_C 
\int \prod_{v\in V}\frac{d\phib_0^v}{2\pi}
\det M.
\end{split}
\ee
Note here that $\det M$ is a function of $\phib_0^v$,
but degenerated for a generic graph since $M$ contains zero eigenvalues.
So we need a suitable but irrelevant regularization
to define ${\cal N}'_C$.

Using the irrelevant overall constant ${\cal N}_C'$, the vev of the volume operator (\ref{zero mode integral})
reduces to
\be
\begin{split}
\left\langle
e^{i\beta {\cal I}_V(g_v)}
\right\rangle^{g_{0,v}=g_{c,v}}_{k^v}
&={\cal N}_C'\int \prod_{v\in V}\left\{\frac{d\phi_0^v}{2\pi}
\prod_{l=1}^h d\lambda_{0,l}^v d\lambdab_{0,l}^v\right\} \, 
\frac{1}{\prod_{e\in E}(-i\phi_0^v {L_v}^e)^{k^v{L_v}^e+\frac{1}{2}\chi_h}}\\
&\qquad\qquad\qquad \times
\exp\left[-2\pi i \beta \sum_{v\in V}\phi_0^v
B^v
-i\sum_{l=1}^h \lambda_{0,l}^v\Omega_{vv'} \lambdab_{0,l}^{v'}\right]\\
&={\cal N}_C'\int \prod_{v\in V}\frac{d\phi_0^v}{2\pi}
\frac{(\det \Omega)^h}{\prod_{e\in E}(-i\phi_0^v {L_v}^e)^{k^v{L_v}^e+\frac{1}{2}\chi_h}}
e^{-2\pi i \beta \sum_{v\in V}\phi_0^v
B^v},
\end{split}
\label{contour integral}
\ee
after integrating out the zero modes $\lambda_0^v$ and $\lambdab_0^v$
with a suitable measure.

The essential part of this integral formula agrees with the formulae developed in \cite{Closset:2015rna} for $S^2$
and \cite{Benini:2016hjo} for the generic Riemann surfaces, if one 
chooses an Abelian gauge group and suitable matter contents and 
turns off the $\Omega$-backgrounds.
The formulae in \cite{Closset:2015rna,Benini:2016hjo} utilize the JK residue formula
to pick up the suitable poles, but the condition of the selection depends only on the linear combinations of FI parameters.

In our formula (\ref{contour integral}), unlike the previous localization formulae,
an extra
exponential factor $e^{-2\pi i \beta \sum_{v\in V}\phi_0^v B^v}$,
which comes from the volume operator $e^{i\beta {\cal I}_V(g_v)}$, is inserted.
The inserted operator controls the choice of the poles (residues)
according to the positive or negative values of the linear combinations of $B^v$ to assure the convergence of the contour integral.
The parameters $B^v$ contain not only the FI parameters but also the vorticity, gauge couplings and area of the Riemann
surface.
In this sense, we can regard the contour integral as a generalization of the JK residue formula.

Thus we finally can express the volume of the vortex moduli space
as simple line (contour) integrals over $\phi_0^v$,
without any explicit information on the metric of the moduli space.
In order to evaluate the integral (\ref{contour integral}),
we need to choose suitable integral path of $\phi_0^v$,
which determines the condition for the Bradlow bounds and wall crossing (the generalization of the JK residue formula).
We will see this phenomenon for concrete examples in the following sections.

\section{Volume of the Quiver Vortex Moduli Space}
\label{Volume of the Quiver Vortex Moduli Space}

In this section, we apply the integral formula 
(\ref{contour integral}) for the volume of the vortex moduli space 
to some Abelian quiver gauge theory. We consider the universal 
coupling case, although we will keep unconstrained 
gauge couplings in many places. 
All the computations should be useful in other cases 
($g_v\not=g_{v'}$) as well, and the universal coupling case is 
obtained by taking the limit $g_v\to g$ at the end. 

\subsection{Two Abelian vertices}

\begin{figure}
\begin{center}
\includegraphics{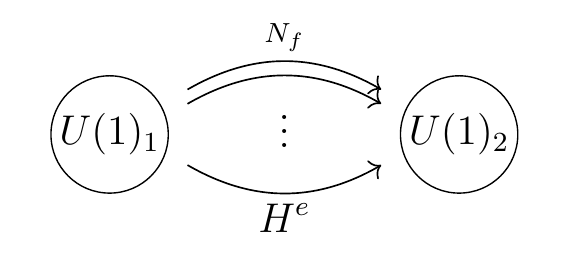}
\end{center}
\caption{The quiver diagram of two Abelian vertices with $N_f$ flavors.}
\label{Abelian Nf flavors}
\end{figure}

We first start with a quiver which has only two vertices with 
Abelian gauge groups.
There exist $N_f$ edges (arrows) from one vertex to the other.
The quiver diagram is depicted in Fig.~\ref{Abelian Nf flavors}.

Each edge corresponds to the bi-fundamental matters. So we have 
$N_f$ kinds of the matters (Higgs fields).
For the Abelian theory, this means that the matter has a positive 
charge under one $U(1)_1$ gauge group 
and a negative charge under the other $U(1)_2$.
The incidence matrix ${L_v}^e$ is a $2\times N_f$ matrix and
represents the charges of the matters by
\be
L =
\overbrace{
\left(
\begin{array}{rrcr}
1 & 1 & \cdots & 1\\
-1 & -1 & \cdots & -1
\end{array}
\right)}^{N_f}.
\ee

In this model, the BPS vortex equation becomes
\be
\begin{split}
\mu^1 &= F^1 - \frac{g_1^2}{2}\left( \zeta^1 - \sum_{e=1}^{N_f}H^e \Hb^e\right)
\omega=0, \\
\mu^2 &= F^2 - \frac{g_2^2}{2}\left( \zeta^2 + \sum_{e=1}^{N_f}\Hb^e H^e\right)\omega=0,\\
\nu^e &=2\del_A \Hb^e = 0,\\
\nub^e & = 2\delb_A H^e =0.
\end{split}
\ee
Let us consider linear combinations of $\mu^1$ and $\mu^2$
\bea
\frac{1}{g_1^2}\mu^1 + \frac{1}{g_2^2}\mu^2
&=&\frac{1}{g_1^2}F^1+\frac{1}{g_2^2}F^2
-\frac{1}{2}\left(\zeta^1+\zeta^2\right)\omega=0,
\label{mu sum}\\
\frac{1}{g_1^2}\mu^1 - \frac{1}{g_2^2}\mu^2
&=&\frac{1}{g_1^2}F^1-\frac{1}{g_2^2}F^2
-\frac{1}{2}\left(\zeta^1-\zeta^2
-2\sum_{e=1}^{N_f}H^e \Hb^e\right)\omega=0.
\label{mu diff}
\eea

Integrating (\ref{mu sum}) on $\Sigma_h$, we find
\be
\frac{k^1}{g_1^2}+\frac{k^2}{g_2^2} = \frac{\zeta^1+\zeta^2}{4\pi}\A.
\ee
However this equation can not be satisfied
for generic value of $g_v$ and $\zeta^v$
since the magnetic fluxes $k_v$ are integer valued.
If $\zeta^1+\zeta^2=0$, there exist the vacuum ($k^1=k^2=0$)
at least, but no BPS vortices is allowed for the generic couplings.
If the gauge couplings of two $U(1)$'s coincide with each other
$g_1=g_2$,
there are infinitely many BPS vortices when $\zeta^1+\zeta^2=0$
and $k^1+k^2=0$.
In this case, 
$U(1)$ of the difference of the generators
in $U(1)_1$ and $U(1)_2$;
\be
A' = \frac{1}{2}(A^1 - A^2)
\ee
is isomorphic to a single $U(1)$ theory with $N_f$ flavors,
and eq.~(\ref{mu diff}) is equivalent to
the BPS vortex equation of $N_f$ flavors with  the
flux $(k^1-k^2)/2=k^1$ and FI parameter $(\zeta^1-\zeta^2)/2=\zeta^1$.

On the other hand, integrating (\ref{mu diff}) on $\Sigma_h$, we get
\be
\frac{\zeta^1-\zeta^2}{4\pi}\A - \frac{k^1}{g_1^2}+\frac{k^2}{g_2^2}
=\frac{1}{2\pi}\sum_{e=1}^{N_f}\int_{\Sigma_h} H^e \Hb^e\omega
\geq 0.
\label{relative Bradlow bound}
\ee
This is a Bradlow bound for the relative charges of the  vortex.
If $g_1=g_2$, then we need to set $\zeta^1 + \zeta^2 = 0$
and $k^1+k^2=0$ and (\ref{relative Bradlow bound})
reduces to the Bradlow bound for the Abelian theory
with the single $U(1)$
\be
\frac{\zeta^1 \A}{4\pi} - \frac{k^1}{g_1^2}
\geq 0.
\ee
So there exists an upper bound for the vorticity $k^1$ 
on $\Sigma_h$ with the finite area $\A$.

Applying the formula (\ref{contour integral}), we obtain
\be
\left\langle
e^{i\beta {\cal I}_V(g_v)}
\right\rangle^{g_{0,v}=g_{c,v}}_{k^1,k^2}
={\cal N}_C'\int \frac{d\phi_0^1}{2\pi}\frac{d\phi_0^2}{2\pi}
\frac{(\det \Omega)^h}{\left(-i(\phi_0^1-\phi_0^2) \right)^{N_f\left(k^1-k^2+\frac{1}{2}\chi_h\right)}}
e^{-2\pi i \beta \left( \phi_0^1 B^1 + \phi_0^2 B^2
\right)},
\ee
where
\be
\begin{split}
\det \Omega &= 
\det\begin{pmatrix}
\frac{\beta}{g_1^2} + \frac{1}{2\pi}\frac{N_f}{-i(\phi_0^1-\phi_0^2)} & - \frac{1}{2\pi} \frac{N_f}{-i(\phi_0^1-\phi_0^2)}\\
-  \frac{1}{2\pi}\frac{N_f}{-i(\phi_0^1-\phi_0^2)} & \frac{\beta}{g_2^2} + \frac{1}{2\pi} \frac{N_f}{-i(\phi_0^1-\phi_0^2)}
\end{pmatrix}\\
&=\frac{\beta}{g_1^2g_2^2}
\left(\beta +  \frac{g_1^2+g_2^2}{2\pi}\frac{N_f}{-i(\phi_0^1-\phi_0^2)}\right).
\end{split}
\ee

Now changing the variables to
\be
\begin{split}
\phi_0^c &\equiv \frac{1}{2}(\phi_0^1+\phi_0^2),\\
\hat{\phi}_0 & \equiv \phi_0^1-\phi_0^2,\\
\hat{k} & \equiv k^1 - k^2,
\end{split}
\ee
we can write
\be
\begin{split}
\left\langle
e^{i\beta {\cal I}_V(g_v)}
\right\rangle^{g_{0,v}=g_{c,v}}_{\hat{k}}
&={\cal N}_C'\int \frac{d\phi_0^c}{2\pi}\frac{d\hat{\phi}_0}{2\pi}
\frac{\left(\frac{\beta}{g_1^2g_2^2}\right)^h
\left(\beta+\frac{g_1^2+g_2^2}{2\pi}\frac{N_f}{-i\hat{\phi}_0}\right)^h}{\left(-i\hat{\phi}_0 \right)^{N_f\left(\hat{k}+\frac{1}{2}\chi_h\right)}}
e^{-2\pi i \beta \left( 2\phi_0^c B^c + \frac{1}{2}\hat{\phi}_0\hat{B}
\right)}\\
&={\cal N}_C'\int \frac{d\phi_0^c}{2\pi}
\left(\frac{\beta}{g_1^2g_2^2}\right)^h
e^{-4\pi i \beta \phi_0^c B^c}
\int\frac{d\hat{\phi}_0}{2\pi}
\frac{\left(\beta+\frac{g_1^2+g_2^2}{2\pi}\frac{N_f}{-i\hat{\phi}_0}\right)^h}{\left(-i\hat{\phi}_0 \right)^{N_f\left(\hat{k}+\frac{1}{2}\chi_h\right)}}
e^{-\pi i \beta \hat{\phi}_0\hat{B}},
\end{split}
\label{U(1)^2 contour integral}
\ee
where
\be
\begin{split}
B^c &\equiv \frac{1}{2}(B^1 + B^2)
= \frac{1}{2}\left(
\frac{\zeta^1+\zeta^2}{4\pi}\A
-\frac{k^1}{g_1^2}-\frac{k^2}{g_2^2}
\right),\\
\hat{B} & \equiv B^2-B^2
=\frac{\zeta^1-\zeta^2}{4\pi}\A
-\frac{k^1}{g_1^2}+\frac{k^2}{g_2^2}.
\end{split}
\ee
The former integral in (\ref{U(1)^2 contour integral}) gives
\be
{\cal N}_C'\int \frac{d\phi_0^c}{2\pi}\left(\frac{\beta}{g_1^2g_2^2}\right)^h
e^{-4\pi i \beta \phi_0^c B^c}
=\frac{{\cal N}_C'}{2\pi}
\left(\frac{\beta}{g_1^2g_2^2}\right)^h\delta(2 \beta B^c),
\ee
which gives a constraint $B^c=0$ as we found\footnote{
$\delta(B^c)$ diverges at $B^c=0$, but we absorb and regularize 
this divergence with
the degenerate normalization ${\cal N}_C'$ at the same time.
So we expect a finite constraint $B^c=0$ from this part.} 
from (\ref{mu sum}).

The latter integral in (\ref{U(1)^2 contour integral});
\be
\int\frac{d\hat{\phi}_0}{2\pi}
\frac{\left(\beta+\frac{g_1^2+g_2^2}{2\pi}\frac{N_f}{-i\hat{\phi}_0}\right)^h}{\left(-i\hat{\phi}_0 \right)^{N_f\left(\hat{k}+\frac{1}{2}\chi_h\right)}}
e^{-\pi i \beta \hat{\phi}_0\hat{B}},
\label{latter integral}
\ee
is nothing but the integral expression for
the volume of the vortex moduli space in $U(1)$ gauge theory
with $N_f$ flavors \cite{Miyake:2011yr,Ohta:2018leq} up to a redefinition of the parameter
$\beta$.

To evaluate the integral (\ref{latter integral}),
we introduce a small twisted mass.
Turning on the twisted mass $\epsilon^e$ for $H^e$,
the supersymmetric transformations
are modified; e.g.
\be
QH^e = \psi^e, \quad Q\psi^e = i\phi^v{L(H)_v}^e - \epsilon^e H^e,
\ee
where we do not sum the repeated index $e$.
This modification by the twisted mass also modifies the
cohomological volume operator into
\be
{\cal I}_V^\epsilon(g_v)
=\int_{\Sigma_h}\left[ \phi_v\mu^v(g_v)
+\frac{i}{2}\sum_{e\in E}\epsilon^e H^e \Hb^e
-\lambda_v \wedge \lambdab^v
+\frac{i}{2}\psi_e\psib^e\omega\right].
\ee

\begin{figure}
\begin{center}
\begin{tabular}{cp{2cm}c}
\includegraphics[scale=0.45]{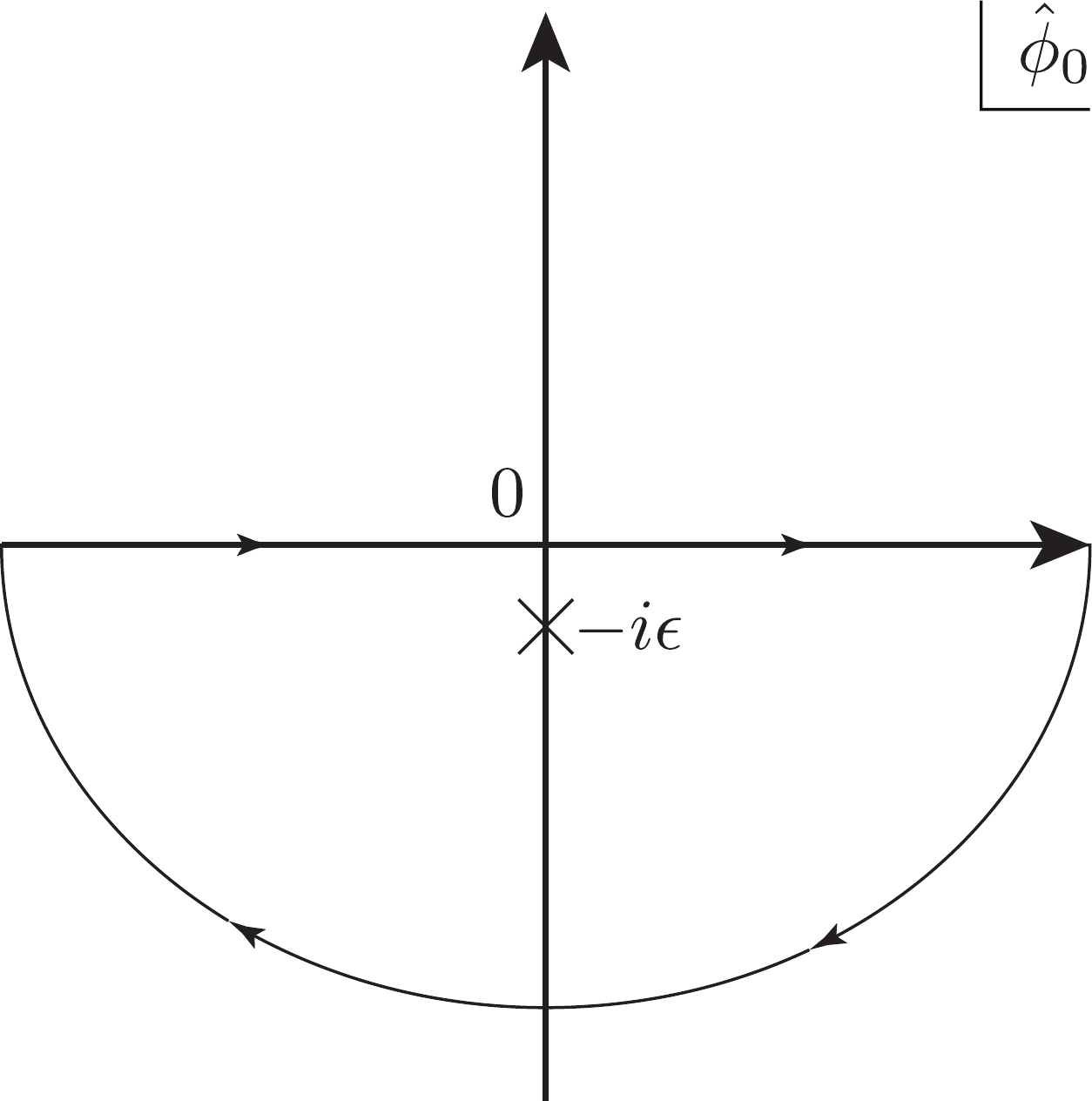}
&&
\includegraphics[scale=0.45]{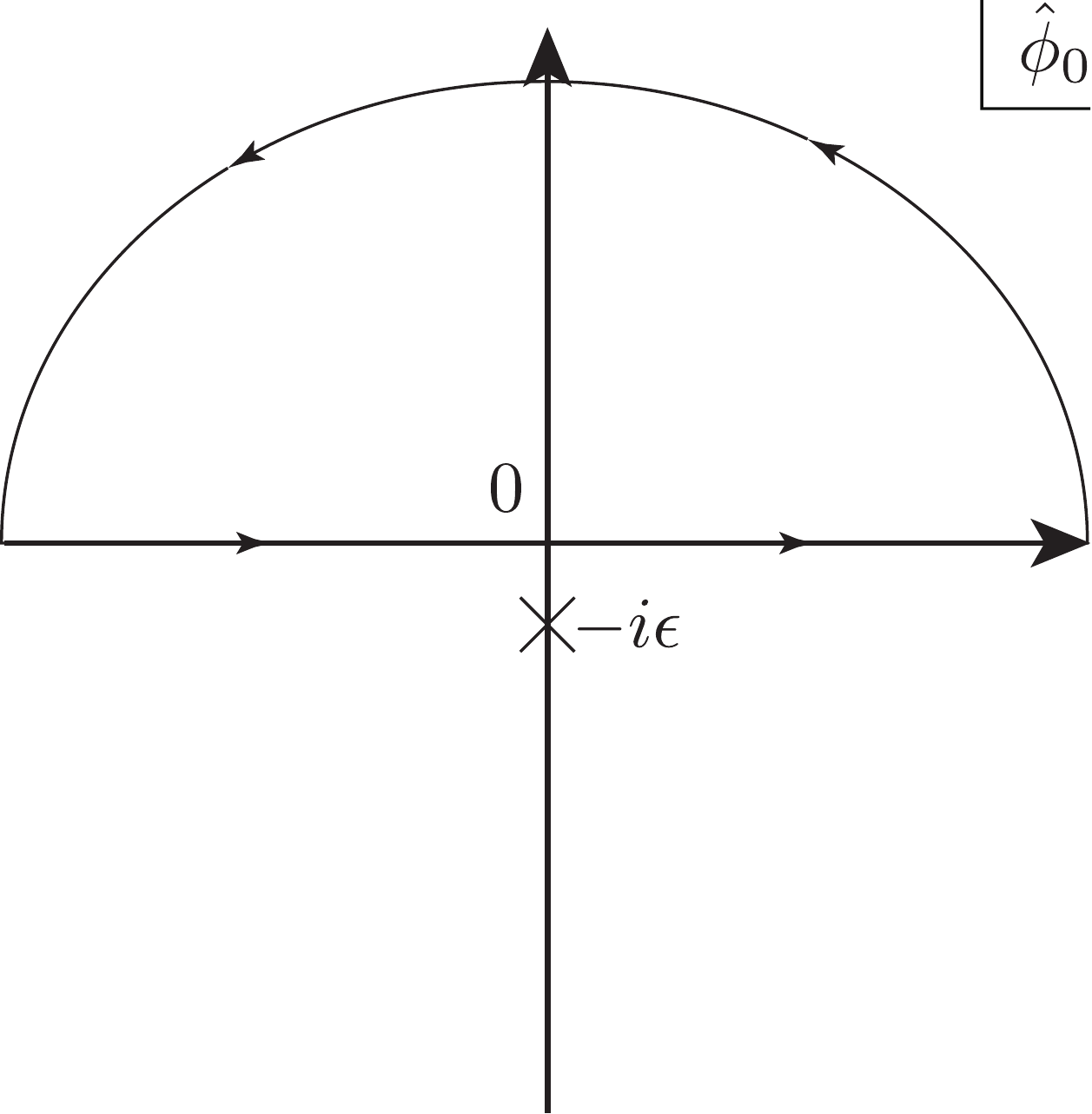}\\
(a) $\hat{B}> 0$ && (b) $\hat{B}< 0$
\end{tabular}
\end{center}
\caption{The integral contours of $\hat{\phi}_0$.
The pole exists at  $\hat{\phi}_0=-i\epsilon$.
For convergence of the integral, we should choose closed circle
on the lower half plane (a) if $\hat{B}> 0$ or on
the upper half plane (b) if $\hat{B}< 0$.
The contour (a) includes the pole inside.  }
\label{lower-upper contour}
\end{figure}

Indeed, we can shift the integral path above the real axis
without any divergences from the integral of the matter fields,
then the integral contour should be closed
on the lower half plane (Fig.~\ref{lower-upper contour}(a)) if $\hat{B}> 0$
or on the upper half plane (Fig.~\ref{lower-upper contour}(b))
if $\hat{B}< 0$.

The contour includes the pole at $\hat{\phi}_0=-i\epsilon$ if $\hat{B}> 0$. So the integral gives
a non-vanishing value.
This is related to the Bradlow bound condition (\ref{relative Bradlow bound}).
Evaluating the integral (\ref{latter integral}),
we obtain
\be
\begin{split}
\int\frac{d\hat{\phi}_0}{2\pi}
\frac{\left(\beta+\frac{g_1^2+g_2^2}{2\pi}\frac{N_f}{-i\hat{\phi}_0}\right)^he^{-\pi i \beta \hat{\phi}_0\hat{B}}}{\left(-i\hat{\phi}_0 +\epsilon\right)^{N_f\left(\hat{k}+\frac{1}{2}\chi_h\right)}}
&=
\sum_{l=0}^h\binom{h}{l}\beta^l
\left(\frac{g_1^2+g_2^2}{2\pi}N_f\right)^{h-l}\\
&\qquad\qquad\times
\int\frac{d\hat{\phi}_0}{2\pi}
\frac{e^{-\pi i \beta \hat{\phi}_0\hat{B}}}{\left(-i\hat{\phi}_0 +\epsilon\right)^{N_f\left(\hat{k}+\frac{1}{2}\chi_h\right)+h-l}}\\
&=\beta^d
\sum_{l=0}^h\binom{h}{l}
\left(\frac{g_1^2+g_2^2}{2\pi}N_f\right)^{h-l}
\frac{(2\pi\hat{B})^{d-l}
e^{-\epsilon \pi\beta \hat{B}}
}
{(d-l )!},
\label{non-vanishing volume}
\end{split}
\ee
where $d\equiv \hat{k}N_f+(N_f-1)(1-h)$.
So we find, in the $\epsilon \to 0$ limit,
the volume of the moduli space is proportional to
\be
\beta^{d}
\sum_{l=0}^h\binom{h}{l}
\left(\frac{g_1^2+g_2^2}{2\pi}N_f\right)^{h-l}
\frac{(2\pi\hat{B})^{d-l}
}{(d-l)!}
\ee

This is the volume of the moduli space of the Abelian vortex
with $N_f$ flavor on $\Sigma_h$.
The dimension of the moduli space is expressed in the power of $\beta$,
i.e.~$d=\hat{k}N_f+(N_f-1)(1-h)$.

On the other hand, the integral vanish if $\hat{B}<0$
since the pole at $\hat{\phi}_0=0$ is not enclosed inside the contour.
This means that there is no BPS vortex solution for $\hat{B}<0$.

To summarize, the volume of the moduli space of the quiver
vortex does not vanish if and only if the condition;
\be
B_c= 0 \quad \text{and} \quad \hat{B}> 0,
\label{quiver vortex condition}
\ee
are satisfied, and takes a value of (\ref{non-vanishing volume}).
The quiver vortex could exist on $\Sigma_h$
with charges which satisfy the condition (\ref{quiver vortex condition}).

\subsection{Non-compact moduli space}

\begin{figure}
\begin{center}
\includegraphics{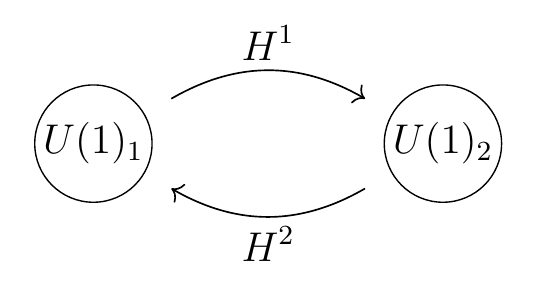}
\end{center}
\caption{The quiver diagram of two Abelian vertices with a loop.}
\label{two Abelian vertices with a loop}
\end{figure}

We now consider a model with two vertices, i.e.~$G=U(1)_1\times U(1)_2$
quiver gauge theory.
In contrast with the prior model, we have only two matters with opposite
charges. Two edge arrows makes a loop between two vertices.
The quiver diagram is depicted in Fig.~\ref{two Abelian vertices with a loop}.

The incidence matrix is given by
\be
L = \left(
\begin{array}{rr}
1 & -1\\
-1 & 1
\end{array}
\right)
\ee
The associated BPS equations are given by
\be
\begin{split}
\mu^1 &= F^1 - \frac{g_1^2}{2}\left( \zeta^1 - H^1 \Hb^1 + \Hb^2 H^2\right)
\omega=0, \\
\mu^2 &= F^2 - \frac{g_2^2}{2}\left( \zeta^2  + \Hb^1 H^1  -  H^2 \Hb^2\right)\omega=0,\\
\nu^e &=2\del_A \Hb^e = 0,\\
\nub^e & = 2\delb_A H^e =0.
\end{split}
\ee

Similar to the prior model, we can consider sum and difference of
$\mu^1$ and $\mu^2$;
\bea
\frac{1}{g_1^2}\mu^1 + \frac{1}{g_2^2}\mu^2
&=&\frac{1}{g_1^2}F^1+\frac{1}{g_2^2}F^2
-\frac{1}{2}\left(\zeta^1+\zeta^2\right)\omega=0,
\label{mu sum 2}\\
\frac{1}{g_1^2}\mu^1 - \frac{1}{g_2^2}\mu^2
&=&\frac{1}{g_1^2}F^1-\frac{1}{g_2^2}F^2
-\frac{1}{2}\left(\zeta^1-\zeta^2
-2H^1\Hb^1+2\Hb^2H^2\right)\omega=0.
\label{mu diff 2}
\eea
From the sum (\ref{mu sum 2}), we have
\be
B^c = \frac{1}{2}(B^1 + B^2) = 0,
\label{eq:constraint_noncompact}
\ee
as a constraint for the couplings and FI parameters.
From the difference (\ref{mu diff 2}), we obtain
\be
\hat{B} = B^1 - B^2 = 
\frac{1}{2\pi}\int_{\Sigma_h} H^1\Hb^1\omega
-\frac{1}{2\pi}\int_{\Sigma_h} H^2\Hb^2\omega.
\label{non-compact B}
\ee
So $\hat{B}$ can take any positive and negative values.
Thus the moduli space of this model should be non-compact
since there are infinitely many combinations of the vev of $H^1$ and $H^2$,
which give the same difference $\hat{B}$.
In particular, if we consider the vacuum ($k^1=k^2=0$)
and assume the covariantly constant equations $\nu^e=\nub^e=0$
have non-trivial and normalizable solutions, the vev of $H^1$
and $H^2$ is given by the difference of the FI parameters
\be
|H^1|^2 - |H^2|^2 = \zeta^1 - \zeta^2,
\label{noncompact moduli}
\ee
which represents a non-compact moduli space
(hyperbolic plane).

Applying the integral formula (\ref{contour integral}),
we obtain
\be
\left\langle
e^{i\beta {\cal I}_V(g_v)}
\right\rangle^{g_{0,v}=g_{c,v}}_{\hat{k}}
=(-1)^{\frac{1}{2}\chi_h-\hat{k}}{\cal N}_C' \left(\frac{\beta}{g_1g_2}\right)^{2h}\int \frac{d\phi_0^c}{2\pi}
e^{-4\pi i \beta \phi_0^c B^c}
\int
\frac{d\hat{\phi}_0}{2\pi}
\frac{e^{-\pi i \beta \hat{\phi}_0 \hat{B}}}
{\left(-i\hat{\phi}_0 \right)^{\chi_h}}
.
\ee
The former integral gives the constraint $B^c = 0$ as expected,
but the integral does not depend on the magnetic flux
$\hat{k}$ except for the overall sign.
Hence the integral of $\hat{\phi}_0$ is highly degenerate 
and the choice of the contour is not well-defined.

This is because the poles associated with $H^1\neq 0$ ($\hat{k}>0$) and
$H^2 \neq 0$ ($\hat{k}<0$) are merged.
To avoid the degeneration, we modify the model by introducing twisted masses
($\Omega$-backgrounds)
for the matter $H^1$ and $H^2$.
The introduction of the twisted masses changes the supersymmetric transformations to
\be
\begin{split}
Q\psi^1 &= i(\phi^1-\phi^2 + i\epsilon^1)H^1 = i(\hat{\phi}+i\epsilon^1)H^1,\\
Q\psi^2 &= i(\phi^2-\phi^1 + i\epsilon^2)H^2 = i(-\hat{\phi}+i\epsilon^2)H^2,
\end{split}
\ee
where $\epsilon^1$ and $\epsilon^2$ are real and positive parameters.
The fixed point equation $Q\psi^1=Q\psi^2=0$
means that $H^1\neq 0$ (or $H^2 \neq 0$) contributes  
near the pole at $\hat{\phi}_0+i\epsilon^1\approx 0$
(or $-\hat{\phi}_0+i\epsilon^2\approx 0$).

Thus, using the separation of the poles,
we can distinguish two branches of the non-compact moduli space,
which are $\hat{B}>0$ and $H^1\neq 0$, or  $\hat{B}<0$ and $H^2\neq 0$.
Turning on the twisted mass does not admit the mixed branch
$H^1\neq 0$ and $H^2 \neq 0$.

The integral formula for the volume is also modified by the twisted mass
into
\be
\begin{split}
\left\langle
e^{i\beta {\cal I}_V(g_v)}
\right\rangle^{g_{0,v}=g_{c,v}}_{\hat{k}}
&={\cal N}_C' \int \frac{d\phi_0^c}{2\pi}
e^{-4\pi i \beta \phi_0^c B^c}\\
&\qquad \times
\int
\frac{d\hat{\phi}_0}{2\pi}
\frac{(\det \Omega)^h \, e^{-\pi i \beta \hat{\phi}_0 \hat{B}}}
{\left(-i\hat{\phi}_0+\epsilon^1 \right)^{\hat{k}+\frac{1}{2}\chi_h}
\left(i\hat{\phi}_0+\epsilon^2 \right)^{-\hat{k}+\frac{1}{2}\chi_h}}
,
\end{split}
\ee
where
\be
\begin{split}
\det \Omega 
&=\frac{\beta}{g_1^2g_2^2}
\left(
\beta+\frac{g_1^2+g_2^2}{2\pi}\frac{1}{-i\hat{\phi}_0+\epsilon^1}
+\frac{g_1^2+g_2^2}{2\pi}\frac{1}{i\hat{\phi}_0+\epsilon^2}
\right)\\
&=\frac{\beta}{g_1^2g_2^2}
\left(
\beta+\frac{g_1^2+g_2^2}{2\pi}\frac{\epsilon^1+\epsilon^2}
{\left(-i\hat{\phi}_0+\epsilon^1\right)\left(i\hat{\phi}_0+\epsilon^2\right)}
\right).
\end{split}
\ee
The former integral gives the constraint $B^c=0$ again.
If $\hat{B}>0$, we need to choose the contour on the lower half plane.
Then we obtain the volume of the moduli space as
\be
\left\langle
e^{i\beta {\cal I}_V(g_v)}
\right\rangle^{g_{0,v}=g_{c,v}}_{\hat{k}}
=
{\cal N}_C''
\beta^{\hat{k}}
\frac{1}{(\epsilon^1+\epsilon^2)^{1-h-\hat{k}}}
\sum_{l=0}^h
\binom{h}{l}
\left(
\frac{g_1^2+g_2^2}{2\pi}
\right)^{h-l}
\frac{(2\pi\hat{B})^{\hat{k}-l}}{(\hat{k}-l)!},
\ee
and $\hat{k}$ should be positive,
where  
\be
{\cal N}_C'' \equiv 
{\cal N}_C' \left(\frac{\beta}{g_1^2g_2^2}\right)^h \int \frac{d\phi_0^c}{2\pi}
e^{-4\pi i \beta \phi_0^c B^c},
\ee
includes the irrelevant constants and constraint.
If $\hat{B}<0$, we need to choose the contour on the upper-half plane
and get
\be
\left\langle
e^{i\beta {\cal I}_V(g_v)}
\right\rangle^{g_{0,v}=g_{c,v}}_{\hat{k}}
=
{\cal N}_C''
\beta^{-\hat{k}}
\frac{1}{(\epsilon^1+\epsilon^2)^{1-h+\hat{k}}}
\sum_{l=0}^h
\binom{h}{l}
\left(
\frac{g_1^2+g_2^2}{2\pi}
\right)^{h-l}
\frac{(-2\pi\hat{B})^{-\hat{k}-l}}{(-\hat{k}-l)!},
\ee
and $\hat{k}$ should be negative.

After regularizing the volume by introducing the twisted masses $\epsilon^1$
and $\epsilon^2$, we find the volume is proportional to
$(\epsilon^1+\epsilon^2)^{h+|\hat{k}|-1}$.
So the volume diverges in the limit of $\epsilon^1\to 0$ and $\epsilon^2\to 0$
if $h=0$ and $\hat{k}=0$.
This reflects the fact that the moduli space of the vacuum on $S^2$, which is given by
a non-trivial and normalizable solution to the covariantly constant equations $\nu^e=\nub^e=0$
and eq.~(\ref{noncompact moduli}),
is non-compact with unit complex dimension.
The regularization causes the separation of the branch of the moduli space.
Each branch contributes to the volumes as Abelian BPS vortices of $N_f=1$.
We can see this results from the equation (\ref{non-compact B})
if the moduli space is separated by two branches
of $\hat{B}>0$, $\hat{k}>0$ and $H^1\neq 0$, or
$\hat{B}<0$, $\hat{k}<0$ and $H^2\neq 0$

For $h=1$ (torus), the volume of the vacuum moduli space takes a constant value and is independent of $\epsilon^1$
and $\epsilon^2$. This means that the moduli space shrinks to isolated points. (The dimension of the moduli space
also vanishes in this case.)
For the higher genus case ($h>1$),
the volume of the vacuum moduli space vanishes in the limit of  $\epsilon^1\to 0$ and $\epsilon^2\to 0$.
In this case, we expect that
there is no solution to the differential equations which determine the moduli space of vacua. 

The non-existence of vacuum solution on higer genus Riemann surface 
is not a strange phenomenon. 
It has been found already in more familiar models with compact 
moduli spaces, such as the $U(1)$ gauge theory with $N_f$ flavors 
of charged Higgs scalars, by means of cohomologies and divisors 
\cite{Baptista:2010rv}, and also by means of localization 
\cite{Miyake:2011yr}. 
In the simplest solution ($g_1=g_2$) of the constraint 
\eqref{eq:constraint_noncompact}, the model is equivalent to a 
single $U(1)$ gauge theory with two opposite charge 
scalar fields as given in \eqref{non-compact B}. 
A similar model for compact moduli space is the $U(1)$ gauge 
theory with two same charge scalar fields, whose moduli space 
has been found to have similar features 
\cite{Baptista:2010rv,Miyake:2011yr} : vacuum sector ($k=0$) 
does not exist for higher genus case ($h>1$), and the moduli space 
has zero dimension for $h=1$, and unit complex dimension for 
$h=0$.

\subsection{Three Abelian vertices}

\subsubsection*{\underline{Non-unidirectional chain}}

\begin{figure}
\begin{center}
\includegraphics{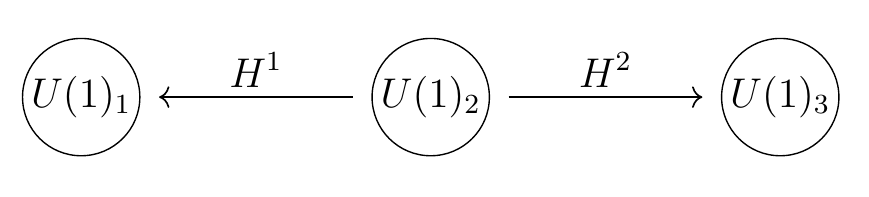}
\end{center}
\caption{The quiver diagram of three Abelian vertices of the non-unidirectional chain.}
\label{non-unidirectional chain}
\end{figure}

We next consider  a quiver diagram with three Abelian vertices.
The first example is two matter fields (edges) between three vertices.
Orientations of the edges are from the second to the first and
from the second to the third; i.e.~the two arrows are emitted from the
second vertex and oriented
in opposite directions to each other.
The quiver diagram is depicted in Fig.~\ref{non-unidirectional chain}.

The incidence matrix is given by
\be
L = \left(
\begin{array}{rr}
-1 & 0 \\
1 & 1 \\
0 & -1
\end{array}
\right),
\ee
and the BPS equations are
\be
\begin{split}
\mu^1 &= F^1 - \frac{g_1^2}{2}\left( \zeta^1 + \Hb^1H^1 \right)
\omega=0, \\
\mu^2 &= F^2 - \frac{g_2^2}{2}\left( \zeta^2  - H^1\Hb^1 - H^2\Hb^2 \right)\omega=0,\\
\mu^3 &= F^3 - \frac{g_3^2}{2}\left( \zeta^3  + \Hb^2 H^2 \right)\omega=0,\\
\nu^e &=2\del_A \Hb^e = 0,\\
\nub^e & = 2\delb_A H^e =0.
\end{split}
\ee
Integrating $\mu^v$ on $\Sigma_h$, we find a constraint and the Bradlow bounds
\be
\begin{split}
&B^1+B^2+B^3 = 0,\\
&B^1\leq 0, \quad B^2\geq 0, \quad B^3 \leq 0.
\end{split}
\ee

The volume of the moduli space is expressed by an integral
over $\phi_0^1$, $\phi_0^2$ and $\phi_0^3$
\be
\left\langle
e^{i\beta {\cal I}^\epsilon_V(g_v)}
\right\rangle^{g_{0,v}=g_{c,v}}_{k^1,k^2,k^3}
={\cal N}_C'
\int \frac{d\phi_0^1}{2\pi}\frac{d\phi_0^2}{2\pi}\frac{d\phi_0^3}{2\pi}
J_\epsilon(\phi_0^1,\phi_0^2,\phi_0^3),
\label{integral without loop-nonunidir}
\ee
where the integrand $J_\epsilon(\phi_0^1,\phi_0^2,\phi_0^3)$
is a rational function of $\phi_0^1$, $\phi_0^2$ and $\phi_0^3$
with poles.
Introducing notations
\be
\phi_0^{vv'} \equiv \phi_0^v - \phi_0^{v'}, \quad 
k^{vv'}  \equiv k^v - k^{v'},
\ee
the integrand is given by
\be
J_\epsilon(\phi_0^1,\phi_0^2,\phi_0^3)
\equiv\frac{(\det \Omega)^h \,  e^{-2\pi i \beta\left(
\phi_0^1B^1+\phi_0^2 B^2+\phi_0^3 B^3
\right)}}
{\left(-i\phi_0^{21}+\epsilon^1\right)^{k^{21}+\frac{1}{2}\chi_h}
\left(-i\phi_0^{23}+\epsilon^2\right)^{k^{23}+\frac{1}{2}\chi_h}
},
\ee
for this model after turning on the twisted masses $\epsilon^1$ and $\epsilon^2$ for each edge, where
\be
\det \Omega = \frac{\beta}{g_1^2g_2^2g_3^2}
\Bigg(
\beta^2 + \frac{\beta}{2\pi}\left(
\frac{g_1^2+g_2^2}{-i\phi_0^{21}+\epsilon^1}
+\frac{g_2^2+g_3^2}{-i\phi_0^{23}+\epsilon^2}
\right)\\
+\frac{1}{(2\pi)^2}
\frac{g_1^2g_2^2+g_2^2g_3^2+g_3^2g_1^2}{\left(-i\phi_0^{21}+\epsilon^1\right)\left(-i\phi_0^{23}+\epsilon^2\right)}
\Bigg).
\ee

Integrating $\phi_0^3$ and $\phi_0^1$ first, we obtain
\be
\left\langle
e^{i\beta {\cal I}^\epsilon_V(g_v)}
\right\rangle^{g_{0,v}=g_{c,v}}_{k^1,k^2,k^3}
={\cal N}_C'
\int \frac{d\phi_0^2}{2\pi}
\Res_{\phi_0^1=\phi_0^2+i\epsilon_1}
\Res_{\phi_0^3=\phi_0^2+i\epsilon_2}J_\epsilon(\phi_0^1,\phi_0^2,\phi_0^3),
\ee
and the condition $B^1<0$ and $B^3<0$
(and $k^{21}+\frac{1}{2}\chi_h>0$ and $k^{23}+\frac{1}{2}\chi_h>0$) is needed
to contain poles inside the contour and get non-vanishing value.
The final integral depends only on $\phi_0^2$
such as $e^{-2\pi i \beta \phi_0^2(B^1+B^2+B^3)}$,
which reduces to the constraint $B^1+B^2+B^3=0$
as expected.
(So we also have $B^2>0$.)

More concretely, if we consider the case on the sphere ($h=0$),
we find
\be
\left\langle
e^{i\beta {\cal I}^\epsilon_V(g_v)}
\right\rangle^{g_{0,v}=g_{c,v}}_{k^1,k^2,k^3}
={\cal N}_C'
\int \frac{d\phi_0^2}{2\pi}
e^{-2\pi i \beta \phi_0^2(B^1+B^2+B^3)}\,
\frac{(-2\pi B^1)^{k^{21}}}{k^{21}!}
\frac{(-2\pi B^3)^{k^{23}}}{k^{23}!}
e^{-2\pi \beta (\epsilon^1 B^1+\epsilon^2 B^2)},
\ee
which is finite in the limit of $\epsilon^e\to 0$
and proportional to a product of two volumes of the moduli space
of ani-vortices with $N_f=1$.

For higher genus case $h\geq 0$, we can also perform the integral in the
similar way by expanding $(\det \Omega)^h$.

\subsubsection*{\underline{Unidirectional chain}}

\begin{figure}
\begin{center}
\includegraphics{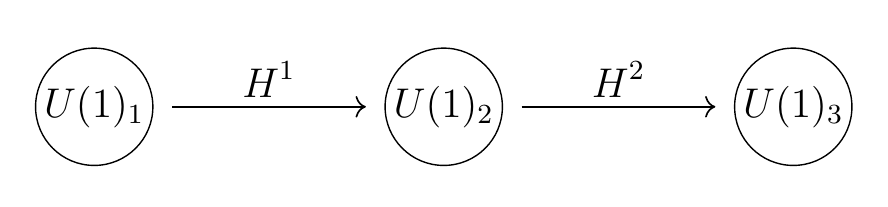}
\end{center}
\caption{The quiver diagram of three Abelian vertices of the unidirectional chain
(oriented arrows).}
\label{unidirectional chain}
\end{figure}

Next we consider a quiver chain with three vertices and oriented (unidirectional)
arrows.
The quiver diagram is depicted in Fig.~\ref{unidirectional chain}.

The incidence matrix and associated BPS vortex equations
are given by
\be
L = \left(
\begin{array}{rr}
1 & 0 \\
-1 & 1 \\
0 & -1
\end{array}
\right),
\ee
and
\be
\begin{split}
\mu^1 &= F^1 - \frac{g_1^2}{2}\left( \zeta^1 - H^1 \Hb^1 \right)
\omega=0, \\
\mu^2 &= F^2 - \frac{g_2^2}{2}\left( \zeta^2  +\Hb^1H^1 - H^2\Hb^2 \right)\omega=0,\\
\mu^3 &= F^3 - \frac{g_3^2}{2}\left( \zeta^3  + \Hb^2 H^2 \right)\omega=0,\\
\nu^e &=2\del_A \Hb^e = 0,\\
\nub^e & = 2\delb_A H^e =0.
\end{split}
\ee
Expected constraint and Bradlow bounds from the BPS equations
are
\be
\begin{split}
&B^1+B^2+B^3=0,\\
&B^1\geq 0, \quad B^3\leq 0.
\end{split}
\ee

The volume of the moduli space is given by
\be
\left\langle
e^{i\beta {\cal I}^\epsilon_V(g_v)}
\right\rangle^{g_{0,v}=g_{c,v}}_{k^1,k^2,k^3}
={\cal N}_C'
\int \frac{d\phi_0^2}{2\pi}
\Res_{\phi_0^1=\phi_0^2-i\epsilon_1}
\Res_{\phi_0^3=\phi_0^2+i\epsilon_2}
J_\epsilon(\phi_0^1,\phi_0^2,\phi_0^3),
\ee
where
\be
J_\epsilon(\phi_0^1,\phi_0^2,\phi_0^3)
\equiv\frac{(\det \Omega)^h \,  e^{-2\pi i \beta\left(
\phi_0^1B^1+\phi_0^2 B^2+\phi_0^3 B^3
\right)}}
{\left(-i\phi_0^{12}+\epsilon^1\right)^{k^{12}+\frac{1}{2}\chi_h}
\left(-i\phi_0^{23}+\epsilon^2\right)^{k^{23}+\frac{1}{2}\chi_h}
},
\ee
and
\be
\det \Omega = \frac{\beta}{g_1^2g_2^2g_3^2}
\Bigg(
\beta^2 + \frac{\beta}{2\pi}\left(
\frac{g_1^2+g_2^2}{-i\phi_0^{12}+\epsilon^1}
+\frac{g_2^2+g_3^2}{-i\phi_0^{23}+\epsilon^2}
\right)\\
+\frac{1}{(2\pi)^2}
\frac{g_1^2g_2^2+g_2^2g_3^2+g_3^2g_1^2}
{\left(-i\phi_0^{12}+\epsilon^1\right)\left(-i\phi_0^{23}+\epsilon^2\right)}
\Bigg).
\ee

Only the difference from the previous case, we obtain the bounds and 
constraint as $B^1>0$, $B^3<0$ and $B^1+B^2+B^3=0$.
For the sphere ($h=0$), we also find the volume is
proportional to a product of two finite moduli space of vortex and
anti-vortex as
\be
\frac{(2\pi B^1)^{k^{12}}}{k^{12}!}
\frac{(-2\pi B^3)^{k^{23}}}{k^{23}!}.
\ee
This means that total moduli space is compact and determined by
the vortex from $\mu^1=0$ and anti-vortex from $\mu^3=0$
with $N_f=1$.
As a result, the moduli space determined from $\mu^2=0$
still remains finite.

\subsubsection*{\underline{Non-unidirectional closed loop}}

\begin{figure}
\begin{center}
\includegraphics{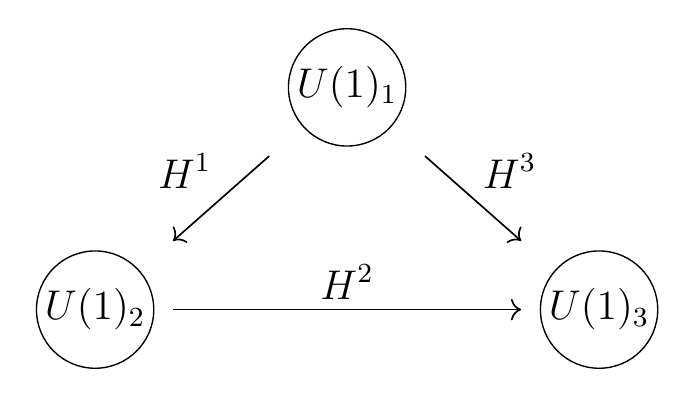}
\end{center}
\caption{The quiver diagram of three Abelian vertices with a
non-unidirectional loop.}
\label{three vertices with non-unidirectional loop}
\end{figure}

There are arrows (edges) between the vertices, but
two arrows are emitted from the first vertex $U(1)_1$
and one arrow is started from $U(1)_2$ and ended at $U(1)_3$.
So there is a loop without one way (unidirectional) arrows.
We depicted the quiver diagram in Fig.~\ref{three vertices with non-unidirectional loop}.
 
The incidence matrix is given by
\be
L=\left(
\begin{array}{rrr}
1 & 0 & 1 \\
-1 & 1 & 0 \\
0 & -1 & -1
\end{array}
\right).
\ee
The BPS vortex equations are
\be
\begin{split}
\mu^1 &= F^1 - \frac{g_1^2}{2}\left( \zeta^1 - H^1 \Hb^1 -  H^3\Hb^3\right)
\omega=0, \\
\mu^2 &= F^2 - \frac{g_2^2}{2}\left( \zeta^2  -  H^2 \Hb^2 + \Hb^1 H^1  \right)\omega=0,\\
\mu^3 &= F^3 - \frac{g_3^2}{2}\left( \zeta^3  +  \Hb^3 H^3+ \Hb^2 H^2   \right)\omega=0,\\
\nu^e &=2\del_A \Hb^e = 0,\\
\nub^e & = 2\delb_A H^e =0.
\end{split}
\ee
Integrating $\mu^v$ on $\Sigma_h$, we get a constraint and bounds
\be
\begin{split}
& B^1 + B^2 + B^3 = 0,\\
& B^1 \geq 0, \quad B^3 \leq 0,\\
& B^1-B^3 \geq 0.
\end{split}
\ee

The integral formula of the volume is given by
\be
\left\langle
e^{i\beta {\cal I}^\epsilon_V(g_v)}
\right\rangle^{g_{0,v}=g_{c,v}}_{k^1,k^2,k^3}
={\cal N}_C'
\int \frac{d\phi_0^1}{2\pi}\frac{d\phi_0^2}{2\pi}\frac{d\phi_0^3}{2\pi}
J_\epsilon(\phi_0^1,\phi_0^2,\phi_0^3),
\label{integral without loop}
\ee
where
\be
\begin{split}
J_\epsilon(\phi_0^1,\phi_0^2,\phi_0^3)
&\equiv\frac{(\det \Omega)^h \,  e^{-2\pi i \beta\left(
\phi_0^1B^1+\phi_0^2 B^2+\phi_0^3 B^3
\right)}}
{\left(-i\phi_0^{12}+\epsilon^1\right)^{k^{12}+\frac{1}{2}\chi_h}
\left(-i\phi_0^{23}+\epsilon^2\right)^{k^{23}+\frac{1}{2}\chi_h}
\left(-i\phi_0^{13}+\epsilon^3\right)^{k^{13}+\frac{1}{2}\chi_h}},\\
\phi_0^{vv'} &\equiv \phi_0^v - \phi_0^{v'},\\ 
k^{vv'} & \equiv k^v - k^{v'},
\end{split}
\ee
and
\begin{multline}
\det \Omega = \frac{\beta}{g_1^2g_2^2g_3^2}
\Bigg(
\beta^2 + \frac{\beta}{2\pi}\left(
\frac{g_1^2+g_2^2}{-i\phi_0^{12}+\epsilon^1}
+\frac{g_2^2+g_3^2}{-i\phi_0^{23}+\epsilon^2}
+\frac{g_1^2+g_3^2}{-i\phi_0^{13}+\epsilon^3}
\right)\\
+\frac{g_1^2g_2^2+g_2^2g_3^2+g_3^2g_1^2}{(2\pi)^2}
\frac{-2i\phi_0^{13}+\e^1+\e^2+\e^3}{\left(-i\phi_0^{12}+\epsilon^1\right)
\left(-i\phi_0^{23}+\epsilon^2\right)
\left(-i\phi_0^{13}+\epsilon^3\right)}\Bigg).
\end{multline}

Integrating $\phi_0^1$ and $\phi_0^3$
of (\ref{integral without loop}) first in turn,
we obtain
\be
\begin{split}
\left\langle
e^{i\beta {\cal I}^\epsilon_V(g_v)}
\right\rangle^{g_{0,v}=g_{c,v}}_{k^1,k^2,k^3}
&={\cal N}_C'\int \frac{d\phi_0^2}{2\pi}
\Big[
\Res_{\phi^1_0=\phi_0^2-i\epsilon^1}
\Res_{\phi^3_0=\phi_0^2+i\epsilon^2}\\
&\qquad\qquad\qquad
+\Res_{\phi^1_0=\phi_0^2-i(\epsilon^2-\epsilon^3)}\Res_{\phi^3_0=\phi_0^2+i\epsilon^2}\\
&\qquad\qquad\qquad\quad+\Res_{\phi^1_0=\phi_0^2-i\epsilon^1}\Res_{\phi^3_0=\phi_0^1+i\epsilon^3}\\
&\qquad\qquad\qquad\qquad+\Res_{\phi^1_0=\phi_0^2-i(\epsilon^2-\epsilon^3)}\Res_{\phi^3_0=\phi_0^1+i\epsilon^3}
\Big]
J_\epsilon(\phi_0^1,\phi_0^2,\phi_0^3).
\end{split}
\ee
To pick up the residues of the poles inside the contour
and obtain a non-vanishing volume,
we need to assume that $B^1>0$ and $B^3<0$, 
which agrees with the Bradlow bound.
In the final integral of $\phi_0^2$, the integrand depends
only on $\phi_0^2$ through the factor
$e^{-2\pi i \beta \phi_0^2 (B^1+B^2+B^3)}$.
So the integral by $\phi_0^2$ gives the constraint
$B^1+B^2+B^3=0$.

For simplicity, let us consider a vacuum on the sphere
($k^v=0$ and $h=0$).
In this case, we can ignore the contribution from $\det \Omega$.
The volume of the moduli space is proportional to
\be
\frac{e^{-\beta(\epsilon^1
\zeta^1+(\epsilon^1-\epsilon^3)\zeta^3) \A/2}}{\epsilon^1+\epsilon^2-\epsilon^3}
\left(
1
-e^{\beta(\epsilon^1+\epsilon^2-\epsilon^3) \zeta^3 \A/2}
\right),
\ee
which takes a finite value $-  \beta \zeta^3 \A/2$
 in the limit of $\epsilon^e \to 0$.
So we can see the moduli space of vacua is compact at least.

\subsubsection*{\underline{Unidirectional closed loop}}

\begin{figure}
\begin{center}
\includegraphics{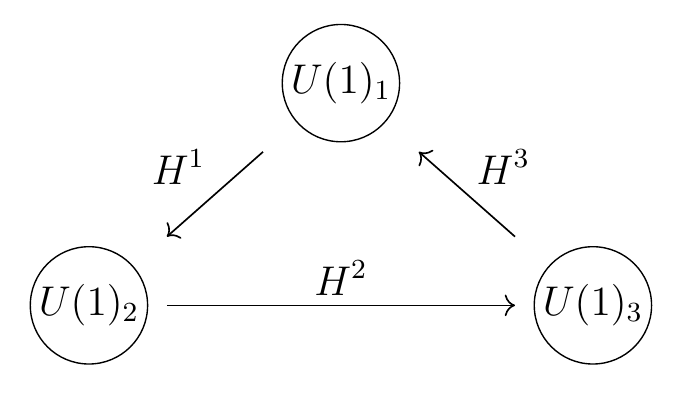}
\end{center}
\caption{The quiver diagram of three Abelian vertices with a
unidirectional loop.}
\label{three vertices with unidirectional loop}
\end{figure}

Let us consider one more case of the three vertices.
In contrast with the previous case, all arrows are aligned in
one direction (unidirectional) on the loop.
The quiver diagram is depicted in Fig.~\ref{three vertices with unidirectional loop}.

The incidence matrix is given by
\be
L=\left(
\begin{array}{rrr}
1 & 0 & -1\\
-1 & 1 & 0\\
0 & -1 & 1
\end{array}
\right).
\ee
The associated BPS vortex equations are
\be
\begin{split}
\mu^1 &= F^1 - \frac{g_1^2}{2}\left( \zeta^1 - H^1 \Hb^1+\Hb^3 H^3 \right)
\omega=0, \\
\mu^2 &= F^2 - \frac{g_2^2}{2}\left( \zeta^2  -  H^2 \Hb^2+ \Hb^1 H^1 \right)\omega=0,\\
\mu^3 &= F^3 - \frac{g_3^2}{2}\left( \zeta^3   - H^3 \Hb^3 + \Hb^2 H^2  \right)\omega=0,\\
\nu^e &=2\del_A \Hb^e = 0,\\
\nub^e & = 2\delb_A H^e =0.
\end{split}
\ee
Each $\mu^v$ contains both of positive and negative charge
matters.
So the moduli space of vacua at least is non-compact.

The volume of the moduli space is given by the following integral
\be
\left\langle
e^{i\beta {\cal I}^\epsilon_V(g_v)}
\right\rangle^{g_{0,v}=g_{c,v}}_{k^1,k^2,k^3}
={\cal N}_C'
\int \frac{d\phi_0^1}{2\pi}\frac{d\phi_0^2}{2\pi}\frac{d\phi_0^3}{2\pi}
J_\epsilon(\phi_0^1,\phi_0^2,\phi_0^3),
\label{integral with unidirectional loop}
\ee
where
\be
J_\epsilon(\phi_0^1,\phi_0^2,\phi_0^3)
\equiv\frac{(\det \Omega)^h \,  e^{-2\pi i \beta\left(
\phi_0^1B^1+\phi_0^2 B^2+\phi_0^3 B^3
\right)}}
{\left(-i\phi_0^{12}+\epsilon^1\right)^{k^{12}+\frac{1}{2}\chi_h}
\left(-i\phi_0^{23}+\epsilon^2\right)^{k^{23}+\frac{1}{2}\chi_h}
\left(-i\phi_0^{31}+\epsilon^3\right)^{k^{31}+\frac{1}{2}\chi_h}},
\ee
and
\begin{multline}
\det \Omega = \frac{\beta}{g_1^2g_2^2g_3^2}
\Bigg(
\beta^2 + \frac{\beta}{2\pi}\left(
\frac{g_1^2+g_2^2}{-i\phi_0^{12}+\epsilon^1}
+\frac{g_2^2+g_3^2}{-i\phi_0^{23}+\epsilon^2}
+\frac{g_3^2+g_1^2}{-i\phi_0^{31}+\epsilon^3}
\right)\\
+\frac{g_1^2g_2^2+g_2^2g_3^2+g_3^2g_1^2}{(2\pi)^2}
\frac{\e^1+\e^2+\e^3}{\left(-i\phi_0^{12}+\epsilon^1\right)
\left(-i\phi_0^{23}+\epsilon^2\right)
\left(-i\phi_0^{31}+\epsilon^3\right)}\Bigg).
\end{multline}

Thus the volume of the moduli space is given by residues of $J_\epsilon$
\begin{multline}
\left\langle
e^{i\beta {\cal I}^\epsilon_V(g_v)}
\right\rangle^{g_{0,v}=g_{c,v}}_{k^1,k^2,k^3}\\
=\begin{cases}
{\cal N}_C'\int \frac{d\phi_0^1}{2\pi}
\Res_{\phi^2_0=\phi_0^1-i(\epsilon^2+\epsilon^3)}\Res_{\phi^3_0=\phi_0^1-i\epsilon^3}J_\epsilon(\phi_0^1,\phi_0^2,\phi_0^3)
& \text{if } B^3>0 \text{ and } B^2>0\\
{\cal N}_C'\int \frac{d\phi_0^1}{2\pi}
\Res_{\phi^2_0=\phi_0^1+i\epsilon^1}
\Res_{\phi^3_0=\phi_0^1-i\epsilon^3}J_\epsilon(\phi_0^1,\phi_0^2,\phi_0^3)
& \text{if } B^3>0 \text{ and } B^2<0\\
{\cal N}_C'\int \frac{d\phi_0^1}{2\pi}\Res_{\phi^2_0=\phi_0^1-i(\epsilon^2+\epsilon^3)}\Res_{\phi^3_0=\phi_0^2+i\epsilon^2}J_\epsilon(\phi_0^1,\phi_0^2,\phi_0^3)
& \text{if } B^3<0 \text{ and } B^2>0\\
{\cal N}_C'\int \frac{d\phi_0^1}{2\pi}\Res_{\phi^2_0=\phi_0^1+i\epsilon^1}\Res_{\phi^3_0=\phi_0^2+i\epsilon^2}
J_\epsilon(\phi_0^1,\phi_0^2,\phi_0^3)
& \text{if } B^3<0 \text{ and } B^2<0.
\end{cases}
\label{each residues}
\end{multline}
In particular, if we consider the vacuum on the sphere ($k^v=0$ and $h=0$),
we obtain
\be
\left\langle
e^{i\beta {\cal I}^\epsilon_V(g_v)}
\right\rangle^{g_{0,v}=g_{c,v}}_{k^1,k^2,k^3}
=-{\cal N}_C'\int \frac{d\phi_0^1}{2\pi}
e^{-i \beta \phi_0^1(\zeta^1+\zeta^2+\zeta^3)\A/2}
\frac{e^{-\beta\left((\epsilon^2+\epsilon^3)\zeta^2+\epsilon^3 \zeta^3\right)\A/2}}
{\epsilon^1+\epsilon^2+\epsilon^3},
\ee
if $B^3>0$ and $B^2>0$.
Including all other cases,
each volume is proportional to $1/(\epsilon^1+\epsilon^2+\epsilon^3)$ and diverges in the limit of $\epsilon^e \to 0$.
This means that the volume of the moduli space of vacua is non-compact.

Finally, we would like to comment on an interesting fact.
Each residues in (\ref{each residues}) diverges in the limit
of $\epsilon^e \to 0$ as mentioned, but the total sum of the volumes
of the vacua
for each region
becomes
\be
{\cal N}_C'\int \frac{d\phi_0^1}{2\pi}
e^{- i \beta \phi_0^1(\zeta^1+\zeta^2+\zeta^3)\A/2}
\frac{e^{\beta(\epsilon^1 \zeta^2-\epsilon^3 \zeta^3)\A/2}
\left(e^{\beta(\epsilon^1+\epsilon^2+\epsilon^3)\zeta^3\A/2}-1\right)}
{\epsilon^1+\epsilon^2+\epsilon^3},
\ee
which is finite in the limit of $\epsilon^e \to 0$.
This is similar to the previous case of the compact moduli space.
The contributions of the divergence from each region
seem to be complementary to each other.

\section{Application to Vortex in Gauged Non-linear Sigma Model}
\label{GNLSM}

In this section, we would like to consider vortices in 
gauged non-linear sigma model on a generic Riemann
surface $\Sigma_h$ with a genus $h$.
If we assume that a target space is a K\"ahler manifold $X$,
the (anti-)BPS vortex equation is defined by
\be
\begin{split}
\mu &= F -\frac{g^2}{2}\left(\zeta + ||Z||^2\right)=0,\\
\nu^i &= 2\del_A \bar{Z}^i = 0,\\
\nub^i &= 2\delb_A Z^i = 0,
\label{GNLSM BPS}
\end{split}
\ee
where 
$Z^i$ are the (inhomogeneous) coordinates of $X$ and $||Z||^2$ 
is a positive definite mapping function from $X$ to $\R$ (moment map), 
which is invariant under a part of $U(1)$ isometries of $X$.
The $U(1)$ gauge symmetry is regarded as a gauging of the 
$U(1)$ isometry of $X$, under which the moment map 
$||Z||^2$ is invariant. 
For later convenience, we here consider the anti-BPS equation;
i.e.~the flux $k=\frac{1}{2\pi}\int_{\Sigma_h}F$ 
and FI parameter $\zeta$ should be negative. 
We will consider only the case of $X=\C P^N$ 
as an example in the following. 

According to \cite{Romao:2018egg}, 
the above vortex system in gauged non-linear sigma model 
can be obtained in a strong coupling limit of a certain gauged 
{\it linear} sigma model (GLSM). 
The GLSM contains two $U(1)$ gauge groups 
and two kinds of matter (Higgs) fields, 
whose total number is $N+1$ ($N+1$ arrows in total).
$n$ of the matter fields are 
charged with respect to both $U(1)$ 
and denoted as $H^e$ ($e=1,2,\ldots,n$). 
The other $N-n+1$ matter fields have positive charges only on 
one $U(1)$ and are denoted as $H^{e'}$ ($e'=n+1,n+2,\ldots,N+1$).
We call this model as a parent GLSM following \cite{Romao:2018egg}.

The parent GLSM is expressed in terms of the quiver gauge theory.
The generic quiver gauge theory contains only the bi-fundamental 
matters (charged under two $U(1)$ vertices), and no fields in 
the fundamental representation. 
However, we can introduce a decoupled vertex, which is defined 
as a  vertex with decoupled gauge fields; i.e.~the gauge coupling 
on that vertex is taken in the weak coupling limit. 
Since the decoupled vertex $U(N)$ stands for a global $U(N)$ 
symmetry instead of a local symmetry, a matter associated with 
an arrow between a $U(N_c)$ vertex and a decoupled vertex becomes 
$N$ flavors of fields in the fundamental representation of 
$U(N_c)$ (charged fields if $N_c=1$). 
We will denote the decoupled vertex in terms of box vertex. 

\begin{figure}
\begin{center}
\includegraphics{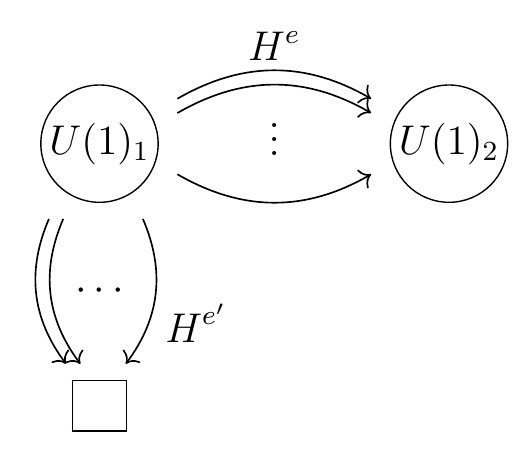}
\end{center}
\caption{The quiver diagram with two Abelian vertices
which is a parent of the gauged non-linear $\sigma$-model.
The box stands for a decoupled vertex.}
\label{parent GLSM}
\end{figure}

The quiver diagram of the parent GLSM is depicted in Fig.~\ref{parent GLSM}.
There is $n$ arrows between two $U(1)$ vertices, which correspond to $H^e$.
We also have $N-n+1$ arrows from one $U(1)$ to the fixed vertex,
which represent the matter fields $H^{e'}$.

Taking care with the charges (representations) of the matter fields,
we can write down the BPS vortex equations 
for the parent GLSM as
\be
\begin{split}
\mu^1 &= F^1 - \frac{g_1^2}{2}\left( \zeta^1 - \sum_{e=1}^{n}H^e \Hb^e
- \sum_{e'=n+1}^{N+1} H^{e'}\Hb^{e'}\right)
\omega=0, \\
\mu^2 &= F^2 - \frac{g_2^2}{2}\left( \zeta^2  
+ \sum_{e=1}^{n}\Hb^e H^e  \right)\omega=0,\\
\nu^e & = 2\left(\del - iA^1+iA^2\right)\Hb^e=0,\\
\nub^e &  = 2\left(\delb + i\Ab^1-i\Ab^2\right)H^e=0,\\
\nu^{e'} & = 2\left(\del - iA^1\right)\Hb^{e'}= 0,\\
\nub^{e'} & = 2\left(\delb + i\Ab^1\right)H^{e'}=0.
\end{split}
\ee
Integrating $\mu^v$ on $\Sigma_h$, we can see Bradlow bounds
\be
B^1 \geq 0, \quad B^2 \leq 0.
\ee
In addition to the above, if we consider a linear combination $\mu^1+\mu^2$,
then we particularly have
\be
B^1+B^2 \geq 0.
\ee
In contrast with the generic quiver model in the previous section,
there is no constraint on $B^v$'s,
reflecting the existence of the decoupled vertex.

In the parent GLSM, we have two gauge couplings $g_1$ and $g_2$
associated with two $U(1)$'s.
If we take a strong coupling limit of one gauge coupling $g_1\to \infty$,
the matter fields are captured on a constraint
\be
 \sum_{e=1}^{n} H^e \Hb^e + \sum_{e'=n+1}^{N+1} H^{e'}\Hb^{e'} = \zeta^1.
\ee
Using this constraint and quotient by $U(1)$ gauge symmetry,
we can regard $H^e$ as a set of the inhomogeneous coordinate of $\C P^N$.
Thus we expect that
\be
\mu^2=\nu^e=\nub^e=0\quad (e=1,\ldots,n),
\ee
express the BPS equation (\ref{GNLSM BPS}) of
the (anti-)vortex of the gauged non-linear sigma model with the target $\C P^N$
in the strong coupling limit $g_1\to \infty$.

We are interested in the volume of the moduli space of the vortex in the
gauged non-linear sigma model with the target $\C P^N$.
However we first would like
to derive the volume of the moduli space of the parent quiver theory
by using the integral formula in the Coulomb branch,
since we can obtain the non-linear sigma model in the strong coupling limit.

The incidence matrix of the parent quiver theory is given by
\be
\raisebox{-1.0ex}{$L=\,$}
\begin{blockarray}{rcrrcr}
 \BAmulticolumn{3}{r}{\overbrace{\hspace{60pt}}^{n}}
& \BAmulticolumn{3}{c}{\overbrace{\hspace{48pt}}^{N-n+1}}\\
\begin{block}{(rcrrcr)}
1 & \cdots & 1 & 1 & \cdots &1\\
-1 & \cdots & -1 & 0 & \cdots &0\\
\end{block}
\end{blockarray}
\ ,
\ee
The $N-n+1$ right-most columns  represent the matters (arrows) 
from one $U(1)$ vertex to the decoupled vertex and contains 
only the positive charge $+1$.
This point is rather special than the usual incidence matrix of the 
oriented graph.

Using the generic integral formula for the quiver theory,
the volume of the moduli space of the vortices
in the parent GLSM is given by
\be
\left\langle
e^{i\beta {\cal I}^\epsilon_V(g_v)}
\right\rangle^{g_{0,v}=g_{c,v}}_{k^1,k^2}
={\cal N}_C'
\int \frac{d\phi_0^1}{2\pi}\frac{d\phi_0^2}{2\pi}
J_\epsilon(\phi_0^1,\phi_0^2).
\ee
Turning on twisted masses $\e$ and $\e'$ for $H^e$ and $H^{e'}$,
respectively, the integrand becomes
\be
J_\epsilon(\phi_0^1,\phi_0^2)
=\frac{(\det\Omega_\e)^h\,e^{-2\pi i \beta\left(\phi_0^1B^1+\phi_0^2 B^2\right)}}
{(-i\phi_0^{12}+\epsilon)^{n\left(k^{12}+\frac{1}{2}\chi_h\right)}(-i\phi_0^1+\epsilon')^{(N-n+1)\left(k^1+\frac{1}{2}\chi_h\right)}},
\ee
where
\be
\begin{split}
\det \Omega_\e &= \frac{1}{g_1^2g_2^2}\left(\beta^2
+\frac{\beta}{2\pi}\left(\frac{n(g_1^2+g_2^2)}{-i\phi_0^{12}+\epsilon}
+\frac{(N-n+1)g_1^2}{-i\phi_0^1+\epsilon'}\right)
+\frac{1}{(2\pi)^2}\frac{n(N-n+1)g_1^2 g_2^2}{(-i\phi_0^{12}+\epsilon)(-i\phi_0^1+\epsilon')}
\right)\\
&=\left(\frac{\beta}{g_2^2}+\frac{1}{2\pi}\frac{n}{-i\phi_0^{12}+\epsilon}\right)
\left(\frac{\beta}{g_1^2}+\frac{\beta}{g_2^2}
+\frac{1}{2\pi}\frac{N-n+1}{-i\phi_0^1+\epsilon'}\right)-\frac{\beta^2}{g_2^4}\,.
\end{split}
\ee
Here we rearranged $\det \Omega_\e$ in the final form for later convenience.

We first expand $(\det \Omega_\e)^h$ by using the binomial theorem
as
\be
\begin{split}
(\det \Omega_\e)^h
&=\sum_{l=0}^h\binom{h}{l}\left(-\frac{\beta^2}{g_2^4}\right)^l
\left(\frac{\beta}{g_2^2}+\frac{1}{2\pi}\frac{n}{-i\phi_0^{12}+\epsilon}\right)^{h-l}
\left(\frac{\beta}{g_1^2}+\frac{\beta}{g_2^2}
+\frac{1}{2\pi}\frac{N-n+1}{-i\phi_0^1+\epsilon'}\right)^{h-l}\\
&=\sum_{l=0}^h\binom{h}{l}\left(-\frac{\beta^2}{g_2^4}\right)^l
\left\{\sum_{j^{12}=l}^{h}
\binom{h-l}{j^{12}-l}\left(\frac{\beta}{g_2^2}\right)^{j^{12}-l}
\left(\frac{1}{2\pi}\frac{n}{-i\phi_0^{12}+\epsilon}\right)
^{h-j^{12}}\right\}\\
&\qquad\qquad\qquad\quad
\times
\left\{\sum_{j^{1}=l}^{h}
\binom{h-l}{j^{1}-l}\left(\frac{\beta}{g_1^2}+\frac{\beta}{g_2^2}\right)^{j^{1}-l}
\left(\frac{1}{2\pi}\frac{N-n+1}{-i\phi_0^{1}+\epsilon'}\right)
^{h-j^{1}}\right\}\,. 
\end{split}
\ee
Then the integrand $J_\e(\phi_0^1,\phi_0^2)$ also can be expanded
as follows
\be
\begin{split}
J_\epsilon(\phi_0^1,\phi_0^2)
&=
\sum_{l=0}^h \frac{h!(h-l)!}{(-1)^ll!}\left(\frac{\beta}{g_2^2}\right)^{2l}\\
&\qquad\qquad
\times
\left\{\sum_{j^{12}=l}^{h}
\frac{
\left(\frac{n}{2\pi}\right)^{h-j^{12}}
\left(\frac{\beta}{g_2^2}\right)^{j^{12}-l}
}
{(j^{12}-l)!(h-j^{12})!}
\frac{1}{(-i\phi_0^{12}+\epsilon)^{d^{12}-j^{12}+1}}
\right\}\\
&\qquad\qquad
\times
\left\{\sum_{j^{1}=l}^{h}
\frac{
\left(\frac{N-n+1}{2\pi}\right)^{h-j^1}
\left(\frac{\beta}{g_1^2}+\frac{\beta}{g_2^2}\right)^{j^{1}-l}
}
{(j^1-l)!(h-j^1)!}
\frac{1}{(-i\phi_0^{1}+\epsilon')^{d^1-j^1+1}}
\right\}\\
&\qquad\qquad
\times
e^{-2\pi i \beta (\phi_0^1 B^1+\phi_0^2 B^2)},
\end{split}
\ee
where we have defined
 $d^{12}\equiv n k^{12}+(n-1)(1-h)$ and
$d^1\equiv (N-n+1)k^1+(N-n)(1-h)$.

Using this expansion, we can integrate $\phi_0^2$ and $\phi_0^1$
in turn. The volume is expressed in terms of the residues
for each term
\be
\begin{split}
\left\langle
e^{i\beta {\cal I}^\epsilon_V(g_v)}
\right\rangle^{g_{0,v}=g_{c,v}}_{k^1,k^2}
&={\cal N}_C'\Res_{\phi_0^1=-i\e'}\Res_{\phi_0^2=\phi_0^1+i\e}
J_\epsilon(\phi_0^1,\phi_0^2)\\
&=
{\cal N}_C'\frac{(2\pi\beta)^{d^{12}+d^1}}{(2\pi)^{2h}}
\sum_{l=0}^h \frac{h!(h-l)!}{(-1)^ll!}\left(\frac{1}{g_2^2}\right)^{2l}\\
&\qquad\qquad
\times
\left\{\sum_{j^{12}=l}^{h}
\frac{n^{h-j^{12}}\left(\frac{1}{g_2^2}\right)^{j^{12}-l}
\left(-B^2\right)^{d^{12}-j^{12}}}
{(j^{12}-l)!(h-j^{12})!(k^{12}-j^{12})!}
\right\}\\
&\qquad\qquad
\times
\left\{\sum_{j^{1}=l}^{h}
\frac{
(N-n+1)^{h-j^1}
\left(\frac{1}{g_1^2}+\frac{1}{g_2^2}\right)^{j^{1}-l}
(B^1+B^2)^{d^1-j^1}}
{(j^1-l)!(h-j^1)!(d^1-j^1)!}
\right\}\\
&\qquad\qquad
\times
e^{-2\pi \e B^1 -2\pi(\e'-\e)B^2}.
\end{split}
\label{Volume of parent GLSM}
\ee

We need to require at least 
\be
d^{12}= n k^{12}+(n-1)(1-h)\geq 0
\quad \text{and} \quad
d^1= (N-n+1)k^1+(N-n)(1-h)\geq 0.
\ee
So we have
\be
k^{12} \geq \max\left\{0,\frac{(n-1)(h-1)}{n}\right\}
\quad \text{and} \quad
k^{1} \geq \max\left\{0,\frac{(N-n)(h-1)}{N-n+1}\right\}.
\ee

In integrating $\phi_0^2$, we get a bound
\be
B^2<0,
\ee
to include the pole at $\phi_0^2=\phi_0^1+i\e$ inside the contour.
The integral of $\phi_0^1$ also gives
\be
B^1+B^2>0,
\ee
as a consequence of the pole at $\phi_0^1=-i\e$.
These agree with the Bradlow bounds.

The volume is always finite in the limit of $\e\to 0$ and 
$\e'\to 0$ (removing the regularization). 
Our result in Eq.~\eqref{Volume of parent GLSM} with $\e=\e'=0$ 
gives the moduli space volume of BPS vortices in the 
$U(1)_1\times U(1)_2$ GLSM with $n$ bifundamental and $N-n+1$ 
fundamental scalar fields. 
When restricted to $N=1$ and $n=1$ (corresponding to $X=\C P^1$ 
case), our GLSM reduces to that studied in \cite{Romao:2018egg}. 
The results precisely agree with each other provided different 
conventions are appropriately translated. 
Our field theoretical derivation is based on a scheme that 
is entirely different from that in \cite{Romao:2018egg}. 
Moreover, our results include the more general cases for generic 
$N$ and $n$ (corresponding to $X=\C P^N$ case with $n$ flavors 
of charged scalars) and are obtained 
without any concrete knowledge of the moduli space
including the metric.

The dimension of the moduli space is given by the overall power of $\beta$.
So we can see the dimension of the total moduli space is $d^{12}+d^1$,
which is the sum of the dimension of the moduli space
of the Abelian vortex with $n$ and $N-n+1$ flavors.
And also the volume of the moduli space (\ref{Volume of parent GLSM})
is an almost direct product of each Abelian vortex moduli space
with $n$ and $N-n+1$ matters, except for the combinatorial factor and
sums.

We found the volume of the vortex moduli space of the parent model.
As we explained above, we can expect that the volume of the vortex moduli space
of the non-linear sigma model with the target space $\C P^N$
can be obtained by the strong coupling limit of $g_1\to \infty$.

In this limit, we find
\be
B^1 \to \frac{\zeta^1 \A}{4\pi}.
\ee
Thus we obtain the volume of the vortex moduli space
of the $U(1)$ GNLSM with the target space $\C P^N$ and with 
$n$ flavors as 
\be
\begin{split}
\lim_{\substack{\e,\e'\to 0\\ g_1\to\infty}}\left\langle
e^{i\beta {\cal I}^\epsilon_V(g_v)}
\right\rangle^{g_{0,v}=g_{c,v}}_{k^1,k^2}
&=
{\cal N}_C'\frac{(2\pi\beta)^{d^{12}+d^1}}{(2\pi)^{2h}}
\sum_{l=0}^h \frac{h!(h-l)!}{(-1)^ll!}\\
&\qquad\qquad
\times
\left\{\sum_{j^{12}=l}^{h}
\frac{n^{h-j^{12}}\left(\frac{1}{g_2^2}\right)^{j^{12}}
\left(-B^2\right)^{d^{12}-j^{12}}}
{(j^{12}-l)!(h-j^{12})!(k^{12}-j^{12})!}
\right\}\\
&\qquad\qquad
\times
\left\{\sum_{j^{1}=l}^{h}
\frac{
(N-n+1)^{h-j^1}
\left(\frac{1}{g_2^2}\right)^{j^{1}}
\left(B^2+\frac{\zeta^1 \A}{4\pi}\right)^{d^1-j^1}}
{(j^1-l)!(h-j^1)!(d^1-j^1)!}
\right\}.
\end{split}
\label{Volume of GSM}
\ee

\section{Non-Abelian Generalization}
\label{Non-Abelian Generalization}

\subsection{Action and integral formula}

We now generalize the Abelian quiver gauge theory to quiver gauge theory with non-Abelian vertices.
There exist $U(N)$ non-Abelian gauge groups on each vertex. So we have quiver gauge theory with
a gauge symmetry $G=\prod_{v\in V} U(N_v)$,
where $N_v$ is a rank of gauge group on a vertex $v\in V$.

There are also directed arrows (edges) connecting between vertices, which represents
matter fields in bi-fundamental representations.
If we pick up one edge $e\in E$, the matter field transform as a fundamental representation
under the gauge group $U(N_{s(e)})$ at the source of the edge
and anti-fundamental representation under $U(N_{t(e)})$ at the target of the edge.
So the BPS quiver vortex equations for this non-Abelian theory are given by
\be
\mu^v = \nu^e = \nub^e =0.
\label{NA BPS equations}
\ee
 

Next, we consider an embedding of the above BPS vortex system into a supersymmetric gauge theory.
To define the supersymmetric gauge theory, we introduce vector multiplets and chiral multiplets.
The vector multiplets exist on each vertex $v$
and contain 0-form scalar fields $\Phi^v$, 1-form vector fields $(A^v,\Ab^v)$, 2-form auxiliary fields $Y^v$,
and their fermionic super partners $\eta^v$, $(\lambda^v,\lambdab^v)$, $\chi^v$.
All fields are $N_v\times N_v$ matrices and belong to the adjoint representation.

The supersymmetry transformations of the vector multiplets are given by
\be
\begin{array}{lcl}
Q\Phi^v = 0, &&\\
Q\Phib^v = 2\eta^v, && Q\eta^v = \frac{i}{2}[\Phi^v,\Phib^v],\\
QA^v = \lambda^v, && Q\lambda^v = -\del_A\Phi^v,\\
Q\Ab^v = \lambdab^v, && Q\lambdab^v = -\delb_A\Phi^v,\\
QY^v = i[\Phi^v,\chi^v], && Q\chi^v = Y^v,
\end{array}
\ee
where $\del_A \Phi^v\equiv \del \Phi^v+i[A^v,\Phi^v]$
and $\delb_A \Phi^v\equiv \delb \Phi^v+i[\Ab^v,\Phi^v]$.
We can see $Q^2=\delta_\Phi$, which is a gauge transformation with respect to a parameter $\Phi^v$.

The chiral multiplets correspond to each arrow on the graph.
We can devide the chiral multiplets into two sets.
One of the sets contains $N_{s(e)}\times N_{t(e)}$ matrix of 0-form bosons and fermions,
which we denote by $(H^e,\psi^e)$, and $(0,1)$-forms $(\Tb^e,\rhob^e)$.
For this part of the chiral multiplets, we can define the supersymmetry by
\be
\begin{array}{lcl}
QH^e = \psi^e, && Q\psi^e = i\Phi^v\cdot {L(H)_v}^e,\\
Q\Tb^e = i \Phi^v\cdot {L(\rhob)_v}^e, && Q \rhob^e = \Tb^e,
\end{array}
\ee
where 
\be
\begin{split}
\Phi^v\cdot {L(H)_v}^e &\equiv
\Phi^{s(e)}H^e - H^e\Phi^{t(e)},\\
\Phi^v\cdot {L(\rhob)_v}^e &\equiv
\Phi^{s(e)}\rhob^e - \rhob^e\Phi^{t(e)},
\end{split}
\ee
i.e.~${L(H)_v}^e$ is a non-Abelian generalization of the (covariant) incidence matrix,
and $\Phi^v$ is acting on the bi-fundamental representation $H^e$
in a suitable way. Again we can see $Q^2=\delta_\Phi$.

Another set of the chiral multiplets is a conjugate of the above.
We have $N_{t(e)}\times N_{s(e)}$ 0-form bosons and fermions $(\Hb^e,\psib^e)$
and $(1,0)$-forms $(T^e,\rho^e)$
\be
\begin{array}{lcl}
Q\Hb^e = \psib^e, && Q\psib^e = -i{L^T(\Hb)^e}_v\cdot\Phi^v,\\
QT^e = -i {L^T(\rho)^e}_v\cdot \Phi^v , && Q \rho_e = T_e,
\end{array}
\ee
where
\be
\begin{split}
{L^T(\Hb)^e}_v\cdot\Phi^v &\equiv
\Hb^e \Phi^{s(e)}-\Phi^{t(e)} \Hb^e,\\
{L^T(\rho)^e}_v\cdot \Phi^v &\equiv
\rho^e \Phi^{s(e)}-\Phi^{t(e)} \rho^e.
\end{split}
\ee

Using these multiplets and supersymmetry transformations, 
we can define the supersymmetric action
in a $Q$-exact form
\be
S = Q\Xi_V + Q\Xi_C,
\ee
where
\bea
\Xi_V&=&  \Tr\left[
\langle \lambda_v, \del_A \Phi^v \rangle
+\langle \lambdab_v, \delb\Phi^v\rangle
+\langle \eta_v, \frac{i}{2}[\Phi^v,\Phib^v]\rangle
+\langle\chi_v,Y^v-2\mu^v_0\rangle
\right],\\
\Xi_C &=& \frac{1}{2}\Tr \bigg[
\langle \psi_e ,i\Phi^v\cdot {L(H)_v}^e\rangle
-\langle \psib_e,i{L^T(\Hb)^e}_v \cdot \Phib^v \rangle\\
&&
\qquad\qquad
-\frac{1}{2}\langle \rho_e, T^e-2\nu^e \rangle
-\frac{1}{2}\langle \rhob_e,  \Tb_e - 2\nub^e \rangle
\bigg].
\eea
The action contains the constraints
\bea
\mu^v_{0} &=& F^v -\frac{g_{0,v}^2}{2}\Big(\zeta^v {\bf 1}_{N_v}- \sum_{e:s(e)=v} H^e\Hb^e + \sum_{e:t(e)=v} \Hb^e H^e\Big)
\omega,\\
\nu^e & =& 2\del_A \Hb^e,\\
\nub^e &=& 2\delb_A H^e,
\eea
which come from the D-term and F-term conditions.
The constraint 
$\mu^v_{0}$ in the supersymmetric action have the same form
as $\mu^v$ of the BPS vortex equations (\ref{original moment map 1}),
but are written in terms of different (controllable) coupling $g_{0,v}$.

Since the action is written in the $Q$-exact form, we can show that
the path integral is independent of an overall coupling $t$ of an action rescaling
\be
S \to t S.
\ee
So we can evaluate exactly the path integral of the supersymmetric theory by
the WKB (1-loop) approximation in the limit of $t\to \infty$.

In the 1-loop approximation, the path integral is localized at fixed points which are determined by the following equations
\bea
&&\mu_0^v = \nu^e=\nub^e=0,\label{NA fixed point eq 1}\\
&&\del_A \Phi^v = \delb_A\Phi^v = 0,\label{NA fixed point eq 2}\\
&&[\Phi^v,\Phib^v] =  0,\label{NA fixed point eq 3}\\
&& \Phi^v\cdot {L(H)_v}^e = {L^T(\Hb)^e}_v\cdot \Phi^v=0.\label{NA fixed point eq 4}
\eea

If we exclude the possibility of a mixed branch, Eqs.~(\ref{NA fixed point eq 1})-(\ref{NA fixed point eq 4})
have two kind of solutions; the Higgs branch $\langle H^e \rangle \neq0$ and
$\langle \Phi^v \rangle =0$, or the Coulomb branch $\langle H^e \rangle =0$ and
$\langle \Phi^v \rangle \neq0$.

In the Higgs brach, the solution to the fixed point equation is generically given by
the vortex solution at the coupling $g_{0,v}$.
On the other hand, a solution in the Coulomb branch breaks the gauge symmetry
at each vertex to $U(1)^{N_v}$ and $\Phi^v$ are diagonalized into
\be
\Phi^v = \diag(\phi_0^{v,1}, \phi_0^{v,2},\ldots,\phi_0^{v,N_v}),
\ee
where $\phi_0^{v,a}$ ($a=1,\ldots,N_v$) are constant on $\Sigma_h$.

If we tune the controllable coupling to $g_{0,v}=g_v$ in the Higgs branch,
the path integral is localized at the solution to the original BPS equations
and reduces to an integral over the vortex moduli space.
However, the path integral itself vanishes due to the existence of the fermion zero mode.

To save this, we need to insert a compensator of the fermion zero modes (volume operator)
\be
e^{i\beta{\cal I}_V(g_v)} = \exp\left\{ i\beta\int_{\Sigma_h}
\Tr\bigg[
\Phi_v\mu^v(g_v)
-\lambda_v\wedge \lambdab^v
+\frac{i}{2}\psi_e\psib^e \omega
\bigg]\right\}.
\ee
So we expect that the vev of the volume operator in the Higgs branch gives the volume of the moduli space 
of the BPS vortex by turning the coupling $g_{0,v}\to g_v$.

It is difficult to evaluate the above vev in the Higgs branch since we do not know the metric of the moduli space.
So we next try to evaluate the vev in the Coulomb branch picture.
Using the coupling independence, we can adjust the controllable coupling $g_{0,v}$ to a critical value $g_{c,v}$,
without changing the vev of the volume operator.

At the critical coupling $g_{c,v}$ in the Coulomb branch, after fixing a suitable gauge,
the path integral reduces to contour integrals\footnote{
For $\chi_h<0$ ($h>1$), the Vandermonde determinant of the non-Abelian gauge groups
provides extra poles.
The JK residue formula says that the singular hyper plane arrangement
should be ``projective'' for the extra poles
\cite{Szenes,Benini:2013xpa}.
To choose the correct poles from the Vandermonde determinant, we need to introduce the regularization
mass for the vector multiplet, for instance, by ``gauging'' a part of the $R$-symmetry \cite{Benini:2016hjo}.
}
\be
\begin{split}
\left\langle
e^{i\beta {\cal I}_V(g_v)}
\right\rangle^{g_{0,v}=g_{c,v}}_{\vec{k}^v}
&={\cal N}_C'\int \prod_{v\in V}\prod_{a=1}^{N_v}\frac{d\phi_0^{v,a}}{2\pi}
\frac{\prod_{v\in V}\prod_{a<b}\left(-i\left(\phi_0^{v,a}-\phi_0^{v,b}\right)\right)^{\chi_h}(\det \Omega)^h}
{\prod_{e\in E}\prod_{a,b}\left(-i\left(\phi_0^{s(e),a} - \phi_0^{t(e),b}\right)\right)^{k^{s(e),a}-k^{t(e),b}+\frac{1}{2}\chi_h}}\\
&\qquad\qquad\qquad\qquad\qquad\qquad
\times e^{-2\pi i \beta \sum_{v\in V}\sum_{a=1}^{N_v}\phi_0^{v,a}
B^{v,a}},
\end{split}
\label{Non-Abelian integral formula}
\ee
where $\vec{k}^v=(k^{v,1}, k^{v,2},\ldots,k^{v,N_v})\in \Z^{N_v}$
 is magnetic fluxes of the Cartan part of $U(N_v)$, and
\be
B^{v,a} \equiv \frac{\zeta^v \A}{4\pi} -\frac{k^{v,a}}{g_v^2}.
\ee

In the above integral formula, we should be careful with the calculation of $\det\Omega$.
To be precise, we need to calculate the exact 1-loop contribution from the matter fields,
but we will take a simplified approach here.
The denominator of the integral formula (\ref{Non-Abelian integral formula})
is the contribution from the matter field, which can be written as an Abelianized effective action in the Coulomb branch
\be
S_{\text{eff}} = \frac{1}{2\pi}\int_{\Sigma_h}\left[
\frac{\del W_{\text{eff}}(\phi)}{\del \phi^{v,a}}F^{v,a}-\frac{\del^2 W_{\text{eff}}(\phi)}{\del \phi^{v,a}\del \phi^{v',b}}\lambda^{v,a} \wedge \lambdab^{v',b}
\right],
\ee
where we need bi-linear terms of $\lambda^{v,a}$ and $\lambdab^{v,a}$, which are Abelian (Cartan) parts of
$\lambda^v$ and $\lambdab^v$,
to preserve the supersymmetry.

The superpotential $W_{\text{eff}}(\phi)$ is given by
\be
W_{\text{eff}}(\phi) = \sum_{e \in E}\sum_{a=1}^{N_{s(e)}}\sum_{b=1}^{N_{t(e)}}
\left(\phi^{s(e),a} - \phi^{t(e),b}\right)\left[\log \left(-i\left(\phi^{s(e),a} - \phi^{t(e),b}\right)\right)-1\right],
\ee
and using the supersymmetric transformation
\be
Q\lambda^{v,a} = -\del \phi^{v,a}, \quad Q\lambdab^{v,a} = -\delb \phi^{v,a},
\ee
we can see $QS_{\text{eff}} =0$.
For this Abelian effective theory, if we consider the localization again,
$\phi^{v,a}$ reduces to the constant zero modes $\phi_0^{v,a}$
and it reproduces the denominator in the integral formula (\ref{Non-Abelian integral formula}).

Since the volume operator originally contains the bi-linear term of $\lambda^{v,a}$ and $\lambdab^{v,a}$,
combining with the contribution from the 1-loop effective action and integrating out the fermion zero modes,
we obtain the determinant of $(\sum_{v\in V} N_v)\times (\sum_{v\in V} N_v)$ matrix
\be
\Omega_{(v,a),(v',b)} = \frac{\beta}{g_v^2}\delta_{vv'}\otimes \delta_{ab}
+\frac{i}{2\pi}\left.\frac{\del^2 W_{\text{eff}}(\phi)}{\del \phi^{v,a}\del \phi^{v',b}}\right|_{\phi=\phi_0}.
\label{Omega formula}
\ee
The determinants come from each handles of $\Sigma_h$. So we obtain $(\det \Omega)^h$.

As we have seen, the non-Abelian groups effectively decompose into the Abelian groups (Abelianization).
So it is useful to consider the decomposition of the non-Abelian vertices into
$\sum_{v\in V} N_v$ Abelian vertices $\tilde{v}=(v,a)$ ($v\in V$ and $a=1,\ldots, N_v$).
This decomposition of the vertices also expands the graph and
increases the number of the edges.
We denote the expanded graph by $\tilde{\Gamma}=(\tilde{V},\tilde{E})$.

\begin{figure}
\begin{center}
\includegraphics{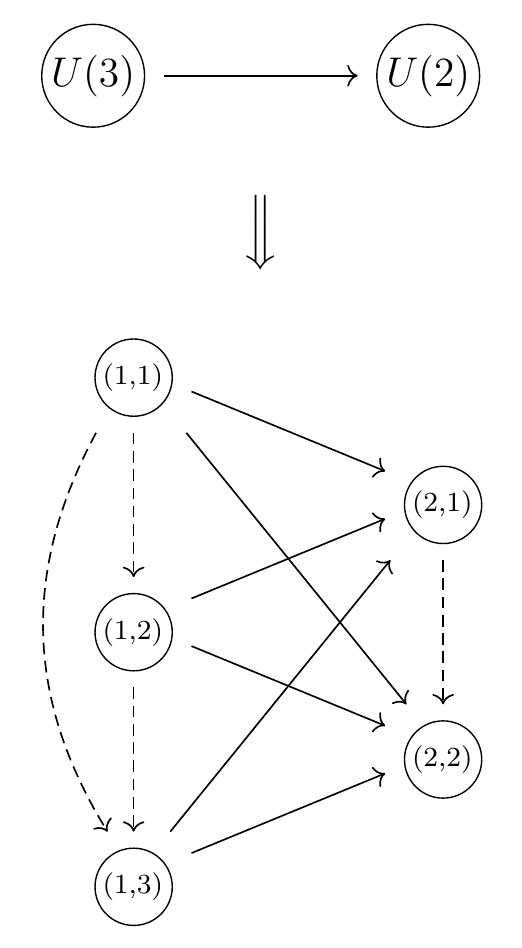}
\end{center}
\caption{An Abelian decomposition of $U(3)\times U(2)$ non-Abelian graph.
$U(3)$ and $U(2)$ non-Abelian vertices are decomposed into three Abelian vertices $\tilde{v}=(1,1), (1,2), (1,3)$ and
two vertices
$\tilde{v}=(2,1), (2,2)$, respectively,
and one edge is multiplexed to six edges.
The dashed arrows represent directed complete graphs inside the non-Abelian vertex,
which gives the Vandermonde determinant in the numerator of the integral formula.}
\label{Abelian decomposition}
\end{figure}

To explain this Abelianization of the graph, let us consider a concrete example
of two non-Abelian vertices ($v=1,2$) and one edge between them.
We assume $N_1=3$ and $N_2=2$; i.e.~$G=U(3)\times U(2)$ quiver gauge theory.
(See the upper in Fig.~\ref{Abelian decomposition}.)

In the original quiver diagram, there is one arrow between two vertices. So the incidence matrix is expressed
by $2\times 1$ matrix.
The Abelian decomposition now expands five vertices and six edges. So the incidence matrix becomes $5\times 6$
matrix as 
\be
{L_v}^e=\left(
\begin{array}{r}
1\\
-1
\end{array}
\right)
\longrightarrow
{\tilde{L}_{\tilde{v}}{}}^{\tilde{e}}=\left(
\begin{array}{rrrrrr}
1 & 1 & 0 & 0 & 0 & 0\\
0 & 0 & 1 & 1 & 0 & 0\\
0 & 0 & 0 & 0 & 1 & 1\\
-1 & 0 & -1 & 0 & -1 & 0\\
0 & -1 & 0 & -1 & 0 & -1
\end{array}
\right),
\ee
where $\tilde{L}$ is the expanded incidence matrix, and $\tilde{v}$ and $\tilde{e}$ are the indices of
decomposed vertices and edges.

Using these expanded vertices, edges and incidence matrix, we can express simply
the integral formula (\ref{Non-Abelian integral formula})
as well as the integral formula of the quiver gauge theory with the Abelian vertices
\begin{multline}
\left\langle
e^{i\beta {\cal I}_V(g_v)}
\right\rangle^{g_{0,v}=g_{c,v}}_{\vec{k}^v}\\
={\cal N}_C'\int \prod_{\tilde{v}\in \tilde{V}}\frac{d\phi_0^{\tilde{v}}}{2\pi}
\frac{\prod_{v\in V}\prod_{a<b}\left(-i\left(\phi_0^{v,a}-\phi_0^{v,b}\right)\right)^{\chi_h}(\det \Omega)^h}
{\prod_{\tilde{e}\in \tilde{E}}
\left(-i\phi_0^{\tilde{v}}{\tilde{L}_{\tilde{v}}{}}^{\tilde{e}}\right)^{k^{\tilde{v}}{\tilde{L}_{\tilde{v}}{}}^{\tilde{e}}
+\frac{1}{2}\chi_h}}
e^{-2\pi i \beta \sum_{\tilde{v}\in \tilde{V}}\phi_0^{\tilde{v}}
B^{\tilde{v}}},
\end{multline}
where
\be
\Omega_{\tilde{v}\tilde{v}'} \equiv \frac{\beta}{g_{\tilde{v}}^2}\delta_{\tilde{v}\tilde{v}'}
+\frac{1}{2\pi}\sum_{\tilde{e}\in \tilde{E}}
{\tilde{L}_{\tilde{v}}{}}^{\tilde{e}}\frac{1}{-i\phi_0^{\tilde{v}''}{\tilde{L}_{\tilde{v}''}{}}^{\tilde{e}}}{{\tilde{L}^T{}}^{\tilde{e}}}_{\tilde{v}'},
\ee
and we have set the gauge coupling to be the same as the original non-Abelian vertices like
$g_{\tilde{v}=(v,a)}=g_v$.

In addition to the Abelian decomposition of the edges,
we can consider extra edges inside the original non-Abelian vertices,
which we have depicted in Fig.~\ref{Abelian decomposition} as the dashed arrows.
These extra edges form a directed complete graph in each non-Abelian vertex
and are regarded as reproducing the Vandermonde determinant 
in the numerator of the integral formula, such that
\be
\prod_{v\in V}\prod_{a<b}\left(-i\left(\phi_0^{v,a}-\phi_0^{v,b}\right)\right)^{\chi_h}
=\prod_{\hat{e}\in \hat{E}}\left(
-i\phi_0^{\tilde{v}}{\hat{L}_{\tilde{v}}{}}^{\hat{e}}
\right)^{\chi_h},
\ee
where $\hat{E}$ is the dashed edges and $\hat{L}$ associated incidence matrix
 of the dashed graph $\hat{\Gamma}=(\tilde{V}, \hat{E})$.

Using the example in Fig.~\ref{Abelian decomposition}, the incidence matrix for $\hat{\Gamma}$
is explicitly given by
\be
{\hat{L}_{\tilde{v}}{}}^{\hat{e}}=
\left(
\begin{array}{rrrr}
1&0&1&0\\
-1&1&0&0\\
0&-1&-1&0\\
0&0&0&1\\
0&0&0&-1
\end{array}
\right).
\ee
Note here that the above incidence matrix is separated into $3\times 3$ and $2\times 1$ blocks,
which come from the two non-Abelian vertices.
 
Thus we can express the integral formula of the non-Abelian quiver gauge theory
in terms of the expanded Abelian graph with two kinds of the graphs $(\tilde{\Gamma},\hat{\Gamma})$.

\subsection{Applications}

\subsubsection*{\underline{Non-Abelian vortex with $N_f$-flavors}}

\begin{figure}
\begin{center}
\includegraphics{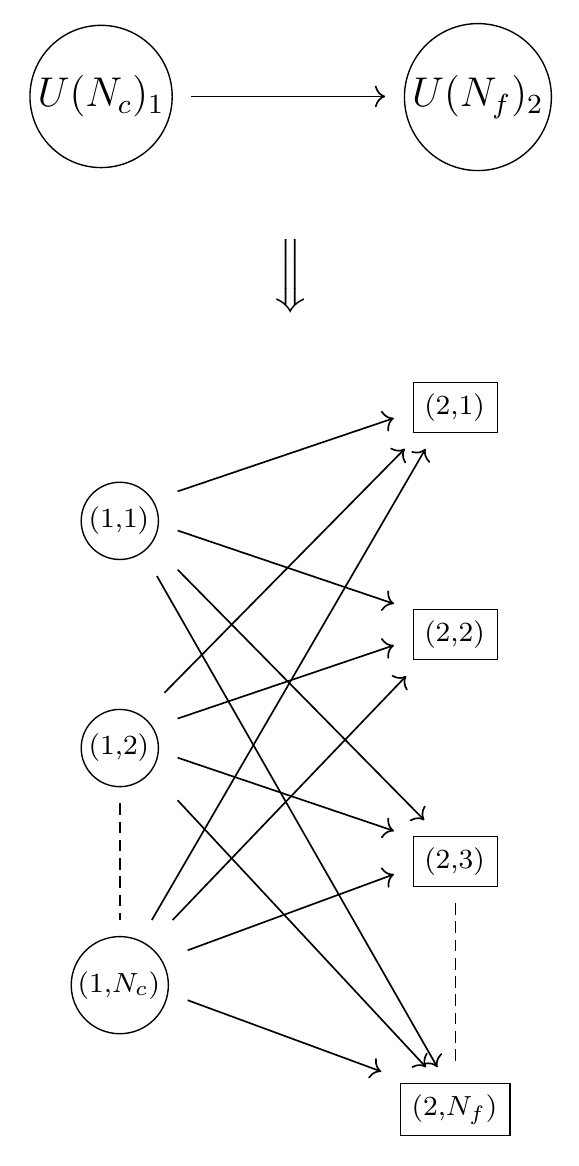}
\end{center}
\caption{The quiver diagram of two non-Abelian vertices, which 
has $U(N_c)_1\times U(N_f)_2$ symmetry.
$U(N_f)$ vertices will be decoupled 
and becomes a global symmetry. }
\label{non-Abelian two vertices}
\end{figure}

In order to check the integral formula for non-Abelian theory,
let us consider the case of two non-Abelian vertices.
One vertex has rank $N_c$ and another has rank $N_f$.
So we have $G=U(N_c)_1\times U(N_f)_2$ non-Abelian gauge theory at first.
The quiver diagram is depicted in the upper of Fig.~\ref{non-Abelian two vertices}.

The integral formula of this quiver gauge theory is given by
\begin{multline}
\left\langle
e^{i\beta {\cal I}_V(g_v)}
\right\rangle^{g_{0,v}=g_{c,v}}_{\vec{k}^1,\vec{k}^2}\\
={\cal N}_C'\int \prod_{a=1}^{N_c}\frac{d\phi_0^{1,a}}{2\pi}
\prod_{i=1}^{N_f}\frac{d\phi_0^{2,i}}{2\pi}
\frac{\prod_{a<b}\left(-i\left(\phi_0^{1,a}-\phi_0^{1,b}\right)\right)^{\chi_h}
\prod_{i<j}\left(-i\left(\phi_0^{2,i}-\phi_0^{2,j}\right)\right)^{\chi_h}(\det \Omega)^h}
{\prod_{a=1}^{N_c}\prod_{i=1}^{N_f}
\left(-i\left(\phi_0^{1,a}-\phi_0^{2,j}\right)\right)^{k^{1,a}-k^{2,j}
+\frac{1}{2}\chi_h}}\\
\times e^{-2\pi i \beta \left(\sum_{a=1}^{N_c}\phi_0^{1,a}
B^{1,a}+\sum_{i=1}^{N_f}\phi_0^{2,i}B^{2,i}\right)},
\end{multline}
where $\Omega$ is given by the formula (\ref{Omega formula}).
This integral formula is written in terms of the decomposed Abelian vertices. This can be obtained from the
quiver diagram shown as the lower of Fig.~\ref{non-Abelian two vertices}.

If we decouple 
one of the vertex by taking $g_2\to 0$,
$\phi_0^{2,i}$ are no longer integral variables, but replaced by a fixed constant,
which is denoted by $m$ (twisted mass for $H$).
The integral formula reduces to
\begin{multline}
\left\langle
e^{i\beta {\cal I}_V(g_v)}
\right\rangle^{g_{0,v}=g_{c,v}}_{\vec{k}^1}\\
={\cal N}_C'\int \prod_{a=1}^{N_c}\frac{d\phi_0^{1,a}}{2\pi}
\prod_{a<b}\left(-i\left(\phi_0^{1,a}-\phi_0^{1,b}\right)\right)^{\chi_h}
\prod_{a=1}^{N_c}\frac{
\left( \frac{\beta}{g_1^2}+\frac{N_f}{2\pi}\frac{1}{-i(\phi_0^{1,a}-m)}\right)^h}{
\left(-i\left(\phi_0^{1,a}-m\right)\right)^{N_f\left(k^{1,a}
+\frac{1}{2}\chi_h\right)}}\\
\times e^{-2\pi i \beta \sum_{a=1}^{N_c}\phi_0^{1,a}
B^{1,a}}.
\end{multline}
This reproduces the volume of the moduli space of the non-Abelian vortex with $N_f$ flavors on $\Sigma_h$
in the limit of $m\to 0$ \cite{Miyake:2011yr,Miyake:2011fq,Ohta:2018leq}.

\subsubsection*{\underline{Non-Abelian vortex in gauged non-linear sigma model}}

Using the above observations,
let us consider a non-Abelian generalization of the vortex in gauged non-linear sigma model
discussed in Sec.~\ref{GNLSM}.

\begin{figure}
\begin{center}
\includegraphics{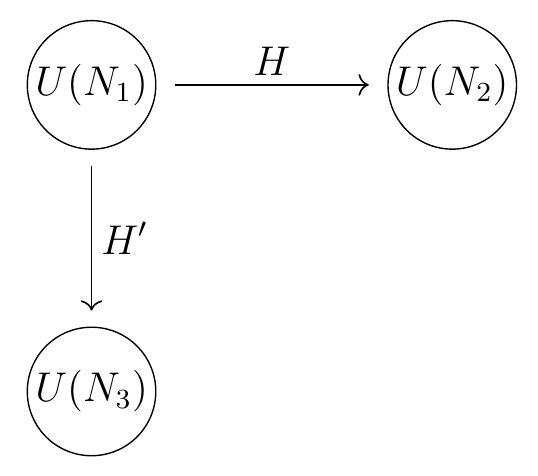}
\end{center}
\caption{The quiver diagram with three non-Abelian vertices
which is a parent of the gauged non-linear sigma model.
The third vertex will be decoupled 
by taking $g_3\to 0$.
Taking the strong coupling limit of the first vertex $g_1\to\infty$,
we get the non-linear sigma model.}
\label{parent NA GLSM}
\end{figure}

We first start with a quiver diagram of three non-Abelian vertices. There are two arrows from
first to second and from first to third vertex.
There exist bi-fundamental matters associated with the arrows. One is denoted by
$H$, which is $N_1\times N_2$ matrix, and another is denoted by $H'$, 
which is $N_1 \times N_3$ matrix. The quiver diagram is depicted in Fig.~\ref{parent NA GLSM}

If we consider a decoupling limit of gauge coupling of the third vertex by taking $g_3\to0$,
the third vertex is decoupled 
and gives $N_3$ flavors of $U(N_1)$ gauge theory at the first vertex.

After decoupling 
the third vertex, we obtain the BPS equations of the parent GLSM of gauged non-linear sigma model
as
\be
\begin{split}
\mu^1 &= F^1 - \frac{g_1^2}{2}\left( \zeta^1{\bf 1}_{N_1} - H \Hb
- H'\Hb'\right)
\omega=0, \\
\mu^2 &= F^2 - \frac{g_2^2}{2}\left( \zeta^2  {\bf 1}_{N_2}
+ \Hb H  \right)\omega=0,\\
\nu & = 2\left(\del \Hb - i\Hb A^1+iA^2\Hb\right)=0,\\
\nub &  = 2\left(\delb H+ i\Ab^1 H-i H \Ab^2\right)=0,\\
\nu' & = 2\left(\del \Hb' - i \Hb' A^1\right) = 0,\\
\nub' & = 2\left(\delb H'+ i\Ab^1 H' \right)=0.
\end{split}
\ee

Furthermore, if we take the strong coupling limit of the first vertex $g_1\to\infty$, we get a constraint
\be
H \Hb
+ H'\Hb'=\zeta^1{\bf 1}_{N_1}.
\ee
$H$ and $H'$ of a solution to the constraint parametrizes the Grassmann coset moduli space of vacua
\be
{\cal M}_{\text{Gr}} = \frac{U(N_2+N_3)}{U(N_1)\times U(N_2+N_3-N_1)}.
\ee
So we expect that the BPS equations $\mu^2=\nu=\nub=0$ represents the vortex equations
 with the target manifold ${\cal M}_{\text{Gr}}$.
 
The volume of the moduli space of the vortex in the parent model is given by the following integral formula
after decoupling 
the third vertex
\begin{multline}
\left\langle
e^{i\beta {\cal I}^\epsilon_V(g_v)}
\right\rangle^{g_{0,v}=g_{c,v}}_{\vec{k}^1,\vec{k}^2}\\
={\cal N}_C'\int \prod_{i=1}^{N_1}\frac{d\phi_0^{1,i}}{2\pi}
\prod_{a=1}^{N_2}\frac{d\phi_0^{2,a}}{2\pi}
\prod_{i<j}\left(-i\left(\phi_0^{1,i}-\phi_0^{1,j}\right)\right)^{\chi_h}
\prod_{a<b}\left(-i\left(\phi_0^{2,a}-\phi_0^{2,b}\right)\right)^{\chi_h}\\
\times
\frac{(\det \Omega_\e)^h \,
e^{-2\pi i \beta \left(\sum_{i=1}^{N_1}\phi_0^{1,i}
B^{1,i}+\sum_{a=1}^{N_2}\phi_0^{2,a}B^{2,a}\right)}
}
{\prod_{i=1}^{N_1}\prod_{a=1}^{N_2}
\left(-i\left(\phi_0^{1,i}-\phi_0^{2,a}\right)+\e\right)^{k^{1,i}-k^{2,a}
+\frac{1}{2}\chi_h}
\prod_{i=1}^{N_1}
\left(-i\phi_0^{1,i}+\e'\right)^{N_3 \left(k^{1,i}
+\frac{1}{2}\chi_h\right)}
},
\end{multline}
where
\be
\det \Omega_\e = \left(\frac{\beta}{g_2^2}
+\frac{1}{2\pi}\sum_{i=1}^{N_1}\sum_{a=1}^{N_2}\frac{1}{-i\left(\phi_0^{1,i}-\phi_0^{2,a}\right)+\e}\right)
\left(
\frac{\beta}{g_1^2}+\frac{\beta}{g_2^2}+\frac{N_3}{2\pi}\sum_{i=1}^{N_1}\frac{1}{-i\phi_0^i+\e'}
\right)-\frac{\beta^2}{g_2^4}.
\ee
It is complicated and difficult to perform explicitly the above contour integral for generic $N_1$, $N_2$ and $h$
due to the existence of the Vandermonde determinants.
We can perform the integral for smaller $N_1$ and $N_2$ or a 
special value of $h$. 

\section{Conclusion and Discussions}
\label{Conclusion and Discussions}

In this paper, we obtain the moduli space volume of the BPS 
vortex in the general quiver gauge theories.
We find that the existence of BPS vortices imposes a stringent 
constraint on possible quiver gauge theories. 
We find two alternative solutions to the constraint: 
universal gauge coupling case, and decoupled vertex case. 
Using localization method, we can express the volume of the 
moduli space by simple contour integrals and exactly evaluate 
the volume in principle. 
A number of examples of Abelian and 
non-Abelian quiver gauge theories are worked out. 
As an application of the quiver gauge theory with a decoupled 
vertex, we obtain the moduli space volume for a GLSM 
which serves as a parent theory for the GNLSM with $\C P^N$ 
target space with $n$ flavors of charge scalar fields. 
When restricted to $N=n=1$, our result agrees with the previous 
result \cite{Romao:2018egg} for $\C P^1$ GNLSM, in spite of the fact 
that our studies are based on a field theoretical scheme 
entirely different from the previous one.

In \cite{Romao:2018egg}, the total scalar curvature (integral of 
the scalar curvature over the moduli space)
is also evaluated by a similar formula to the volume.
This suggests that the total scalar curvature also can be 
evaluated by the localization formula in the
supersymmetric gauge theory.
So it is an interesting question to consider what cohomological operator 
in the supersymmetric gauge theory gives the total scalar curvature.

The volume of the vortex moduli space itself is proportional to 
the thermodynamical partition function
of the vortex. So we expect that free energy and equation of the 
state of the quiver vortex gas can be obtained
in the large vorticity limit with a fixed number density $k/\A$.
It is likely to be able to perform the contour integrals and 
evaluate the volume of the moduli space in this limit,
since we can sum up the contributions from all vorticity without 
violating the Bradlow bound.
The thermodynamics of the quiver vortices is interesting in order 
to explore interactions between various kinds of the vortices 
charged under the quiver gauge group.

The quiver gauge theories frequently appear in superstring theory, 
since they are regarded as the effective theory on the D-branes at 
the orbifold singularity.
The quiver diagram is associated with the Dynkin diagram of the discrete group
of the orbifolding. The gauge coupling of each quiver vertex
are common at the orbifold point.
So the constrains $\sum_{v\in V} B^v=0$ can be solved and there exist
the quiver vortices, which may be regarded as the D-brane bound states.
The analysis of the quiver gauge theory via the localization sheds lights on
the dynamics of the D-branes at the orbifold singularity.

The quiver gauge theory (quiver quantum mechanics) is also useful to interpret 
the D-particle system and multi-centered blackholes \cite{Denef:2002ru}.
The volume of the moduli space is closely related with the 
degeneracy (BPS index) of the BPS bound states \cite{Ohta:2014ria,Ohta:2015fpe}.
So the study of the volume of the moduli space of the BPS solitons 
is also important for understanding the BPS bound state of the 
superstring theory and supergravity.
The understanding of the contour integral in the Coulomb branch is 
closely related to the gravitational picture of the BPS states.
In this sense, it is also interesting to consider the large rank 
(large $N$) limit of the quiver gauge theory to see the holographic 
(supergravitational) interpretation of the volume of the moduli space.
The volume of the moduli space might give important implications 
in the study of the superstring theory and supergravity.

\section*{Acknowledgments}

N.~S.~would like to thank J.~M.~Speight for giving a stimulating 
seminar at Keio University.
This work is supported in part 
by Grant-in-Aid for Scientific Research (KAKENHI) (B) Grant Number 
17K05422 (K.~O.) and
18H01217 (N.~S.).

\appendix

\section{Moduli space metric and path-integral measure }
\label{sc:moduli_spce_metric}

One of the most commonly used definitions of moduli space metric 
is by means of a low-energy effective Lagrangian, on a 
background $\Phi_{\rm cl}^A(x)$ satisfying field equations 
\begin{equation}
\frac{\delta}{\delta \Phi^A(x,t)}\int dt d^dx {\cal L}(\Phi^A(x,t))=0, 
\label{eq:EOM}
\end{equation}
where $A$ denotes species of fields\footnote{
In this Appendix, $\Phi^A$ stands for the generic bosonic field 
in the low-energy effective theory, 
not for the bosonic scalar field in the vector multiplet in the main part. 
}, $\Phi_{\rm cl}^A(x)$ depends 
on $d$-dimensional spatial coordinates $x$, and the kinetic 
term of the Lagrangian density ${\cal L}$ is given in terms 
of the metric $g_{AB}$ 
in field 
space 
\begin{equation}
{\cal L}=\frac{1}{2}g_{AB}(\Phi)\partial_M \Phi^A\partial^M \Phi^B
+\cdots. 
\label{eq:EOM}
\end{equation}
Let us suppose that the solution has moduli parameters $p^i, i=1, \cdots, K$,
and $K$ is the dimension of the moduli space.
Assuming the moduli parameters vary in time slowly, 
we obtain the effective Lagrangian $L_{\rm eff}$ at the 
classical approximation by inserting $\Phi_{\rm cl}^A(x;p(t))$ into 
the action and integrating over $X$. 
We find 
\begin{equation}
L_{\rm eff}
=\int d^d x {\cal L}(\Phi^A_{\rm cl}(x; p^i(t))
=\frac{1}{2} G_{ij} \frac{d p^i}{dt}
\frac{d p^j}{dt}, 
\end{equation}
where the absence of the first order term in derivative 
$\frac{d p^i}{dt}$ is assured by the field equation \eqref{eq:EOM}. 
Since $p^i$ is just one possible parametrization of moduli, 
we have a general coordinate invariance in moduli space, 
corresponding to various choices of coordinate system in moduli space. 
In this sense, this $G_{ij}$ is called the moduli space metric 
and is given by 
\begin{equation}
G_{ij} = \int d^dx 
\frac{\partial \Phi_{\rm cl}^A}{\partial p^i}g_{AB}
\frac{\partial \Phi_{\rm cl}^B}{\partial p^j} 
\equiv \left\langle \frac{\partial \Phi_{\rm cl}}{\partial p^i}, 
\frac{\partial \Phi_{\rm cl}}{\partial p^j}\right\rangle. 
\label{eq:moduli_space_metric}
\end{equation}
We can regard the moduli space metric as a matrix constructed by 
an inner product of $i$-th and $j$-th infinite dimensional vectors 
whose components are labeled by $(A, x)$. 

To define the path-integral measure, we need to 
decompose field $\Phi^A$ into modes 
\begin{equation}
\Phi^A(x, t) = \Phi_{\rm cl}^A(x;p(t)) + \Phi_{\rm massive}^A(x,t), 
\end{equation}
where the massive part $\Phi_{\rm massive}^A$ should give 
higher derivative terms suppressed by powers of the mass gap 
if their path-integral is done. 
The zero mode (moduli) part $\Phi_{\rm cl}^A$ can be explicitly 
expressed in terms of an orthonormal basis of mode functions 
at each point in the moduli space 
\begin{equation}
\Phi_{\rm cl}^A(x; p(t)) = \sum_{a=1}^{K} \phi^a(t) \Phi_a^A(x), 
\label{eq:zero_mode}
\end{equation}
where the $i$-th effective field is denoted as $\phi^a(t), a=1,\cdots,K$, 
and $A$-th component of its mode function is denoted as 
$\Phi_a^A(X)$ which
are normalized as
\begin{equation}
\langle \Phi_a, \Phi_b\rangle \equiv 
\int d^dx \, \Phi_a^A(x) g_{AB}\Phi_b^B(x)
= \delta_{ab}, 
 \label{eq:orthnormal} 
\end{equation}
The effective fields $\phi^a$ gives an orthonormal coordinate 
at each point in moduli space, similarly to the local Lorentz 
frame of reference in general relativity. 
Inserting \eqref{eq:zero_mode} into 
\eqref{eq:moduli_space_metric}, we find 
\begin{equation}
G_{ij} = \frac{\partial \phi^a}{\partial p^i}
\frac{\partial \phi^a}{\partial p^j} 
. 
\label{eq:metric_coordinate_tr}
\end{equation}
It is most convenient to define the path-integral measure for 
zero modes in terms of the orthnormal coordinate system 
in the moduli space 
\begin{equation}
\left[\int D\Phi^A\right]_{\rm zero \; mode}=
\prod_{a=1}^K d\phi^a. 
 \label{eq:path-integral-orthnormal} 
\end{equation}
To use the moduli space parametrization in terms of a general 
parametrization $m(t)$, we introduce the ``vielbein'', namely the transformation 
matrix between the local orthonormal frame to the general 
coordinate frame of reference as 
\begin{equation}
E_i^a=\frac{\partial \phi^a}{\partial p^i}. 
 \label{eq:vielbein} 
\end{equation}
In fact, \eqref{eq:metric_coordinate_tr} shows  
\begin{equation}
E_i^a E_j^a= G_{ij}. 
 \label{eq:matric_vielbein} 
\end{equation}
Hence the determinant of the vielbein matrix gives the Jacobian 
of the change of variables from the 
orthonormal coordinates $\phi^a$ to the general coordinates $p^i$ 
\begin{equation}
\left[\int D\Phi^A\right]_{\rm zero \; mode}=
\prod_{i=1}^K dp^i \left|\det E_i^a\right| 
=
\prod_{i=1}^K dp^i \sqrt{\det G_{ij}}. 
 \label{eq:moduli-path-integral} 
\end{equation}
This is a general derivation of the volume form of the moduli space for the real coordinates.
Thus we obtain \eqref{eq:zero_mode_measure}
if we apply to the complex Hermitian (K\"ahler) manifold.



\begin{thebibliography}{99}

\bibitem{Bogomolny:1975de}
  E.~B.~Bogomolny,
  Sov.\ J.\ Nucl.\ Phys.\  {\bf 24} (1976) 449
  [Yad.\ Fiz.\  {\bf 24}  (1976) 861]; 

\bibitem{Prasad:1975kr}
  M.~K.~Prasad and C.~M.~Sommerfield,
  Phys.\ Rev.\ Lett.\  {\bf 35} (1975) 760.

\bibitem{Bradlow:1990ir} 
  S.~B.~Bradlow,
  Commun.\ Math.\ Phys.\  {\bf 135}, 1 (1990).
  doi:10.1007/BF02097654


\bibitem{Manton:2010sa}
N.~S.~Manton and N.~M.~Romao,
J. Geom. Phys. \textbf{61} (2011), 1135-1155
doi:10.1016/j.geomphys.2011.02.017
[arXiv:1010.0644 [hep-th]].


\bibitem{MantonSutcliffe}
N.~S.~Manton and P.~Sutcliffe, 
``Topological Solitons,'' 
Cambridge University Press (Cabridge, UK), 2004.


\bibitem{Eto:2006pg} 
  M.~Eto, Y.~Isozumi, M.~Nitta, K.~Ohashi and N.~Sakai,
  J.\ Phys.\ A {\bf 39}, R315 (2006)
  doi:10.1088/0305-4470/39/26/R01
  [hep-th/0602170].

\bibitem{Manton:1993tt}
N.~S.~Manton,
Nucl. Phys. B \textbf{400} (1993), 624-632
doi:10.1016/0550-3213(93)90418-O

\bibitem{Manton:1998kq} 
  N.~S.~Manton and S.~M.~Nasir,
  Commun.\ Math.\ Phys.\  {\bf 199}, 591 (1999)
  doi:10.1007/s002200050513
  [hep-th/9807017].

\bibitem{Eto:2007aw} 
  M.~Eto, T.~Fujimori, M.~Nitta, K.~Ohashi, K.~Ohta and N.~Sakai,
  Nucl.\ Phys.\ B {\bf 788}, 120 (2008)
  doi:10.1016/j.nuclphysb.2007.06.020
  [hep-th/0703197].


\bibitem{Fujimori:2010fk} 
  T.~Fujimori, G.~Marmorini, M.~Nitta, K.~Ohashi and N.~Sakai,
  Phys.\ Rev.\ D {\bf 82}, 065005 (2010)
  doi:10.1103/PhysRevD.82.065005
  [arXiv:1002.4580 [hep-th]].


\bibitem{Moore:1997dj}
  G.~W.~Moore, N.~Nekrasov and S.~Shatashvili,
  Commun.\ Math.\ Phys.\  {\bf 209} (2000) 97
  doi:10.1007/PL00005525
  [hep-th/9712241].

\bibitem{Gerasimov:2006zt}
  A.~A.~Gerasimov and S.~L.~Shatashvili,
  Commun.\ Math.\ Phys.\  {\bf 277} (2008) 323
  doi:10.1007/s00220-007-0369-1
  [hep-th/0609024].


\bibitem{Miyake:2011yr}
  A.~Miyake, K.~Ohta and N.~Sakai,
  Prog.\ Theor.\ Phys.\  {\bf 126}, 637 (2011)
  doi:10.1143/PTP.126.637
  [arXiv:1105.2087 [hep-th]].

\bibitem{Miyake:2011fq} 
  A.~Miyake, K.~Ohta and N.~Sakai,
  J.\ Phys.\ Conf.\ Ser.\  {\bf 343}, 012107 (2012)
  doi:10.1088/1742-6596/343/1/012107
  [arXiv:1111.4333 [hep-th]].


\bibitem{Ohta:2018leq}
  K.~Ohta and N.~Sakai,
  PTEP {\bf 2019}, no. 4, 043B01 (2019)
  doi:10.1093/ptep/ptz016
  [arXiv:1811.03824 [hep-th]].


\bibitem{Yang:1998qca}
Y.~Yang,
Phys. Rev. Lett. \textbf{80} (1998), 26-29
doi:10.1103/PhysRevLett.80.26

\bibitem{Baptista:2004rk}
J.~M.~Baptista,
Commun. Math. Phys. \textbf{261} (2006), 161-194
doi:10.1007/s00220-005-1444-0
[arXiv:math/0411517 [math.DG]].

\bibitem{Baptista:2007ap}
J.~M.~Baptista,
JHEP \textbf{02} (2008), 096
doi:10.1088/1126-6708/2008/02/096
[arXiv:0707.2786 [hep-th]].

\bibitem{Baptista:2008ex}
J.~M.~Baptista,
Commun. Math. Phys. \textbf{291} (2009), 799-812
doi:10.1007/s00220-009-0838-9
[arXiv:0810.3220 [hep-th]].

\bibitem{Baptista:2010rv}
J.~M.~Baptista,
Nucl. Phys. B \textbf{844} (2011) 308,
doi:10.1016/j.nuclphysb.2010.11.005
[arXiv:1003.1296 [hep-th]].


\bibitem{Romao:2018egg}
  N.~M.~Romao and J.~M.~Speight,
  arXiv:1807.00712 [math.DG].


\bibitem{Schroers:1996zy}
B.~J.~Schroers,
Nucl. Phys. B \textbf{475} (1996), 440-468
doi:10.1016/0550-3213(96)00348-3
[arXiv:hep-th/9603101 [hep-th]].

\bibitem{Kan:2009tu}
N.~Kan, K.~Kobayashi and K.~Shiraishi,
Phys. Rev. D \textbf{80} (2009), 045005
doi:10.1103/PhysRevD.80.045005
[arXiv:0901.1168 [hep-th]].


\bibitem{Witten:1990bs}
  E.~Witten,
  Int.\ J.\ Mod.\ Phys.\ A {\bf 6} (1991) 2775.
  doi:10.1142/S0217751X91001350

\bibitem{Jeffrey-Kirwan}
L.~C.~Jeffrey and F.~C.~Kirwan,
Topology {\bf 34} (1995) 291
doi:10.1016/0040-9383(94)00028-J
[arXiv:alg-geom/9307001].

\bibitem{Bullimore:2019qnt}
M.~Bullimore, A.~E.~V.~Ferrari and H.~Kim,
[arXiv:1912.09591 [hep-th]].

\bibitem{Bullimore:2020nhv}
M.~Bullimore, A.~E.~V.~Ferrari, H.~Kim and G.~Xu,
[arXiv:2007.11603 [hep-th]].


\bibitem{Blau:1994et}
M.~Blau and G.~Thompson,
Nucl. Phys. B \textbf{439} (1995), 367-394
doi:10.1016/0550-3213(95)00058-Z
[arXiv:hep-th/9407042 [hep-th]].


\bibitem{Blau:1995rs}
M.~Blau and G.~Thompson,
J. Math. Phys. \textbf{36} (1995), 2192-2236
doi:10.1063/1.531038
[arXiv:hep-th/9501075 [hep-th]].

\bibitem{Closset:2015rna}
C.~Closset, S.~Cremonesi and D.~S.~Park,
JHEP \textbf{06} (2015), 076
doi:10.1007/JHEP06(2015)076
[arXiv:1504.06308 [hep-th]].

\bibitem{Benini:2016hjo}
F.~Benini and A.~Zaffaroni,
Proc. Symp. Pure Math. \textbf{96} (2017), 13-46
[arXiv:1605.06120 [hep-th]].


\bibitem{Szenes}
A.~Szenes and M.~Vergne,
Invent. Math. \textbf{158} no. 3, (2004) 453–495,
[arXiv:math/0306311 [math.AT]].


\bibitem{Benini:2013xpa}
F.~Benini, R.~Eager, K.~Hori and Y.~Tachikawa,
Commun. Math. Phys. \textbf{333} (2015) no.3, 1241-1286
doi:10.1007/s00220-014-2210-y
[arXiv:1308.4896 [hep-th]].


\bibitem{Denef:2002ru}
  F.~Denef,
  JHEP {\bf 0210} (2002) 023
  doi:10.1088/1126-6708/2002/10/023
  [hep-th/0206072].

\bibitem{Ohta:2014ria}
K.~Ohta and Y.~Sasai,
JHEP \textbf{1411} (2014), 123
doi:10.1007/JHEP11(2014)123
[arXiv:1408.0582 [hep-th]].
  
\bibitem{Ohta:2015fpe}
  K.~Ohta and Y.~Sasai,
  JHEP {\bf 1602} (2016) 106
  doi:10.1007/JHEP02(2016)106
  [arXiv:1512.00594 [hep-th]].





\end{thebibliography}
\end{document}